\documentclass[11pt]{article}

\usepackage{color}
\usepackage{float}
\restylefloat{table}

\usepackage[title,titletoc,toc]{appendix}

\usepackage{bigints}
\usepackage{amsthm}
\usepackage{graphicx}
\usepackage{pdfsync}
\usepackage{dsfont}
\usepackage{mathrsfs}
\usepackage{amssymb}
\usepackage{mathptmx}

\usepackage{pgf,tikz}
\usepackage{pgfplots}
\usetikzlibrary{automata}
\usetikzlibrary{arrows}
\usetikzlibrary{matrix,positioning}
\usetikzlibrary{snakes}
\usetikzlibrary{decorations.pathreplacing}

\usetikzlibrary{decorations.markings}
\usetikzlibrary{decorations.shapes}
\tikzset{decoration={brace}}

\newcommand{\stateOne}{%
\begin{tikzpicture}
  \draw[thick] (2,-2) -- (2,2) -- (-2,2) -- (-2,-2) -- (2,-2);
  \draw[thick] (-1.2,-1) -- (1.2,-1);
  \filldraw (-0.95,-1) circle (0.2);
  \filldraw (0.95,-1) circle (0.2);
  \draw[thick] (-1.2,1) -- (1.2,1);
  \filldraw (-0.95,1) circle (0.2);
  \filldraw (0.95,1) circle (0.2);
\end{tikzpicture}%
}

\newcommand{\stateTwo}{%
\begin{tikzpicture}
  \draw[thick] (2,-2) -- (2,2) -- (-2,2) -- (-2,-2) -- (2,-2);  
  \draw[thick] (-1.2,-1) -- (1.2,-1);
  \filldraw (-0.95,-1) circle (0.2);
  \filldraw (0.95,-1) circle (0.2);
  \draw[thick] (-1.2,1) -- (0,1);
  \filldraw (-0.95,1) circle (0.2);
  \draw[thick] (0,0) -- (1.2,0);
  \filldraw (0.95,0) circle (0.2);
\end{tikzpicture}%
}

\newcommand{\stateThree}{%
\begin{tikzpicture}
  \draw[thick] (2,-2) -- (2,2) -- (-2,2) -- (-2,-2) -- (2,-2);
  \draw[thick] (-1.2,-1) -- (0,-1);
  \filldraw (-0.95,-1) circle (0.2);
  \draw[thick] (-1.2,0.33) -- (0,0.33);
  \filldraw (-0.95,0.33) circle (0.2);
  \draw[thick] (0,-0.33) -- (1.2,-0.33);
  \filldraw (0.95,-0.33) circle (0.2);
  \draw[thick] (0,1) -- (1.2,1);
  \filldraw (0.95,1) circle (0.2);
\end{tikzpicture}%
}

\newcommand{\stateFour}{%
\begin{tikzpicture}
  \draw[thick] (2,-2) -- (2,2) -- (-2,2) -- (-2,-2) -- (2,-2);
  \draw[thick] (-1.2,-1) -- (1.2,-1);
  \filldraw (-0.95,-1) circle (0.2);
  \draw (0.95,-1) node[anchor=center,scale=10]{$\times$};
  \draw[thick] (-1.2,1) -- (0,1);
  \filldraw (-0.95,1) circle (0.2);
\end{tikzpicture}%
}

\newcommand{\stateFive}{%
\begin{tikzpicture}
  \draw[thick] (2,-2) -- (2,2) -- (-2,2) -- (-2,-2) -- (2,-2);
  \draw[thick] (-1.2,-1) -- (1.2,-1);
  \draw (-0.95,-1) node[anchor=center,scale=10]{$\times$};
  \filldraw (0.95,-1) circle (0.2);
  \draw[thick] (0,1) -- (1,1);
  \filldraw (0.95,1) circle (0.2);
\end{tikzpicture}%
}

\newcommand{\stateSix}{%
\begin{tikzpicture}
  \draw[thick] (2,-2) -- (2,2) -- (-2,2) -- (-2,-2) -- (2,-2);
  \draw[thick] (-1.2,-1) -- (0,-1);
  \filldraw (-0.95,-1) circle (0.2);
  \draw[thick] (-1.2,1) -- (0,1);
  \filldraw (-0.95,1) circle (0.2);
  \draw[thick] (0,0) -- (1.2,0);
  \draw (0.95,0) node[anchor=center,scale=10]{$\times$};
\end{tikzpicture}%
}

\newcommand{\stateSeven}{%
\begin{tikzpicture}
  \draw[thick] (2,-2) -- (2,2) -- (-2,2) -- (-2,-2) -- (2,-2);
  \draw[thick] (0,-1) -- (1.2,-1);
  \filldraw (0.95,-1) circle (0.2);
  \draw[thick] (-1.2,0) -- (0,0);
  \draw (-0.95,0) node[anchor=center,scale=10]{$\times$};
  \draw[thick] (0,1) -- (1.2,1);
  \filldraw (1,1) circle (0.2);
\end{tikzpicture}%
}

\newcommand{\stateEight}{%
\begin{tikzpicture}
  \draw[thick] (2,-2) -- (2,2) -- (-2,2) -- (-2,-2) -- (2,-2);
  \draw[thick] (0,0.7) -- (1.2,0.7);
  \draw (0.95,0.7) node[anchor=center,scale=10]{$\times$};
  \draw[thick] (-1.2,-0.7) -- (0,-0.7);
  \draw (-0.95,-0.7) node[anchor=center,scale=10]{$\times$};
\end{tikzpicture}%
}

\newcommand{\stateNine}{%
\begin{tikzpicture}
  \draw[thick] (2,-2) -- (2,2) -- (-2,2) -- (-2,-2) -- (2,-2);
  \draw[thick] (-1.2,0) -- (1.2,0);
  \draw (-0.95,0) node[anchor=center,scale=10]{$\times$};
  \draw (0.95,0) node[anchor=center,scale=10]{$\times$};
\end{tikzpicture}%
}

\newcommand{\treeOne}{
\begin{tikzpicture}[scale=0.5]
  \draw (0,2.5) circle(0.5);
  \node at (0,2.5) {A};
  \draw[thick] (0,0) -- (2,3) -- (3,0);
  \draw[thick] (1,0) -- (0.6,0.9);
  \draw[thick] (2,0) -- (1.6,2.4);
\end{tikzpicture}%
}

\newcommand{\treeTwo}{
\begin{tikzpicture}[scale=0.5]
  \draw (0,2.5) circle(0.5);
  \node at (0,2.5) {B};
  \draw[thick] (0,0) -- (0.5,1) -- (1,0);
  \draw[thick] (2,0) -- (2.5,2) -- (3,0);
  \draw[thick] (0.5,1) -- (1.5,3) -- (2.5,2);
\end{tikzpicture}%
}

\newcommand{\treeThree}{
\begin{tikzpicture}[scale=0.5]
  \draw (0,2.5) circle(0.5);
  \node at (0,2.5) {C};
  \draw[thick] (0,0) -- (1,1) -- (2.5,3) -- (3,0);
  \draw[thick] (1,0) -- (1,1) -- (2,0);
\end{tikzpicture}%
}

\newcommand{\treeFour}{
\begin{tikzpicture}[scale=0.5]
  \draw (0,2.5) circle(0.5);
  \node at (0,2.5) {D};
  \draw[thick] (0,0) -- (1,1) -- (2,3) -- (3,0);
  \draw[thick] (1,0) -- (1,1);
  \draw[thick] (2,0) -- (2,3);
\end{tikzpicture}%
}

\newcommand{\treeFive}{
\begin{tikzpicture}[scale=0.5]
  \draw (0,2.5) circle(0.5);
  \node at (0,2.5) {E};
  \draw[thick] (0,0) -- (1.5,3) -- (3,0);
  \draw[thick] (1,0) -- (1.5,3) -- (2,0);
\end{tikzpicture}%
}

\usepackage{bm}
\newtheorem{theorem}{Theorem}[section]

\newtheorem{algorithm}[theorem]{Algorithm}
\newtheorem{definition}[theorem]{Definition}

\theoremstyle{definition}
\newtheorem{example}[theorem]{Example}

\theoremstyle{remark}
\usepackage{tensor}
\usepackage{comment}

\newcommand{\Exp}{\mathds{E}}

\newcommand{\Prob}{\mathds{P}}

\newcommand{\vect}[1]{\vec{#1}}
\renewcommand{\vect}[1]{\boldsymbol{\bm #1}}
\newcommand{\mat}[1]{\boldsymbol{\bm #1}}

\renewcommand{\matrix}[2][ccccccccccccccccccccc]{\left(\begin{array}{#1}#2
      \\ \end{array} \right)}

\begin{document}
\title{Phase-type distributions in population genetics}
\date{\today}
\author{Asger Hobolth$^1$, Arno Siri-J\'egousse$^2$ and Mogens Bladt$^{3}$,\\
\small
1. Aarhus University, Bioinformatics Research Center; asger@birc.au.dk \\
\small
2. UNAM, IIMAS, Departamento de Probabilidad y Estad\'istica;
arno@sigma.iimas.unam.mx \\
\small
3. University of Copenhagen, Department of Mathematical Sciences;
bladt@math.ku.dk}
\maketitle
\section*{Abstract}
Probability modelling for DNA sequence evolution is well established and provides a rich framework for understanding genetic variation between samples of individuals from one or more populations. We show that both classical and more recent models for coalescence (with or without recombination) can be described in terms of the so-called phase-type theory, where complicated and tedious calculations are circumvented by
the use of matrices. 
The application of phase-type theory consists of describing the stochastic model as a Markov model by appropriately setting up a state space and calculating the corresponding intensity and reward matrices. Formulae of interest are then expressed in terms of these aforementioned matrices. We illustrate this by a few examples calculating the mean, variance and even higher order moments of the site frequency spectrum in the multiple merger coalescent models, and by analysing the mean and variance for the number of segregating sites for multiple samples in the two-locus ancestral recombination graph.
We believe that phase-type theory has great potential as a tool for analysing probability models in population genetics. The compact matrix notation is useful for clarification of current models, in particular their formal manipulation (calculation), but also for further development or extensions.  
\section*{Keywords} Coalescent theory, multiple merger, phase-type theory, recombination.  
\section{Introduction}
Queueing and (collective) risk theories, as we know them today, both originates around the same year of 1909 with the works of Erlang and Lundberg, but it was not until 1961 that Prabhu~\cite{Prabhu-1961} recognized the connection (duality) between the two theories. In spite of this discovery, the interaction between the two theories remained rather limited, and it was not until the 1970's that the development of the theory of phase-type distributions, also based on Erlang's earlier work, was applied to both queueing and risk models in the following decades.
The theory is characterized by the use of matrices instead of performing calculations based on individual states, and expressions for functionals of interest are expressed in terms of functions of matrices which 
are both simpler, more transparent and easier to implement.  Phase-type theory is recently summarized in the monograph \cite{bladt-nielsen-2017}. 

Coalescent theory was formulated in the 1980's by John Kingman. We refer to Chapter~1 in \cite{wakeley2008coalescent} for a brief account of the history of the coalescent. Coalescent theory is a mathematical model for genetic variation within and between species, and is a backward-in-time description of the forward evolutionary model of Sewall Wright and Ronald Fisher (e.g. Chapter 3.1.1 in \cite{wakeley2008coalescent}). The Wright-Fisher model and its extensions describe the evolutionary forces that shape genetic variation. The most important forces include random genetic drift, mutation, recombination, migration and selection. The coalescent with multiple mergers (or $\Lambda$-coalescents) was introduced by Pitman \cite{Pit} and Sagitov \cite{Sag}.
Apart from Kingman's coalescent, which is also an element of this family, they permit more than two lineages to merge at a coalescence event.
Multiple merger coalescent models applies when the variability in the reproduction success is large \cite{MS, EW, Sch03}, and they can also be used for populations under strong selection \cite{Des, NH, Sch17}.
They can also be helpful to integrate uncertainty in a phylogeny.
They have been widely studied during the past years but most of the theoretical results are asymptotic in terms of sample size \cite{BG, Drm, BBS, DDS, Ker, DK}. 

In this paper we demonstrate the similarities between coalescent theory and phase-type theory, and discuss the implications of the close connection. In particular we show that the translation is useful because complex and difficult-to-derive coalescent theory formulae and equations are easy to define and calculate using phase-type theory and matrix notation.
The age of the most recent common ancestor (the height of the tree), the site frequency spectrum (related to the branch lengths and branching pattern of the tree), and the number of segregating sites in two neighbouring loci are examples. 

We develop our method for a variety of examples inspired by different scenarios of evolution.
Explicit sampling formulae for expected frequency spectra are known only  in very few cases (mainly for Kingman's coalescent \cite{FL} and Bolthausen-Sznitman coalescent \cite{NH}).
Recently, \cite{Blath} developed an iterative method to compute cross moments of the site frequency spectrum in coalescent models with multiple mergers.
The phase-type approach provides an alternative tool to obtain those results. Furthermore, we are able to compute the Laplace transform, and therefore we can easily compute higher-order moments. We are also able to derive densities of the height and the total branch length of the coalescent trees.
These applications are similar to the analysis of genealogical histories in structured populations in \cite{KumagaiUyenoyama2015}. 
We illustrate our results by providing explicit formulae for the mean and covariance of the site frequency spectrum for Kingman's coalescent and for two models for populations with a skewed offspring number.

The rest of the paper is organized as follows. In Section \ref{sec:PH} we review the relevant phase-type theory and relate it to basic coalescent models like Kingman's coalescent and a peripatric coalescent \cite{LM} (also known as the seed-bank coalescent \cite{BGKW}), where lineages can be active or inactive and switch from one state to another. In Section \ref{sec:norecomb} we consider examples of coalescent models without recombination, where we provide a detailed account on the construction of the phase-type model for Kingman's coalescent of general order, and the calculations of (joint) moments for the site frequency spectrum of general $\Lambda$-coalescent processes. Phase-type descriptions of coalescent models with recombination is the theme in Section \ref{sec:recomb}, where we obtain explicit formulae for the joint distribution of tree height, and explicit formulae for expected values and covariances of the tree height and total branch lengths. Finally we conclude the paper by a short discussion.

\section{Phase-type distributions}\label{sec:PH}
Phase-type distributions is a rather general class of distributions for positive random variables which includes mixtures and convolutions of exponential distributions. For example, the tree height and total tree length of the ancestral tree in the fundamental coalescent model, both with or without recombination, are examples of phase-type distributions. In the presence of recombination we also identify some more complicated situations involving  joint distributions which naturally fall into a class of multivariate phase-type distributions. 

\subsection{Definition and examples}
The following notational conventions are standard for phase-type distributions and will be used throughout unless otherwise stated. Matrices are written in bold majuscules (e.g. $\mat{S}$ and $\mat{\Lambda}$), column vectors in bold, roman minuscules (e.g. $\mat{s}$ and $\mat{t}$) while row vectors are bold, greek minuscules (e.g. $\vect{\alpha}$ and $\vect{\beta}$).
Elements of vectors and matrices are denoted by their corresponding minuscule letters (e.g. $\vect{\alpha}=(\alpha_i)_i$ and $\mat{S}=\{ s_{ij}\}_{i,j}$). Dimensions are usually not explicitly stated unless needed. In particular, the identity matrix $\mat{I}$, the (column) vector of ones $\vect{e}=(1,1,\dotsc,1)^\prime$ and the $i$th unit (column) vector $\vect{e}_i=(0,\dotsc,0,1,0,\dotsc,0)^\prime$ ($1$ on the
$i$th location) may be of any appropriate dimension which should be clear from the context.

Consider a Markov jump process (continuous time Markov chain) $\{ X_t\}_{t\geq 0}$ with finite state-space $\{ 1,2,...,p,p+1\}$, where states $1,...,p$ are transient and state $p+1$ is absorbing. This means that $\{ X_t\}_{t\geq 0}$ has an intensity (rate) matrix $\mat{\Lambda}$ of the form
\[ \mat{\Lambda} =\begin{pmatrix}
\mat{S} & \vect{s} \\
\vect{0} & 0
\end{pmatrix} , \] 
where we refer to the $p\times p$ sub-matrix of rates between the transient states, $\mat{S} =\{ s_{ij}\}_{i,j=1,...,p}$, 
 as a {\it sub-intensity} matrix, the $p$-dimensional column vector $\vect{s}=(s_i)_{i=1,...,p}$  as an {\it exit rate} vector (since its elements are the intensities for jumping to the absorbing state) and where $\vect{0}$ is a $p$-dimensional row vector of zeros.

Assume that $\{ X_t\}_{t\geq 0}$ can only start in a transient state and let $\vect{\alpha}=(\alpha_1,...,\alpha_p)$ where  $\alpha_i=\Prob (X_0=i)$, $i=1,...,p$. Then $\vect{\alpha}\vect{e}=\sum_{i=1}^p \alpha_i=1$ and $\vect{\alpha}$ is a probability vector on the set of transient states $E=\{ 1,2,\dotsc,p\}$. Since $\mat{\Lambda}$ is an intensity matrix, then its rows must sum to zero (i.e. $\mat{\Lambda}\vect{e}=\vect{0}$ where $\vect{0}$ is now the column vector of zeros) so $\vect{s}=-\mat{S}\vect{e}$. Hence the specification of a sub-intensity matrix $\mat{S}$ implies the form of the exit rate vector $\vect{s}$.   

We recall (from the forward and backward differential equations of Kolmogorov) that the corresponding transition matrix $\mat{P}^t=\{ p_{ij}^t\}_{i,j=1,...,p+1}$ is given by 
\[  \mat{P}^t = e^{\mat{\Lambda}t} = \sum_{n=0}^\infty \frac{\mat{\Lambda}^nt^n}{n!}  . \]
By using the fact that $\vect{s}=-\mat{S}\vect{e}$ it is easily proved that
\begin{equation}
  \mat{P}^t = \begin{pmatrix}
e^{\mat{S}t} & \vect{e}-e^{\mat{S}t}\vect{e} \\
\vect{0} & 1   
\end{pmatrix} . \label{eq:exp-of-Lambda}
 \end{equation}
Hence the restriction of $\mat{P}^t$ to the transient states set $E$ is simply $\exp (\mat{S}t)$.

\begin{definition}[Phase-type distribution]
The time until absorption
 \[  \tau = \inf\{ t>0 : X_t = p+1  \} \]
is said to have a {\it phase-type distribution} of order $p$ with {\it phase-space} $E=\{1,2,\dotsc,p\}$, {\it initial distribution} $\vect{\alpha}$ and {\it sub-intensity} ({\it generator}) matrix $\mat{S}$, and we write 
\[ \tau\sim \mbox{PH}_p(\vect{\alpha},\mat{S}) . \]
The exit rate vector will always be denoted by a bold minuscule letter corresponding to the letter for the generator, here $\vect{s}$.
\end{definition}
Let $0<S_1<S_2<\dotsc $ denote the jump times of $\{ X_t\}_{t\geq 0}$ and $T_n=S_n-S_{n-1}$, $n=1,2,\dots$ ($S_0=0$) the corresponding inter-arrival times. Furthermore we define the discrete time process $Y_n=X_{S_n}$, $n=0,1,....$ which keeps track of the states visited. Then $\{ Y_n\}_{n\in\mathds{N}}$ is a Markov chain on $\{ 1,2,...,p+1\}$ with transition probability matrix $\mat{Q}=\{ q_{ij}\}$, say, and referred to as the {\it embedded} Markov chain. Conditionally on $Y_{n-1}=i$, $T_{n}=S_n-S_{n-1}$ has an exponential distribution with parameter $\lambda_{i}=-\lambda_{ii}$. For $i\neq j$, $i,j=1,...,p$, set
\[   \lambda_{ij}=\lambda_i q_{ij} ,  \]
which suggests the important interpretation of $\lambda_{ij}dx$ being the probability of a jump from $i$ to $j$ during a small time interval $[x,x+dx)$. 

\begin{figure}[H]
\begin{tikzpicture}[scale=0.7,domain=-1:10]
\draw[->] (0,0) -- (15.2,0) node[right] {$t$};
\draw[-] (0,0) --(0,3);
\draw[-,dashed] (0,3)--(0,4);
 \draw[->] (0,4) -- (0,6) node[above] {$X_t$};

\foreach \y/\ytext in {0.5/1, 1/2, 1.5/3, 2/4,2.5/5,4.5/p}
\draw[color=blue,shift={(0,\y)}] (-2pt,0pt) -- (2pt,0pt) node[left] {$\ytext$};
\foreach \y/\ytext in {5/p+1}
\draw[color=purple,shift={(0,\y)}] (-2pt,0pt) -- (2pt,0pt) node[left] {$\ytext$};

\foreach \x/\xtext in {2/S_1,3/S_2,4.5/S_3,6/S_4,9/S_5,10.5/S_6,13/\tau}
\draw[shift={(\x,0)}] (0pt,-2pt) -- (0pt,2pt) node[below] {$\xtext$};

\draw[color=blue,very thick,domain=0:1.9] plot (\x,{1.5});
\draw[color=blue] (2,1.5) circle (3pt);
\draw[color=blue,fill] (2,0.5) circle (3pt);
\draw[color=blue,very thick,domain=2.1:2.9] plot (\x,{0.5});
\draw[color=blue] (3,0.5) circle (3pt);

\draw[color=blue,fill] (3,1) circle (3pt);
\draw[color=blue,very thick,domain=3.1:4.4] plot (\x,{1});
\draw[color=blue] (4.5,1) circle (3pt);

\draw[color=blue,fill] (4.5,2) circle (3pt);
\draw[color=blue,very thick,domain=4.6:5.9] plot (\x,{2});
\draw[color=blue] (6,2) circle (3pt);

\draw[color=blue,fill] (6,4.5) circle (3pt);
\draw[color=blue,very thick,domain=6.1:8.9] plot (\x,{4.5});
\draw[color=blue] (9,4.5) circle (3pt);

\draw[color=blue,fill] (9,0.5) circle (3pt);
\draw[color=blue,very thick,domain=9.1:10.4] plot (\x,{0.5});
\draw[color=blue] (10.5,0.5) circle (3pt);

\draw[color=blue,fill] (10.5,1) circle (3pt);
\draw[color=blue,very thick,domain=10.6:12.9] plot (\x,{1});
\draw[color=blue] (13,1) circle (3pt);

\draw[color=purple,fill] (13,5) circle (3pt);
\draw[->,color=purple,very thick,domain=13.1:15] plot (\x,{5}) node[right] {$\mbox{\small absorption}$};

\draw[decorate,decoration={brace,amplitude=8pt},xshift=-30pt,yshift=-10pt]
(0.5,0.5) -- (0.5,5.0) node [black,midway,xshift=-1.4cm]{$Y_0=X_0\sim \vect{\alpha}$}; 

\draw[decorate,decoration={brace,amplitude=5pt},xshift=0pt,yshift=5pt,rotate=0]
(4.5,2) -- (6,2) node [black,midway,above,yshift=3pt,xshift=0.8cm]{\small $T_4\sim \exp (\lambda_4)$}; 

\draw[decorate,decoration={brace,amplitude=5pt},xshift=0pt,yshift=5pt,rotate=0]
(10.5,1) -- (13,1) node [black,midway,above,yshift=3pt,xshift=0.8cm]{\small $T_7\sim \exp (\lambda_2)$};

\end{tikzpicture}

\caption{\label{fig:PH}\small A Markov process with $p$
  transient states (blue), one absorbing state (purple), times of jumps $S_1<S_2,\dots$ and time to absorption $\tau$.
   The filled and empty
  circles indicate that the process is assumed continuous from the right. The embedded chain $Y_n=X_{S_n}$ here takes the values $Y_0=3$, $Y_1=1$, $Y_2=2$, $Y_3=4$ etc. 
  Holding times between jumps, $T_n=S_n-S_{n-1}$, are exponentially distributed with a parameter which depends on $Y_{n-1}$ only.}
\end{figure}
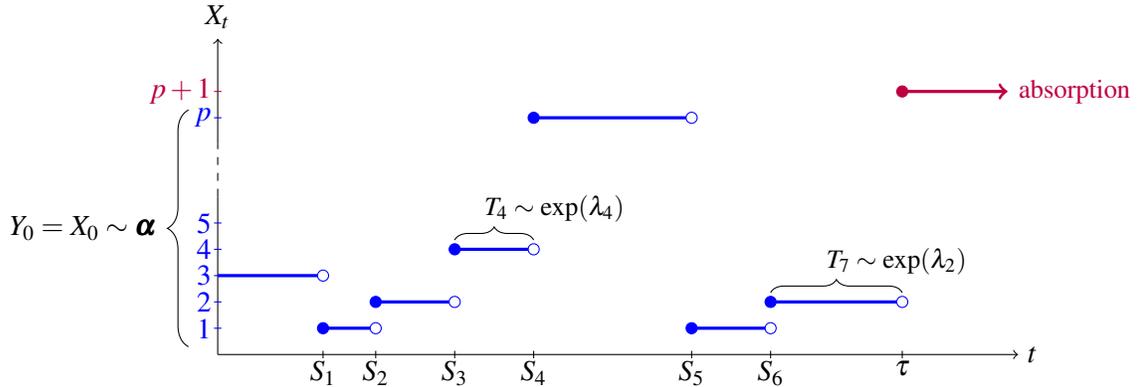

In Figure \ref{fig:PH} we illustrate a sample path of a Markov jump process generating a phase-type distribution. The initial state is chosen according to $\vect{\alpha}$. Given initiation in $Y_0=X_0=3$, the time until the first jump, $S_1$, will then be exponentially distributed with intensity $\lambda_3=-\lambda_{33}=-s_{33}>0$. 
The process then jumps to a state $j\neq 3$, with probability $q_{3j}=\lambda_{3j}/\lambda_3=-s_{3j}/s_{33}$ or to the absorbing state with probability $q_{3,p+1}=\lambda_{3,p+1}/\lambda_3=-s_{3}/s_{33}$.

Of particular interest in population genetics are the so-called generalized Erlang distributions which are convolutions of exponential distributions. Let $T_1,T_2,...,T_n$ be independent random variables with $T_i\sim \mbox{Exp} (\lambda_i)$ for some $\lambda_i>0$, $i=1,...,n$. Then we say that $\tau=T_1+\cdots + T_n$ has a generalized Erlang distribution with parameters $\lambda_1,...,\lambda_n$ and order $n$. If $\lambda_1=\cdots=\lambda_n=\lambda$ then we say that $\tau$ has an Erlang distribution with parameter $\lambda$ and order $n$,  which will be denoted by $\mbox{Er}_n(\lambda )$. 
In particular, the height of a Kingman coalescent with sample size~$n$ has a generalized Erlang distribution with order $n-1$ and parameters $\lambda_i={n-i+1 \choose 2}$. 
Another example is the total branch length, which has a generalized Erlang distribution with order $n-1$ and parameters $\lambda_i=(n-i)/2$. See Example~\ref{Kingmancoalescent1} below. 

\begin{figure}[H]
  \centering
  \begin{tikzpicture}[scale=0.7,domain=-1:10]
\draw[-] (0,0) -- (5.5,0);
\draw[-,dashed] (5.5,0)--(6.5,0);
\draw[->] (6.5,0) -- (10.2,0) node[right] {$t$};
\draw[-] (0,0) --(0,3);
\draw[-,dashed] (0,3)--(0,4);
 \draw[->] (0,4) -- (0,6) node[above] {$X_t$};
\foreach \y/\ytext in {0.5/1, 1/2, 1.5/3, 2/4,2.5/5,4.5/n} 
\draw[color=blue,shift={(0,\y)}] (-2pt,0pt) -- (2pt,0pt) node[left] {$\ytext$};
\draw[color=red] (0,5) node[left] {$n+1$};
\draw[-,color=black] (-0.1,5) -- (0.1,5);

\draw[color=blue,very thick,domain=0:0.90] plot (\x,{0.5});
\draw[color=blue,very thick,domain=1.1:2.4] plot (\x,{1.0});
\draw[color=blue,very thick,domain=2.5:2.9] plot (\x,{1.5});
\draw[color=blue,very thick,domain=3.1:4.1] plot (\x,{2.0});
\draw[color=blue,very thick,domain=4.2:5.5] plot (\x,{2.5});

\draw[color=blue,very thick,domain=6.5:7.0] plot (\x,{4.5});
\draw[->,color=purple,very thick,domain=7.2:9] plot (\x,{5}) node[right] {\small absorption};

\draw[color=blue] (1,0.5) circle (3pt);
\draw[color=blue,fill] (1,1.0) circle (3pt);
\draw[color=blue] (2.5,1.0) circle (3pt);
\draw[color=blue,fill] (2.5,1.5) circle (3pt);
\draw[color=blue] (3.0,1.5) circle (3pt);
\draw[color=blue,fill] (3,2.0) circle (3pt);
\draw[color=blue] (4.2,2.0) circle (3pt);
\draw[color=blue,fill] (4.2,2.5) circle (3pt);
\draw[color=blue] (7.1,4.5) circle (3pt);
\draw[color=purple,fill] (7.1,5) circle (3pt);

\foreach \x/\xtext in {1/S_1,2.5/S_2,3/S_3,4.2/S_4,7.1/\tau}
\draw[shift={(\x,0)}] (0pt,2pt) -- (0pt,-2pt) node[below] {$\xtext$};
\end{tikzpicture}
\caption{\label{fig:erlang} A phase-type representation of the convolution of $n$ exponential distributions. 
 Here
$S_i=T_1+T_2+\cdots +T_i$ is the time of the $i$th jump where $T_i\sim \mathrm{Exp} (\lambda_i)$ and $T_1,\dots,T_n$ are independent.}
\end{figure}
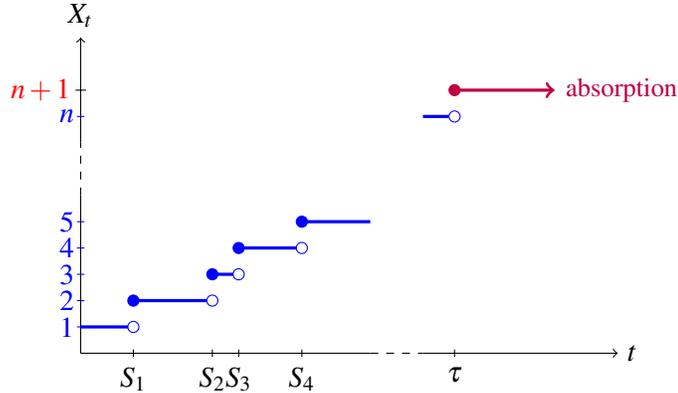

Generalized Erlang distributions are phase-type distributions (Figure~\ref{fig:erlang}). Here, the process initiates in state $1$ with probability $1$ and jumps to state $2$ with probability $1$ after time $T_1\sim \mathrm{Exp} (\lambda_1)$. Continuing this way, from state $n-1$ the process jumps to state $n$ with probability $1$ and remains in this state for the time $T_n\sim \mathrm{Exp} (\lambda_n)$. From here it jumps to the absorbing state. Thus the time $\tau$ it takes the process to reach the absorbing state $n+1$ is exactly the sum of the exponentially distributed random variables. A  phase-type representation is given by
 \[  \vect{\alpha}=(1,0,0,\dotsc ,0), \ \
 \mat{S} =\left(
  \begin{array}{cccccc}
    -\lambda_1 & \lambda_1 & 0 & 0 & \cdots & 0  \\
              0      & -\lambda_2 & \lambda_2 & 0 & \cdots & 0 \\
       0 & 0 & -\lambda_3 & \lambda_3 & \cdots & 0  \\
     \vdots & \vdots & \vdots & \vdots & \vdots & \vdots  \\
    0 & 0 & 0 & 0 & \cdots & -\lambda_n
  \end{array}
 \right) .
  \]
Since it does not matter in which order we sum the random variables in $S=T_1+\cdots + T_n$ we could have chosen any other permutation of $\lambda_1,...,\lambda_n$. Thus phase-type representations are {\it not} unique for a given distribution.

\subsection{Properties}
Let $\tau \sim \mbox{PH}_p(\vect{\alpha},\mat{S})$ and let $\{ X_t\}_{t\geq 0}$ denote its underlying Markov jump process which generates $\tau$. The density $f$ of $\tau$ can then be deduced by a neat probabilistic argument as follows. First, we notice that $f(u)du =\Prob (\tau \in [u,u+du))$ is the probability of $X_t$ jumping to the absorbing state in the interval $[u,u+du)$. 
Conditioning first on $X_0=i$, then on $X_u=j$, we see that
\begin{eqnarray*}
f(u)du&=&\sum_{i=1}^p \Prob (X_0=i)\sum_{j=1}^p \Prob (X_u=j|X_0=i) \Prob (\mbox{jump from } j \mbox{ to } p+1 \mbox{ during } [u,u+du)| X_u=j) .
\end{eqnarray*}
But $\Prob (X_0=i)=\alpha_i$, $\Prob (X_u=j|X_0=i) =p_{ij}^u =\left( e^{\mat{S}u} \right)_{ij}$ (see \eqref{eq:exp-of-Lambda}) and $\Prob (\mbox{jump from } j \mbox{ to } p+1 \mbox{ during } [u,u+du)|X_u=j) = s_jdu$ since $s_j$ is the rate of jumping from $j$ to $p+1$. Thus we have that
\[ f(u)du = \sum_{i=1}^p \alpha_i \sum_{j=1}^p \left( e^{\mat{S}u} \right)_{ij} s_j \ du \]
from which
\[  f(u)=\vect{\alpha}e^{\mat{S}u} \vect{s} . \]
Similarly, the distribution function can be derived even more directly as
\begin{eqnarray*}
1-F(u)&=&\Prob (\tau >u) \\
&=&\Prob (X_u \in \{ 1,2,\dotsc,p\}) \\
&=&\sum_{i=1}^p\alpha_i\sum_{j=1}^p \Prob (X_u=j|X_0=i) \\
&=& \vect{\alpha}e^{\mat{S}u}\vect{e} .
\end{eqnarray*}
The matrix $\mat{S}$ has eigenvalues with strictly negative real parts, hence invertible, and $t\mat{I}-\mat{S}$ is thus a matrix with eigenvalues which have strictly positive real parts whenever $t\geq 0$. The matrix $t\mat{I}-\mat{S}$ is therefore invertible for $t\geq 0$.  
Thus the Laplace transform for $\tau$ can be calculated by
\begin{eqnarray*}
L_\tau (t)&=&\int_0^\infty e^{-tx}\vect{\alpha}e^{\mat{S}x}\vect{s}dx \\
&=&\vect{\alpha}\left( \int_0^{\infty} e^{-(t\mat{I}-\mat{S})x} dx \right)\vect{s} .
\end{eqnarray*}
Here we have used that $e^{(\mat{A}+\mat{B})x}=e^{\mat{A}x}e^{\mat{B}x}$ when the matrices $\mat{A}$ and $\mat{B}$ commute $(\mat{A}\mat{B}=\mat{B}\mat{A})$, and that $\mat{I}$ commutes with $\mat{S}$. Using that
\[ \int e^{\mat{A}x} dx= \mat{A}^{-1}e^{\mat{A}x} , \]
and that $-(t\mat{I}-\mat{S})$ have eigenvalues with strictly negative real parts, we get that
\[ \int_0^{\infty} e^{-(t\mat{I}-\mat{S})x} dx= (t\mat{I}-\mat{S})^{-1}  .  \]
Thus
\begin{equation}
L_\tau(t)=\vect{\alpha}(t\mat{I}-\mat{S})^{-1}\vect{s} .  \label{eq:LPT}
\end{equation}
From the Laplace transform we obtain the moments of $\tau$ to be
\begin{equation}\label{moments}
\mu_n=\Exp (\tau^n) = 
\vect{\alpha}(-\mat{S}^{-1})^n \vect{e} = 
\vect{\alpha}\mat{U}^n \vect{e} ,
\end{equation}
where $\mat{U}=-\mat{S}^{-1}$.
The matrix $\mat{U}=\{ u_{ij} \}$ is the so-called {\it Green} matrix and its elements have the following interpretation: $u_{ij}$ equals the expected time the process $\{ X_t\}$ spends in state $j$ prior to absorption given that $X_0=i$. From this interpretation we can also obtain the formula for $\mu_1$ without using the Laplace transform. 

Phase-type distributions may be heavily over-parametrized. For example, if the exit rate vector
\begin{equation}
\vect{s}= \lambda \vect{e} , \label{eq:PH-exponential-dist}
\end{equation}
for some $\lambda>0$
i.e. the exit rate is the same from all states, then the phase-type distribution with representation $\mbox{PH}_p(\vect{\alpha},\mat{S})$ is simply an exponential distribution with rate $\lambda$. To see this, simply notice that if $\vect{s}=\lambda \vect{e}$, then 
\[ S^\prime (x) = -f(x) = -\lambda \vect{\alpha}e^{\mat{S}x}\vect{e} = \lambda S(x) ,  \] 
where $S(x)=1-F(x)$ is the survival function of $\tau$. Since $S(0)=1$ we then get that $S(x)=\exp (-\lambda x)$, and hence $\tau \sim \exp (\lambda)$.
For parameter estimation this means that a careful investigation of how the model is specified is needed to ensure a minimal representation and parameter identifiability.

\subsection{Rewards}
Let $\tau \sim \mbox{PH}_p(\vect{\alpha},\mat{S})$,  $\{
X_t\}_{t\geq 0}$ its underlying Markov jump process  and $\vect{r}=(r(1),$ $\dotsc,r(p))\in \mathds{R}_+^p$ a vector
of nonnegative numbers (reward rates). Then define the total reward $Y$ earned during the time $\tau$ as
\begin{equation}
Y=\int_0^\tau r(X_t)dt  .  \label{eq:PH-reward}
\end{equation}
If $r(i)\neq 0$ and $T\sim \exp (\lambda_i)$ is a holding time in state $i$, then the reward earned during this holding time is simply $r(i)\cdot T\sim \exp (\lambda_i/r(i))$. Hence, if all $r(i)\neq 0$ and $\mat{\Delta}(\vect{r})$ denotes the diagonal matrix with $\vect{r}$ on the diagonal, we have that
\[  Y \sim \mbox{PH}_p(\vect{\alpha},\mat{\Delta}^{-1}(\vect{r})\mat{S}) . \] 
Observe that equation \eqref{moments} then translates into
\begin{equation}\label{momentsrewards}
\Exp (Y^n) = \vect{\alpha}(\mat{U}\mat \Delta(\vect r))^n\vect{e}.
\end{equation}

\begin{example}
\label{Kingmancoalescent1}
Consider Kingman's $n$-coalescent. Consider independent $T_i\sim \exp \left( \lambda_i \right)$ where $\lambda_i = {i \choose 2}=i(i-1)/2$, $i=2,...,n$. The total tree height (time to the most recent common ancestor) is given by 
 \[  \tau_n = T_2+\cdots +T_n \]
 and the total branch length by
 \[   \mathcal L_n = nT_n+(n-1)T_{n-1}+\cdots + 2T_2 . \]
The total tree height is phase-type distributed $\tau_n\sim \mbox{PH}_{n-1}(\vect{\pi},\mat{T})$ with $\vect{\pi}=(1,0,\dotsc,0)$ and
 \[  \mat{T}=\begin{pmatrix}
 -n(n-1)/2 & n(n-1)/2 & 0 & \cdots & 0 \\
 0 & -(n-1)(n-2)/2 & (n-1)(n-2)/2 & \cdots & 0 \\
 0 & 0 & -(n-2)(n-3)/2 & \cdots & 0 \\
 \vdots & \vdots & \vdots & \vdots\vdots\vdots & \vdots \\
 0 & 0 & 0 & \cdots & -1
 \end{pmatrix} . \]
Defining rewards $\vect{r}=(n,n-1,...,2)$ we see that $\mathcal L_n$ has a phase-type distribution with representation $\mbox{PH}_{n-1}(\vect{\pi},\mat{S})$, where
\begin{eqnarray*}
\mat{S}&=&\mat{\Delta}^{-1}(\vect{r})\mat{T}\\
&=&\frac{1}{2} \begin{pmatrix}
 -(n-1) & (n-1) & 0 & \cdots & 0 \\
 0 & -(n-2) & (n-2) & \cdots & 0 \\
 0 & 0 & -(n-3) & \cdots & 0 \\
 \vdots & \vdots & \vdots & \vdots\vdots\vdots & \vdots \\
 0 & 0 & 0 & \cdots & -1
 \end{pmatrix} .
\end{eqnarray*}
 Since for general constants $a_2,...,a_n$
 \[ -\begin{pmatrix}
 -a_n & a_n & 0 & \cdots & 0 \\
 0 & -a_{n-1} & a_{n-1} & \cdots & 0 \\
 0 & 0 & -a_{n-3} & \cdots & 0 \\
 \vdots & \vdots & \vdots & \vdots\vdots\vdots & \vdots \\
 0 & 0 & 0 & \cdots & -a_2
 \end{pmatrix}^{-1} = \begin{pmatrix}
 \frac{1}{a_n} & \frac{1}{a_{n-1}} & \frac{1}{a_{n-2}} & \cdots & \frac{1}{a_2} \\
 0 & \frac{1}{a_{n-1}} & \frac{1}{a_{n-2}} & \cdots & \frac{1}{a_2} \\
 0 & 0 & \frac{1}{a_{n-2}} & \cdots & \frac{1}{a_2} \\
 \vdots & \vdots & \vdots & \vdots\vdots\vdots & \vdots \\
 0& 0 & 0& \cdots & \frac{1}{a_2} \\
 \end{pmatrix}
   \]
 we see that the mean of $\mathcal L_n$ amounts to the sum of the first row of $-\mat{S}^{-1}$, i.e. 
 \begin{eqnarray*}
  \Exp (\mathcal L_n) \; = \; \vect{\pi}(-\mat{S})^{-1}\vect{e} 
  \;=\; 2\sum_{j=1}^{n-1} \frac{1}{j} 
  \; \sim \; 2\log (n)   
\end{eqnarray*}  
as is well known. See Figure \ref{3moments-KingmanCoalescent} for a graph of the first three moments.

\begin{figure}[H]
  \centering
  \includegraphics[scale=0.30]{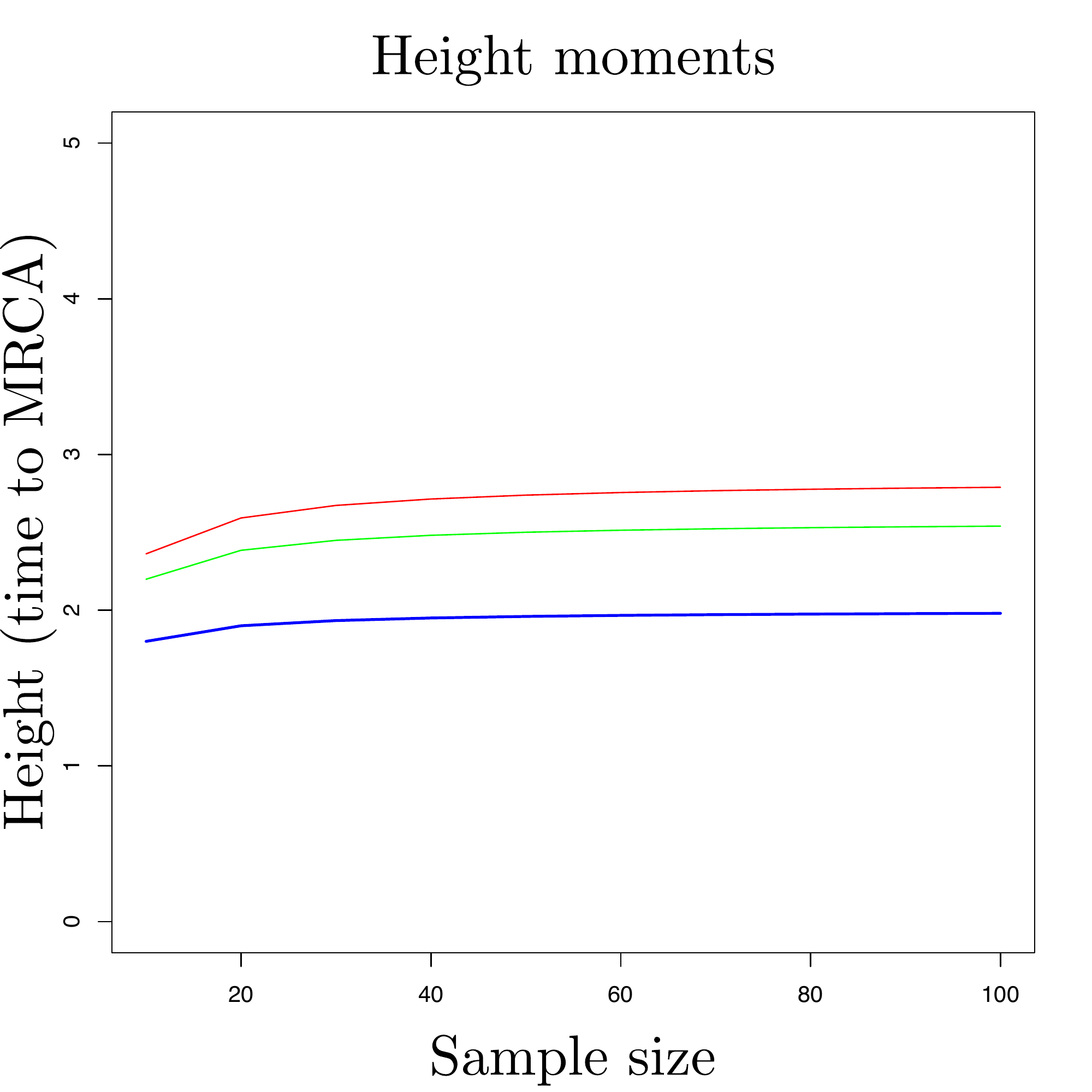}
  \includegraphics[scale=0.30]{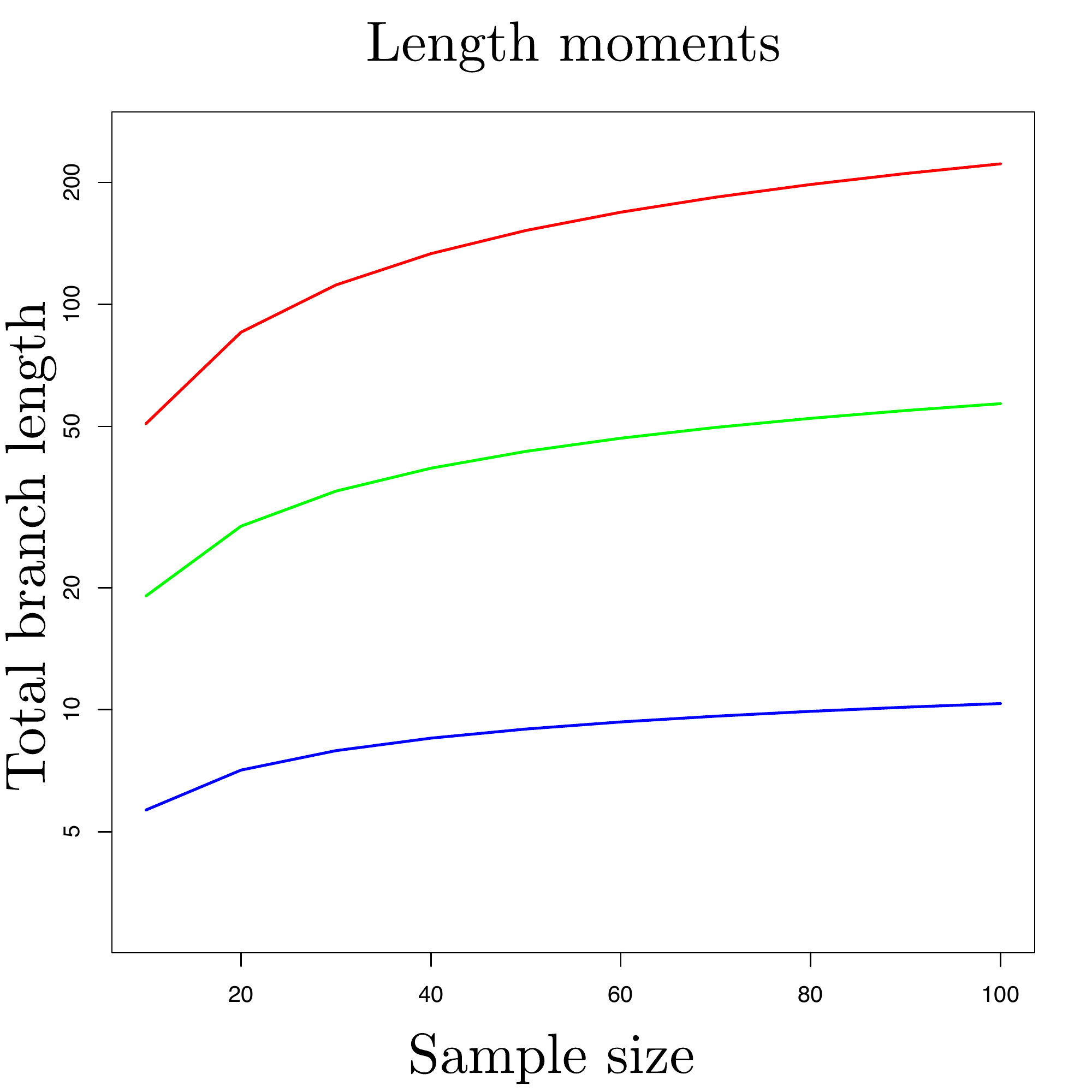}
  \caption{Three first moments of height and length in Kingman's coalescent. Blue:first moment, green:second moment and red:third moment.}
  \label{3moments-KingmanCoalescent}
\end{figure}

\end{example}

Often, and in particular when constructing multivariate phase-type distributions, some rewards will be zero. 
Then the non-zero rewards earned during holding times will still be exponentially distributed obtained by scaling with the appropriate reward, but the embedded chain of the new phase-type distribution will change since going from a state with positive reward to another with positive reward can take place via transitions to zero-reward states in between. 

Define $E^+=\{ i\in E: r(i)>0 \}$ and $E^0 = \{ i\in E: r(i)=0\}$ and decompose accordingly the vector $\vect{\alpha}=(\vect{\alpha}^+,\vect{\alpha}^0)$ and  transition matrix $\mat{Q}$ (of the embedded chain $\{ Y_n\}_{n\in\mathds{N}}$)
\[ \mat{Q} = \begin{pmatrix}
\mat{Q}^{++} & \mat{Q}^{+0} \\
\mat{Q}^{0+} & \mat{Q}^{00}
\end{pmatrix} . \]
Let $d=|E^+|$ be the number of elements in $E^+$ and define
\begin{eqnarray*}
\mat{P}&=&\mat{Q}^{++}+\mat{Q}^{+0}
\left( \mat{I} - \mat{Q}^{00}\right)^{-1}\mat{Q}^{0+} \\
\vect{\pi} &=& \vect{\alpha}^+ + \vect{\alpha}^0
(\mat{I}-\mat{Q}^{00})^{-1}\mat{Q}^{0+} .
\end{eqnarray*}
Then $\mat{P}=\{ p_{ij}\}_{i,j=1,...,d}$ is the transition matrix of the Markov chain which is obtained from $\{ Y_n\}_{n\in \mathds{N}}$ at times when $Y_n\in E^+$. This follows by noticing that the $ij$th element of
$\mat{Q}^{+0}(\mat{Q}^{00})^n\mat{Q}^{0+}$ is the probability of going from $i$ to $j$ by first making a transition to a state in $E^0$, remaining in $E^0$ for the next $n$ jumps and finally jumping from a state in $E^0$ to $j$, and
 since
\[ \left( \mat{I} - \mat{Q}^{00}\right)^{-1} = \sum_{m=0}^{\infty} (\mat{Q}_{00})^m . \]
With a similar argument, $\pi_i$ gives the probability that a Markov process  starts earning rewards from state $i\in E^+$, which can either happen by $X_0=i\in E^+$ or by $X_0\in E^0$ and returning to $E^+$ eventually.  
Since there in general exists the possibility of never entering $E^+$ if the process is started in $E^0$, there will
in general be an atom at zero of size $\pi_{d+1}=1-\vect{\pi}\vect{e}$. Hence we have proved the following:

\begin{theorem}
The random variable $Y$ of (\ref{eq:PH-reward}) is a mixture of an atom at 0 of size $\pi_{d+1}=1-\vect{\pi}\vect{e}$ and a phase-type distribution with
representation $\mbox{PH}_d(\vect{\pi},\mat{T}^*)$ where $\mat{T}^\ast = \{
t_{ij}^\ast:i,j\in E^+\}$ is given by
\begin{eqnarray*}
  t_{ij}^*= -\frac{s_{ii}}{r(i)}p_{ij} \;\; \mathrm{for} \ \ i\neq j 
  \;\; \mathrm{and} \;\;
  t_{ii}^*= \frac{s_{ii}}{r(i)}(1-p_{ii}).
\end{eqnarray*}
\end{theorem}

\begin{example}\label{ex:seedbank}
In this example we consider a genealogical process appearing in peripatric metapopulations \cite{LM} and seed-bank models \cite{BGKW}.
In this model lineages can be active (continent or plants) or inactive (islands or seeds) and they switch from one state to the other at a fixed rate. When they are active, lineages coalesce according to Kingman's coalescent dynamics.
More precisely, let $c$ be the rate for an active branch to unactivate and $K$ be the rate for an inactive branch to re-activate.
Transition rates can be tidied up in the following way.
Let $\lambda_{i,j}=(i-j)(i-j-1)/2+(i-j)c+jK$, $j=0,\ldots,i$.
Define the $(i+1)\times (i+1)$ and $(i+1)\times i$ matrices
\begin{eqnarray*}
\mat{\Lambda}(i)&=&
\matrix{-\lambda_{i,0} & ic & 0 & 0 & \cdots & 0 & 0\\
K & -\lambda_{i,1} & ({i-1})c & 0 & \cdots & 0  & 0\\
0 & 2K & -\lambda_{i,2} & ({i-2})c & \cdots & 0 & 0\\
\vdots & \vdots & \vdots & \vdots\vdots\vdots & \vdots& \vdots & \vdots  \\
0 & 0 & 0 & 0 & \cdots & -\lambda_{i,i-1} & c \\
0 & 0 & 0 & 0 & \cdots & {i}K & -\lambda_{i,i}}\\
\mat{D}(i)&=& \matrix{i(i-1)/2 & 0 & 0 & \cdots & 0 \\
0 & ({i-1})( {i-2})/2 & 0 & \cdots & 0 \\
\vdots & \vdots & \vdots & \vdots\vdots\vdots & \vdots \\
0 & 0 & 0 & \cdots & 0 \\
0 & 0 & 0 & \cdots & 0}.
\end{eqnarray*}
Then the subintensity matrix for the height of the coalescent tree can be represented as
\[  \matrix{\mat{\Lambda}(n) & \mat{D}(n) & \mat{0} & \mat{0} & \cdots & \mat{0} & \mat{0}\\
            \mat{0} & \mat{\Lambda}(n-1) & \mat{D}(n-1) & \mat{0} & \cdots & \mat{0} & \mat{0} \\
              \mat{0} & \mat{0} & \mat{\Lambda}(n-2) & \mat{D}(n-2) & \cdots & \mat{0}& \mat{0} \\
              \vdots & \vdots & \vdots & \vdots & \vdots\vdots\vdots & \vdots \\
              \mat{0} & \mat{0} & \mat{0} & \mat{0} & \cdots & \mat{0} &\mat{\Lambda}(2)  }. \]
The matrix $\mat{\Lambda}(i)$ gives the transition rates when the whole system starts and remains with total size $i$. The matrix $\mat{D}(i) $ gives the transition rates when the whole system loses an element (by coalescence) starting from total size $i$. Row $j$ of $\mat{\Lambda}(i)$, $j=1,...,i+1$, corresponds to the case where out of the remaining $i$ branches, $j-1$ of them are presently inactive. 
The height of the tree has a phase-type distribution. 
Its asymptotics is known to be of order $\log\log n$ (see \cite{BGKW}) where $n$ stands for the initial size of the sample but the precise limit of the variable remains unknown.

In the seed bank model the mutation rate can be inferred from the total length of the active part of the coalescent (because mutations only occur out of the seed bank).
To this end, we use the reward vector $(n, n-1,...,1,0,n-1,n-2,...,0,....,2,1,0)$.
In the peripatric model, the population is separated in continent and islands, hence they can mutate at both stages.
The total number of mutations is in this case related with the total branch length, and 
in this case the reward vector is $(n,\dots,n,n-1,\dots, 3,2,2,2)$.
Results on expected heights and lengths are summarized in Figure \ref{Seedbank}.
Moreover, it is interesting to consider the total number of mutations as the sum of continental mutations and island mutations. 
This problem can be studied in the multivariate phase-type framework.
\begin{figure}[H]
  \centering
  \includegraphics[scale=0.25]{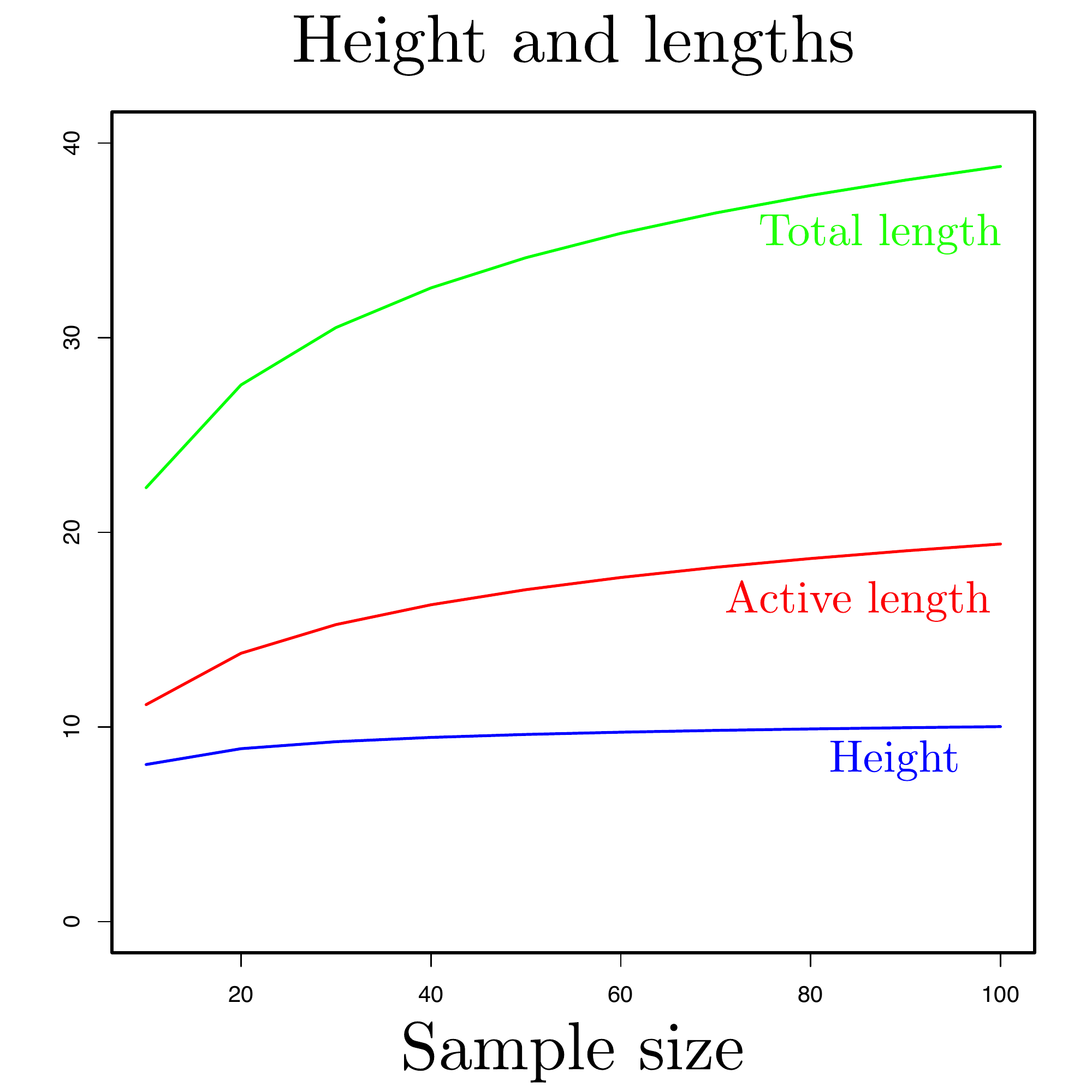}
  \includegraphics[scale=0.25]{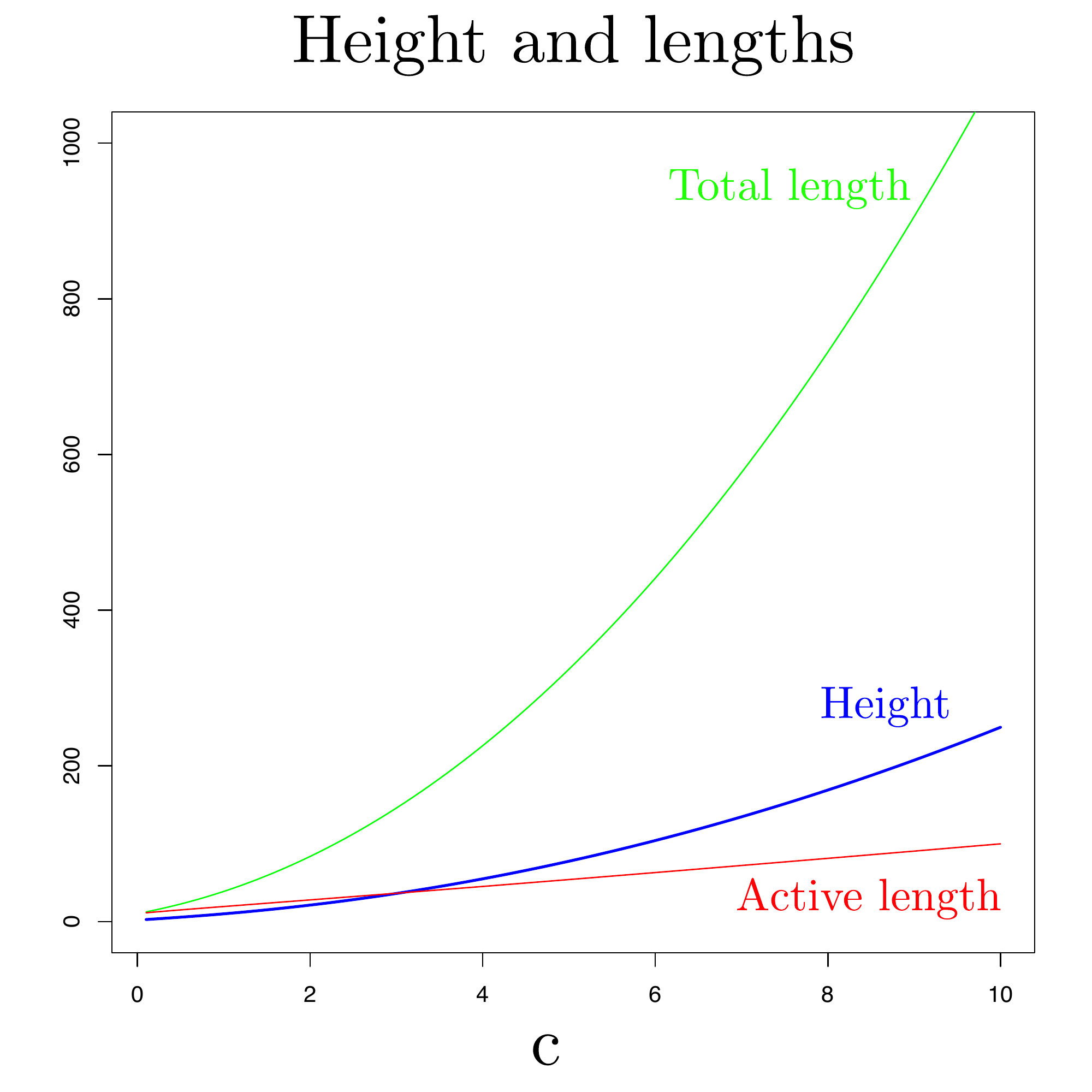}
    \includegraphics[scale=0.25]{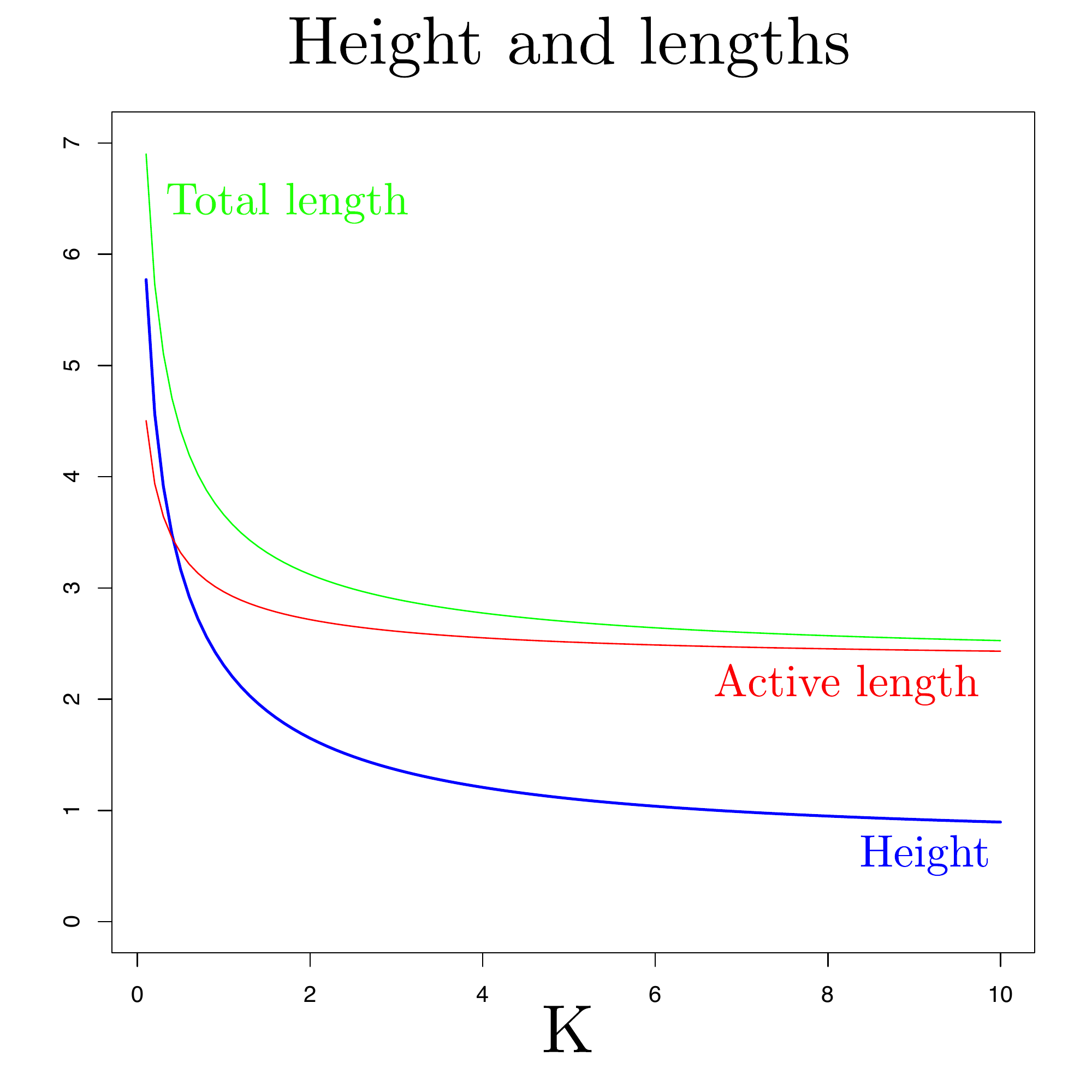}
  \caption{Expected height, active branch length and total branch length of the peripatric/seed-bank coalescent with respect to the sample size. Left: $c=K=1$, Middle: $n=100, K=1$, and Right: $n=100, c=1$.}
  \label{Seedbank}
\end{figure}
\end{example}
\subsection{Multivariate phase-type distributions}\label{sec:multvar}
Let $\tau\sim$PH$_p(\vect{\alpha},\mat{S})$ and let $\{ X_t\}_{t\geq 0}$ denote the underlying Markov jump process which generates $\tau$. 
Let $n$ be a positive integer and let $\mat{R}  = \{R_{ij}\}$ be a $p\times m$ matrix of non-negative constants. Each column $j$ of $\mat{R}$ may be considered to be a function $r_j: \{ 1,2,\dotsc,p\} \rightarrow \mathds{R}_+$ defined by $r_j(i)=R_{ij}$. Then we define
\[  Y_j = \int_0^\tau r_j (X_t)dt = \int_0^\tau R_{X_t,j}\; dt , \]
and say that the random vector $\vect{Y}=(Y_1,\dots,Y_m)$ has a multivariate phase-type distribution parametrized by $\vect{\alpha}$, $\mat{S}$, and $\mat{R}$, and
write $\vect{Y}\sim$MPH$_p^\ast(\vect{\alpha},\mat{S},\mat{R})$. 

\begin{figure}[H]
\begin{tikzpicture}[scale=0.7,domain=-1:10]
\draw[->] (0,0) -- (15.2,0) node[right] {$t$};

 \draw[->] (0,0) -- (0,4) node[above] {$X_t$};

\foreach \y/\ytext in {0.5/1, 1/2, 1.5/3, 2/4,2.5/5,3.0/6}
\draw[shift={(0,\y)}] (-2pt,0pt) -- (2pt,0pt) node[left] {$\ytext$};

\foreach \x/\xtext in {2/S_1,3/S_2,4.5/S_3,6/S_4,9/S_5,10.5/S_6,13/\tau}
\draw[shift={(\x,0)}] (0pt,-2pt) -- (0pt,2pt) node[below] {$\xtext$};

\draw[color=blue,very thick,domain=0:1.9] plot (\x,{1.525});
\draw[color=green,very thick,domain=0:1.9] plot (\x,{1.475});
\draw[color=blue] (2,1.5) circle (3pt);
\draw[color=blue,fill] (2,0.5) circle (3pt);


\draw[color=blue,very thick,domain=2.1:2.9] plot (\x,{0.5});

\draw[color=blue] (3,0.5) circle (3pt);

\draw[color=blue,fill] (3,1) circle (3pt);
\draw [green, xshift=3cm, yshift=1cm,domain=180:360, fill] plot(\x:0.105);
\draw[color=blue,very thick,domain=3.1:4.4] plot (\x,{1.025});
\draw[color=green,very thick,domain=3.1:4.4] plot (\x,{0.975});
\draw[color=blue] (4.5,1) circle (3pt);

\draw[color=black,fill] (4.5,2) circle (3pt);
\draw[color=black,very thick,domain=4.6:5.9] plot (\x,{2});
\draw[color=black] (6,2) circle (3pt);

\draw[color=black,fill] (6,2.5) circle (3pt);
\draw[color=black,very thick,domain=6.1:8.9] plot (\x,{2.5});
\draw[color=black] (9,2.5) circle (3pt);

\draw[color=blue,fill] (9,0.5) circle (3pt);
\draw[color=blue,very thick,domain=9.1:10.4] plot (\x,{0.5});
\draw[color=blue] (10.5,0.5) circle (3pt);

\draw[color=blue,fill] (10.5,1) circle (3pt);
\draw [green, xshift=10.5cm, yshift=1cm,domain=180:360, fill] plot(\x:0.105);
\draw[color=blue,very thick,domain=10.6:12.9] plot (\x,{1.025});
\draw[color=green,very thick,domain=10.6:12.9] plot (\x,{0.975});
\draw[color=blue] (13,1) circle (3pt);

\draw[color=purple,fill] (13,3) circle (3pt);
\draw[->,color=purple,very thick,domain=13.1:15] plot (\x,{3}) node[right] {$\mbox{\small absorption}$};

\end{tikzpicture}
\caption{\label{fig:joint-PH}\small A Markov process with $5$
  transient states (blue, green and black) and one absorbing state (purple). The total time $Y_1$ spent in states $2$ and $3$ prior to absorption (green) and the total time $Y_2$ spent in states $1,2,3$ (blue) prior to absorption defines a bivariate phase-type distribution.}
\end{figure}
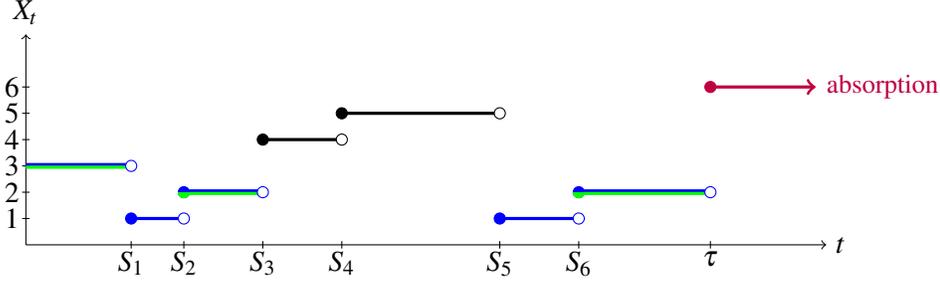

For example, we may consider the joint distribution of the times that the process $\{ X_t\}$ has spent in different (possibly overlapping) subsets of the state-space prior to absorption. This will generate a multivariate phase-type distribution based on rewards which are either zero or one (see Figure \ref{fig:joint-PH} for an example).

The joint distribution of $\mat Y$ can be expressed in a compact form in terms of the joint Laplace transform.
\begin{theorem}[Theorem  8.1.2 in \cite{bladt-nielsen-2017}]
\label{th:MPH}
Let $\vect{Y}\sim \mbox{MPH}^*(\vect{\alpha},\mat{S},\mat{R})$ and 
$\left\langle \cdot,\cdot\right\rangle$ denote the usual dot procuct. 
Then for any vector $\vect{\theta}\geq \vect{0}$, the joint Laplace transform $L_{\vect{Y}}(\vect{\theta})=\Exp (\exp (-\left\langle \vect{Y},\vect{\theta}\right\rangle )) $ is given by 
\begin{equation}
\label{eq:kullpt}
L_{\vect{Y}}(\vect{\theta})
=\vect{\alpha}\left(\mat{\Delta}(\mat{R}\vect{\theta})-\mat{S}  \right)^{-1}\vect{s} .
\end{equation}
\end{theorem}
In general it is not possible to provide explicit formulae for the  joint density function or distribution functions, however, in some important special cases it is possible to derive strikingly simple expressions (see e.g. Section 8.1 of \cite{bladt-nielsen-2017}).
Of special interest are means, variances and covariances between elements of $\vect{Y}$. If $\vect{R}_{\cdot i}$ denotes the $i$th column of $\mat{R}$ and $\mat{U}=-\mat{S}^{-1}$ the Green matrix, then we have that
\begin{eqnarray}
\Exp (Y_i)&=& \vect{\alpha} \mat{U}\vect{R}_{\cdot i} \label{eq:gen-mean} \\
\Exp(Y_i Y_j)&=&  \vect{\alpha} \mat{U}\mat{\Delta}(\mat{R}_{\cdot i})\mat{U}\mat{R}_{\cdot j}  + \vect{\alpha} \mat{U}\mat{\Delta}(\mat{R}_{\cdot j})\mat{U}\mat{R}_{\cdot i} \label{eq:gen-cross-moment}
\end{eqnarray}
for all $i,j$ (including $i=j$) and from which we can calculate the covariance by the well known formula,
\begin{equation}
 \mbox{Cov}(Y_i,Y_j) = \Exp (Y_iY_j)-\Exp (Y_i)\Exp (Y_j) . \label{eq:gen-cov}
\end{equation}
Higher order moments (see Theorem 8.1.5 of \cite{bladt-nielsen-2017}) can be calculated by the formula
\begin{equation}
\Exp\left( \prod _ { j = 1} ^ { p } Y _ { j } ^ { h _ { j } } \right) = \vect{\alpha} \sum _ { \ell = 1} ^ { h ! } \left( \prod _ { i = 1} ^ { h } U \mat{\Delta} \left( \mat { R } _ { \cdot \sigma _ { \ell } ( i ) } \right) \right) \vect{e}, \label{eq:higher-order-moments}
\end{equation}
where $h = \sum _ { j = 1} ^ { n } h _ { j }$ and $\sigma_\ell (i)$ is the index value for entrance $\ell$ of the $i$th permutation. For example, if we want to calculate $\Exp (Y_i Y_j Y_k)$ for $i,j,k$ all different, then we consider all ordered permutations of $(i,j,k)$ which amounts to $(i,j,k)$,$(i,k,j)$,$(j,i,k)$,$(j,k,i)$, $(k,i,j)$ and $(k,j,i)$ resulting in the formula
\begin{eqnarray}
\Exp (Y_iY_jY_k)&=&\vect{\alpha}\mat{U}\mat{\Delta}(\mat{R}_{\cdot i})\mat{U}\mat{\Delta}(\mat{R}_{\cdot j})\mat{U}\mat{\Delta}(\mat{R}_{\cdot k})\vect{e} + \vect{\alpha}\mat{U}\mat{\Delta}(\mat{R}_{\cdot i})\mat{U}\mat{\Delta}(\mat{R}_{\cdot k})\mat{U}\mat{\Delta}(\mat{R}_{\cdot j})\vect{e} \nonumber \\
&&+ \vect{\alpha}\mat{U}\mat{\Delta}(\mat{R}_{\cdot j})\mat{U}\mat{\Delta}(\mat{R}_{\cdot i})\mat{U}\mat{\Delta}(\mat{R}_{\cdot k})\vect{e} + \vect{\alpha}\mat{U}\mat{\Delta}(\mat{R}_{\cdot j})\mat{U}\mat{\Delta}(\mat{R}_{\cdot k})\mat{U}\mat{\Delta}(\mat{R}_{\cdot i})\vect{e}\nonumber \\
&&+\vect{\alpha}\mat{U}\mat{\Delta}(\mat{R}_{\cdot k})\mat{U}\mat{\Delta}(\mat{R}_{\cdot i})\mat{U}\mat{\Delta}(\mat{R}_{\cdot j})\vect{e} + \vect{\alpha}\mat{U}\mat{\Delta}(\mat{R}_{\cdot k})\mat{U}\mat{\Delta}(\mat{R}_{\cdot j})\mat{U}\mat{\Delta}(\mat{R}_{\cdot i})\vect{e}  . \label{eq:third-order-moment}
\end{eqnarray}
For $\Exp (Y_i^2 Y_j Y_k)$ we would have to consider permutations of $(i,i,j,k)$ and summing expressions on the form
\[ \vect{\alpha}\mat{U}\mat{\Delta}(\mat{R}_{\cdot i_1})\mat{U}\mat{\Delta}(\mat{R}_{\cdot i_2})\mat{U}\mat{\Delta}(\mat{R}_{\cdot i_3})\mat{U} \mat{\Delta}(\mat{R}_{\cdot i_4})\vect{e} , \]
where two among the $i_1,i_2,i_3,i_4$ are identical to $i$ while among the remaining two one equals $j$ and the other equals $k$.

\section{Coalescent theory without recombination}\label{sec:norecomb}

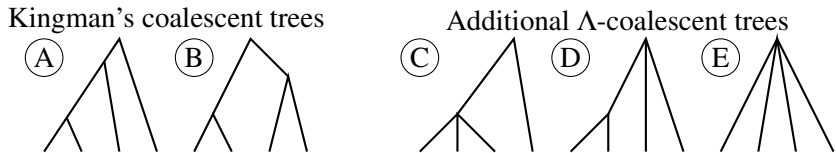
\begin{figure}[H]
\centering
\begin{tikzpicture}
\node at (0,1) {\treeOne};
\node at (2,1) {\treeTwo};
\node at (5,1) {\treeThree};
\node at (7,1) {\treeFour};
\node at (9,1) {\treeFive};
\node at (1,2) {Kingman's coalescent trees}; 
\node at (7,2) {Additional $\Lambda$-coalescent trees};
\end{tikzpicture}
\caption{The five possible $\Lambda$-coalescent topologies for four sequences.}
\label{Fig:TreesForFourSequences}
\end{figure}

The $\Lambda$-coalescent, introduced independently by Pitman \cite{Pit} and Sagitov \cite{Sag}, defines a class of exchangeable coagulation processes including various useful models in population genetics.
Its dynamics is  characterized by a finite measure $\Lambda$ on $[0,1]$.
When the process has $b$ lineages, each subset of $k$ lineages merges at a rate 
\begin{equation}
\label{ratesLambda}\lambda_{b,k}=\int_{[0,1]}x^{k-2}(1-x)^{b-k}\Lambda(dx).\end{equation}
The dynamics of Kingman's coalescent is obtained by taking $\Lambda=\delta_0$, the unit mass at zero, leading to binary mergers only.
In Figure~\ref{Fig:TreesForFourSequences} we show the five possible unlabelled $\Lambda$-coalescent topologies for a sample of size~$n=4$.

In the general case, the height of the tree of a sample of size $n$
is phase-type distributed $\mbox{PH}_{n-1}(\vect{\alpha},\mat{S})$ with $\vect{\alpha}=(1,0,\dotsc,0)$ and
 \[  \mat{S}=\begin{pmatrix}
 -g_n & g_{n,2} & g_{n,3} & \cdots &  g_{n,n-1} \\
 0 & -g_{n-1} & g_{n-1,2} & \cdots &  g_{n-1,n-2} \\
 0 & 0 & -g_{n-2} & \cdots &  g_{n-2,n-3} \\
 \vdots & \vdots & \vdots & \vdots\vdots\vdots & \vdots \\
 0 & 0 & 0 & \cdots & -g_2
 \end{pmatrix}  \]
where $g_{i,k}={i \choose k}\lambda_{i,k}$ for $i=2,\ldots,n$ and $k=2,\ldots,i$, and  $g_{i}=\sum_{k=2}^i g_{i,k}$  (notice that $g_2=1$).
As for Example \ref{Kingmancoalescent1}, the total branch length can be studied using the reward vector $(n, n-1,\dots,2)$.

\begin{example}
A class of interest is the Psi-coalescent that appears as the genealogical process of Moran models with highly skewed offspring distribution \cite{EW}.
Rare reproduction events make that an individual's progeny will replace a proportion $\psi\in(0,1)$ of the next generation.
Here the probability measure $\Lambda$ is the unit mass in $\psi$.
This gives the transition rates
$$\lambda_{b,k}=
\psi^{k-2}(1-\psi)^{b-k}.$$
Note that we vary from the original model of \cite{EW} by a constant $\psi^2$ so that we obtain the Kingman coalescent as $\psi\to0$. See Figures  \ref{3moments-height-psiCoalescent} and \ref{3moments-length-psiCoalescent} for graphs of the first three moments of height and lengths in the Psi-coalescent. 

\begin{figure}[H]
  \centering
  \includegraphics[scale=0.25]{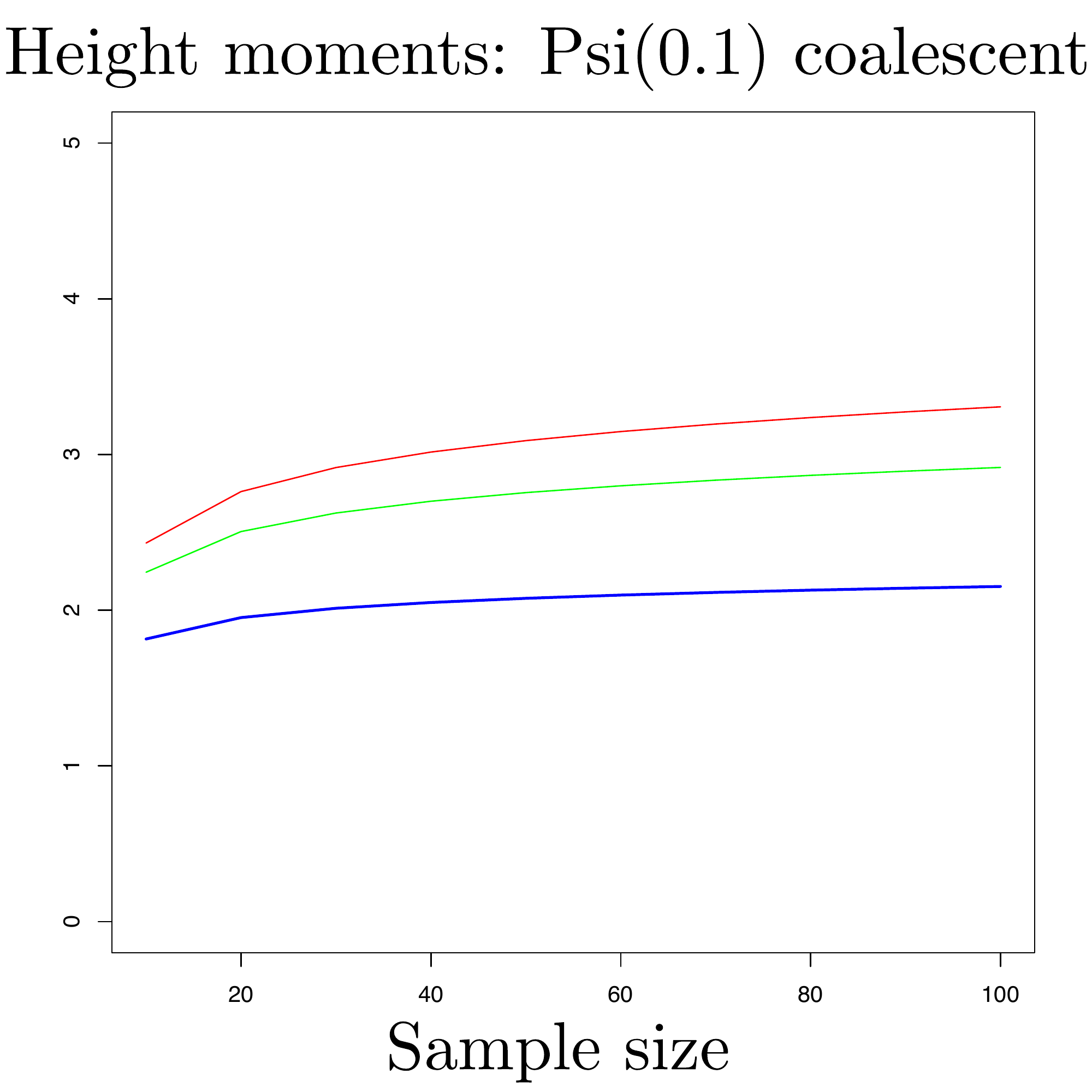} \hspace{1cm}
  \includegraphics[scale=0.25]{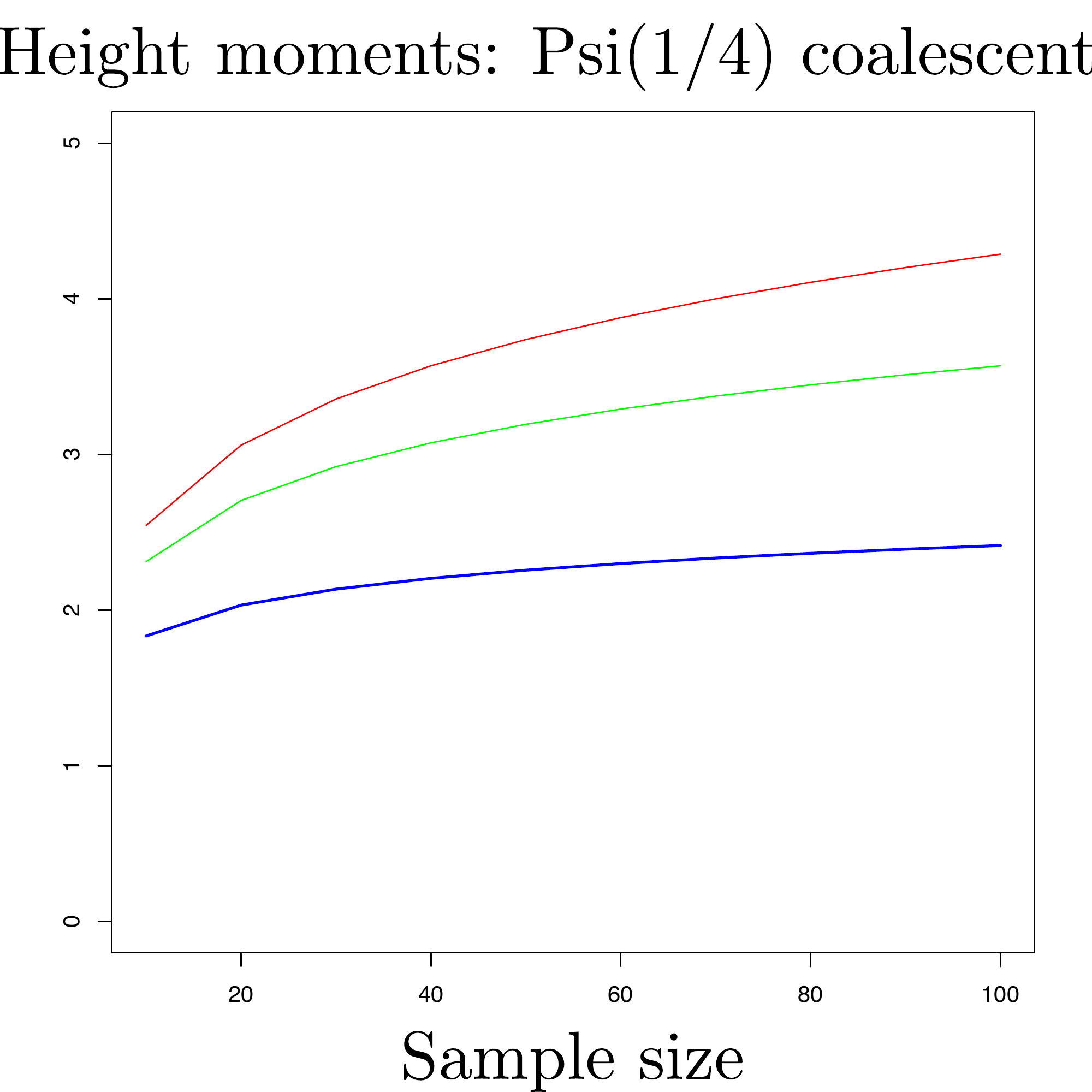}
  \caption{Three first moments of heights in the Psi-coalescent. Blue:first moment, green:second moment and red:third moment.}
  \label{3moments-height-psiCoalescent}
\end{figure}

\begin{figure}[H]
  \centering
  \includegraphics[scale=0.25]{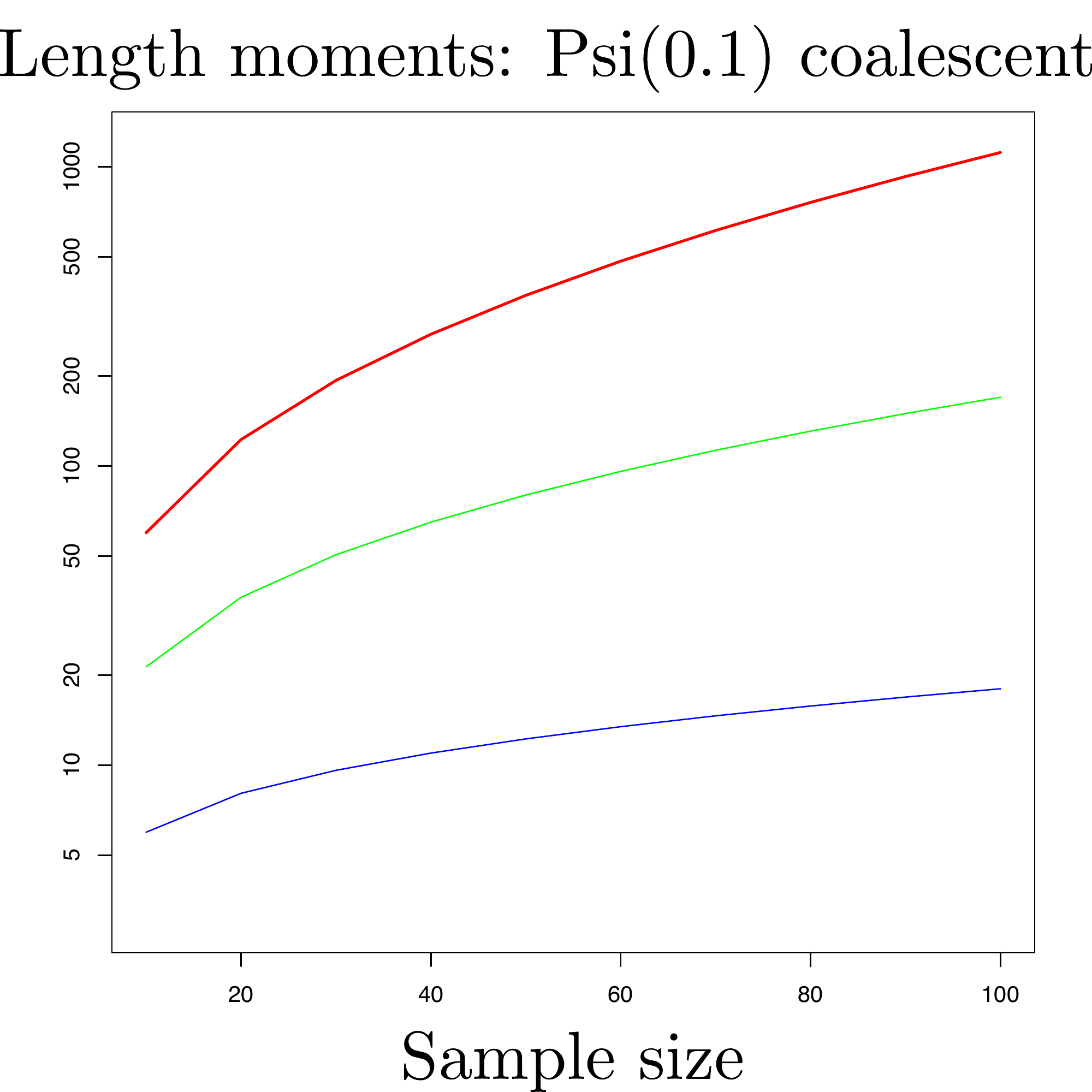} \hspace{1cm}
  \includegraphics[scale=0.25]{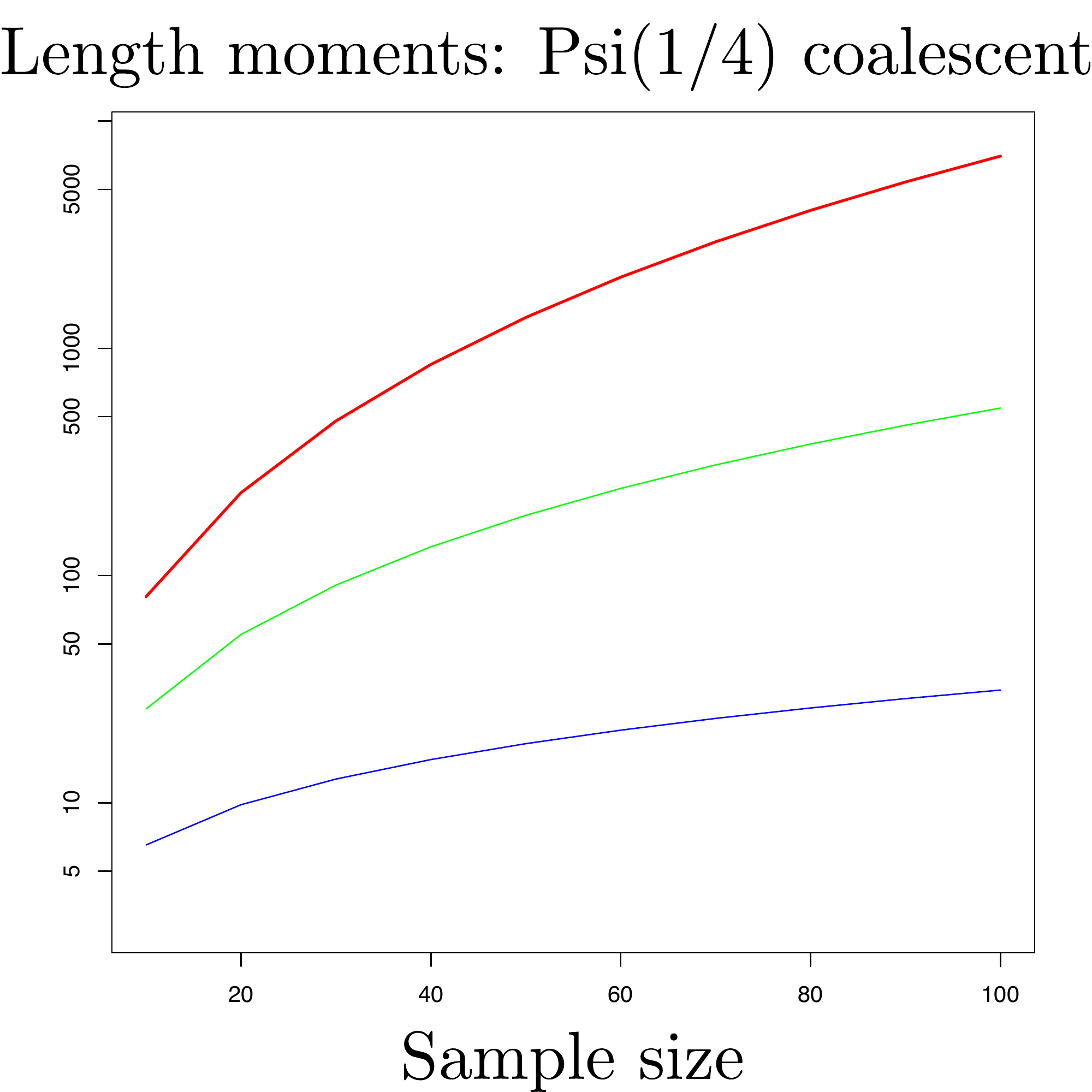}
  \caption{Three first moments of lengths in the Psi-coalescent. Blue:first moment, green:second moment and red:third moment. }
 \label{3moments-length-psiCoalescent}
\end{figure}

\end{example}

\begin{example}\label{ex:3.2}
Another class of interest is the Beta-coalescent that appears as the genealogical process of stable Galton-Watson populations \cite{Sch03} and have been applied to marine populations \cite{Blath}.
Here the probability measure $\Lambda$ is that of a $\mbox{Beta}(2-\alpha,\alpha)$ distribution with $1\leq\alpha<2$, i.e.,
$$\Lambda(dx)=\frac{1}{\Gamma(2-\alpha)\Gamma(\alpha)}x^{1-\alpha}(1-x)^{\alpha-1}dx.$$
This model gives the transition rates
\begin{equation}\label{ratesbeta}
\lambda_{b,k}=
\frac{\beta(k-\alpha,b-k+\alpha)}{\beta(\alpha,2-\alpha)}
\end{equation}
where $\beta$ is the Beta function.
The case $\alpha\to2$ represents the Kingman coalescent, whereas the case $\alpha=1$ gives the Bolthausen-Sznitman coalescent, that appears as the genealogical model of populations under strong selection \cite{Des, NH, Sch17}.
Asymptotic behavior of the height and the length has been studied in \cite{BBS, DDS, Ker} for $\alpha\in(1,2)$. 
Note that the height converges without scaling to a finite random variable whereas the length is of order $n^{2-\alpha}$. When $\alpha=1$, the height is of order $\log\log n$ (see \cite{GM}) and the length is of order $n/\log n$ (see \cite{Drm}). See Figures  \ref{3moments-height-betaCoalescent} and \ref{3moments-length-betaCoalescent} for graphs of the first three moments of height and lengths in the Psi-coalescent. 

\begin{figure}[H]
  \centering
  \includegraphics[scale=0.25]{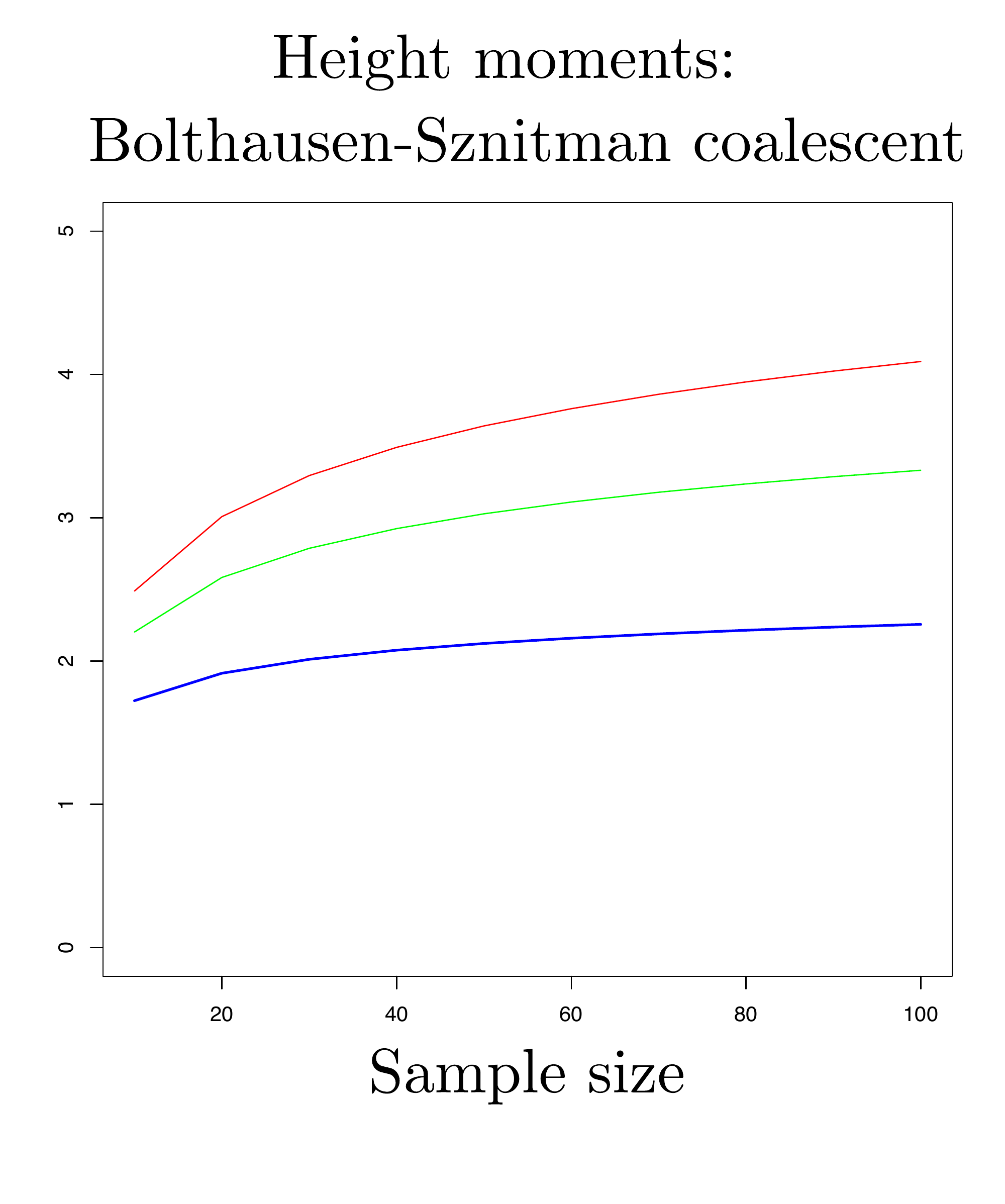} \hspace{1cm}
  \includegraphics[scale=0.25]{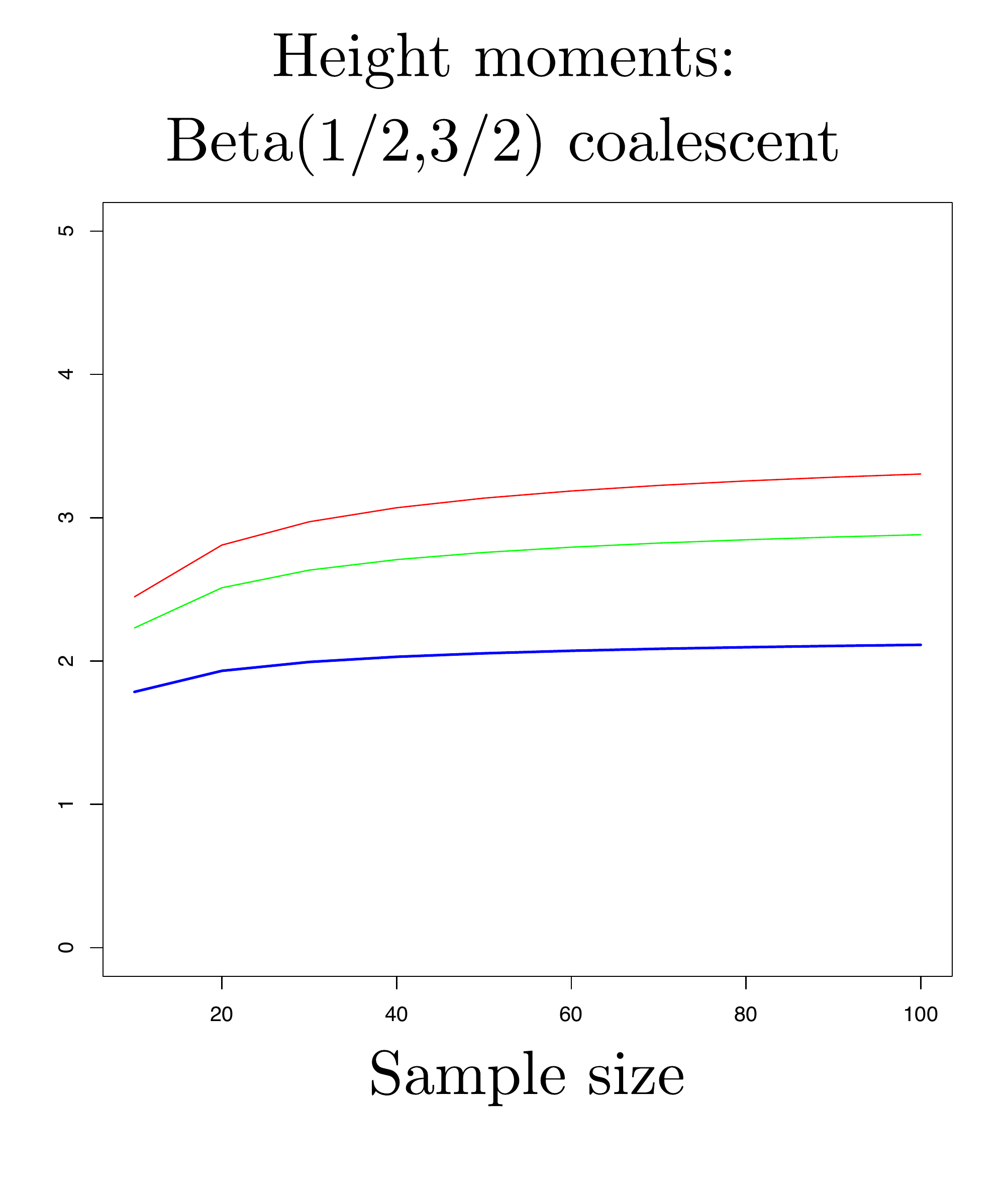}
  \caption{Three first moments of heights in the Beta-coalescent. Blue:first moment, green:second moment and red:third moment.}
  \label{3moments-height-betaCoalescent}
\end{figure}

\begin{figure}[H]
  \centering
  \includegraphics[scale=0.25]{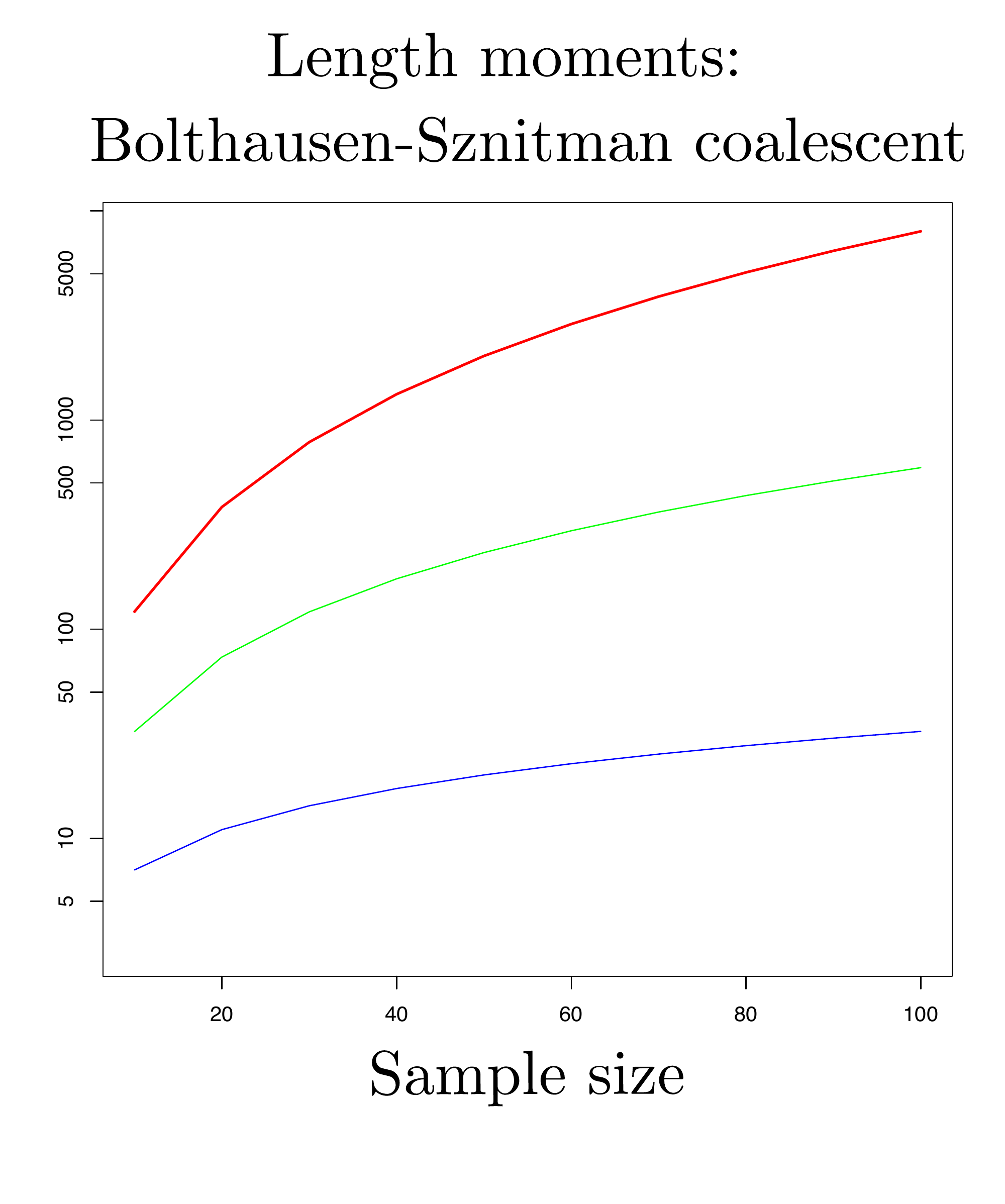}\hspace{1cm}
  \includegraphics[scale=0.25]{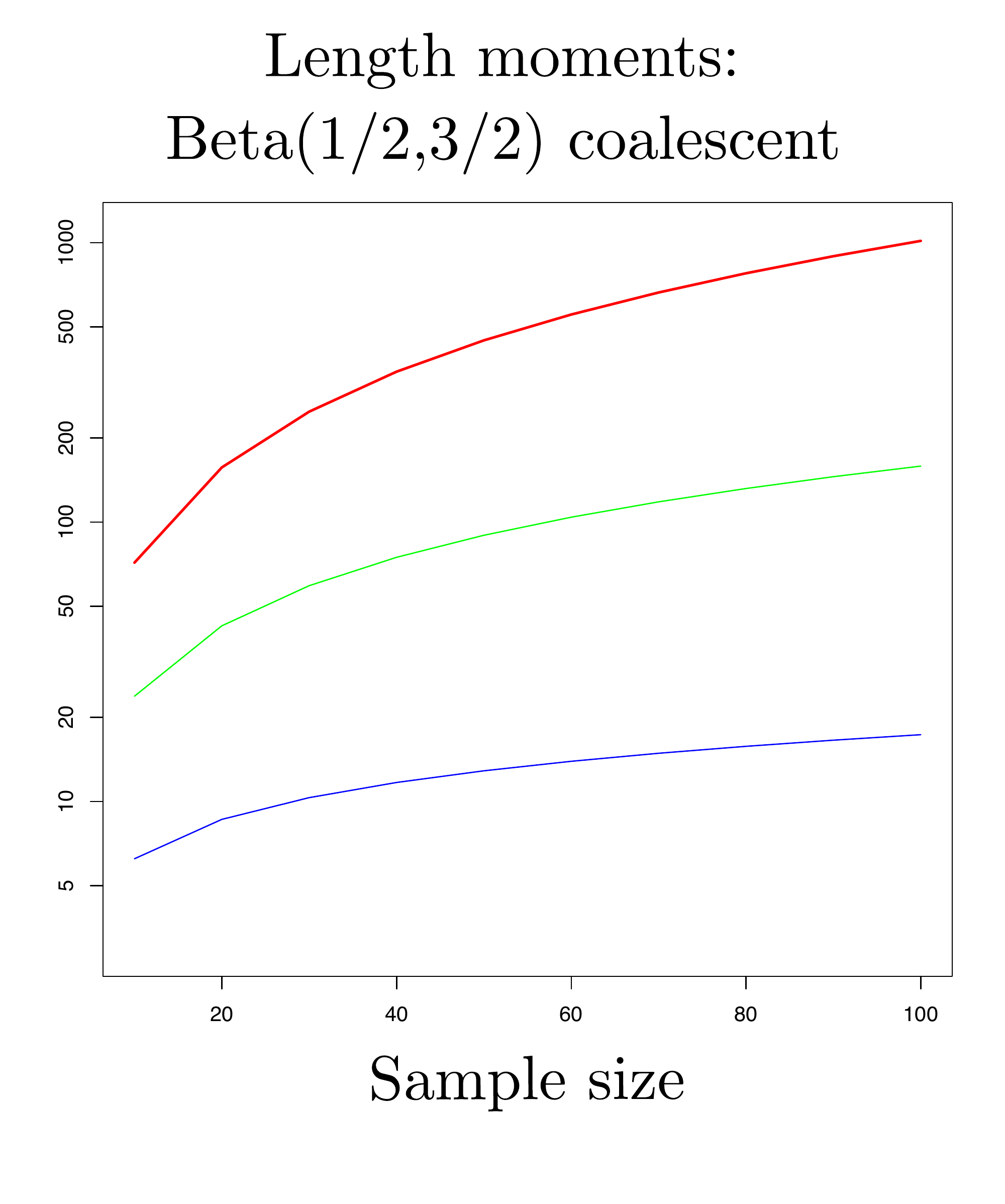}
  \caption{Three first moments of lengths in the Beta-coalescent. Blue:first moment, green:second moment and red:third moment.}
 \label{3moments-length-betaCoalescent}
\end{figure}
\end{example}
In order to study the site frequency spectrum we need to introduce an appropriate state-space and a corresponding reward matrix. For a sample of size $n$, we represent the states by a vector $\vect{a}=(a_1,a_2,...,a_n)$ where $a_i$ denotes the number of branches with $i$ descendants. The state-space is thus given by 
$$\{\vect a=(a_1,\dots,a_n) \in\mathbb Z_+^n \; : \; \sum_{i=1}^n ia_i=n\}.$$
This representation is similar to the summary of a sample of DNA sequences used for the infinite alleles model in Ewens' sampling formula.
For Kingman's coalescent the possible transition are 
\[ (a_1,\dots,a_n) \rightarrow (a_1,\dots,a_i-1,\dots, a_j-1\dots, a_{i+j}+1,\dots,a_n) \]
 with rate $a_ia_j$ for $a_i,a_j\geq 1$, and
 \[ (a_1,\dots,a_n) \rightarrow (a_1,\dots,a_i-2,\dots, a_{2i}+1,\dots,a_n)  \]
 with rate ${a_i \choose 2}$ for $a_i\geq 2$.
The row in the reward matrix corresponding to a state $\vect{a}=(a_1,...,a_n)$ is given by $(a_1,...,a_{n-1})$ because $a_1$ is the number of branches with one descendant, $a_2$ is the number of branches with two descendants etc. (see also Table~\ref{Table:RatesForFourSequences}).
\begin{example}\label{ex:lamda-coalescent-4}
Consider Kingman's coalescent with $n=4$. In Figure~\ref{Fig:KingmanFlowFourSeq} we show the state space and possible transitions.
\begin{figure}[H]
\centering
\begin{tikzpicture}


\node at (0,4) {(4,0,0,0)};
\node at (2.5,4) {(2,1,0,0)};
\node at (8,4) {(0,0,0,1)};
\node at (5,5) {(1,0,1,0)};
\node at (5,3) {(0,2,0,0)};
\draw[->,thick] (0.9,4) -- (1.6,4);
\draw[->,thick] (3.2,4.1) -- (4.2,4.9);
\draw[->,thick] (3.2,3.9) -- (4.2,3.1);
\draw[->,thick] (5.8,3) -- (7.2,3.9);
\draw[->,thick] (5.8,5) -- (7.2,4.2);

\draw (3.5,4.7) circle(0.2); \node at (3.5,4.7) {A};
\draw (3.5,3.3) circle(0.2); \node at (3.5,3.3) {B};
\end{tikzpicture}
\caption{A flow diagram for the case of four sequences in the Kingman's coalescent, where circles refer to the topologies from Figure \ref{Fig:TreesForFourSequences}.}
\label{Fig:KingmanFlowFourSeq}
\end{figure}
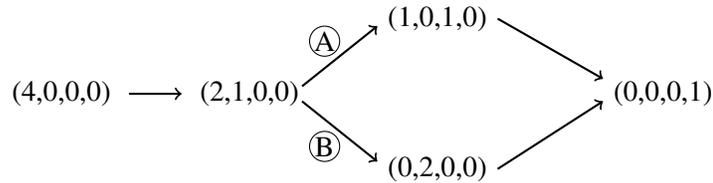
The intensity and reward matrices are given in Table~\ref{Table:RatesForFourSequences}. 
\begin{table}[H]
\renewcommand{\arraystretch}{1.2}
\centering
\begin{tabular}{c|c|ccccc|ccc|c}
 \multicolumn{2}{c|}{State} & \multicolumn{5}{|c|}{Intensity matrix} & \multicolumn{3}{c|}{Reward $\mat{R}$} & Number of \\ 
  Type & Index & 1 & 2 & 3 & 4 & 5 & 
  $\mat{R}_{\cdot 1}$ & $\mat{R}_{\cdot 2}$ & $\mat{R}_{\cdot 3}$ & branches \\ \hline 
 $(4,0,0,0)$   & 1 & $-{4 \choose 2}$ & ${4 \choose 2}$ & 0 & 0 & 0 & 
4 &0 &0 & 4 \\
 $(2,1,0,0)$ & 2 & 0 & $-3$ & 1 & 2 &0& 
 2 & 1 &0& 3 \\
 $(0,2,0,0)$ & 3 & 0& 0& $-1$ & 0& 1 & 0
 & 2 &0& 2 \\
 $(1,0,1,0)$ & 4 & 0& 0&0 & $-1$ & 1 & 
 1 &0& 1 & 2 \\
 $(0,0,0,1)$ & 5 & 0& 0&0 &0 & $0$ & 0
 &0&0& 1 \\
\end{tabular}
\caption{Intensity matrix for Kingman's coalescent and reward matrix for calculating the site frequency spectrum for $n=4$ sequences.}
\label{Table:RatesForFourSequences}
\end{table}
The elements of each row in $\mat{R}$ correspond to the number of branches with one, two or three descendants. The row sums of the reward matrix equals the number of branches, except for the last absorbing state where only one lineage is present. 
\end{example}

We now provide an algorithm for generating the general state-space and corresponding transition rates. 

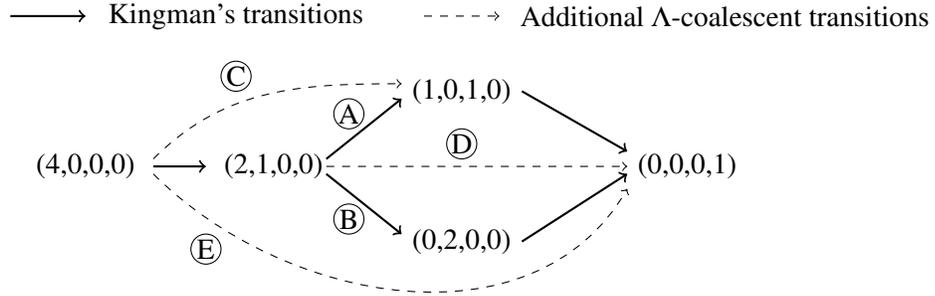
\begin{figure}[H]
\centering
\begin{tikzpicture}
\node at (0,4) {(4,0,0,0)};
\node at (2.5,4) {(2,1,0,0)};
\node at (8,4) {(0,0,0,1)};
\node at (5,5) {(1,0,1,0)};
\node at (5,3) {(0,2,0,0)};
\draw[->,thick] (0.9,4) -- (1.6,4);
\draw[->,thick] (3.2,4.1) -- (4.2,4.9);
\draw[->,thick] (3.2,3.9) -- (4.2,3.1);
\draw[->,thick] (5.8,3) -- (7.2,3.9);
\draw[->,thick] (5.8,5) -- (7.2,4.2);
\draw[->,dashed] (0.9,4.1) to[out=45,in=180] (4.2,5.1);
\draw[->,dashed] (0.9,3.9) to[out=-45,in=-120] (7.2,3.7);
\draw[->,dashed] (3.2,4) -- (7.2,4);
\draw (3.5,4.7) circle(0.2); \node at (3.5,4.7) {A};
\draw (3.5,3.3) circle(0.2); \node at (3.5,3.3) {B};
\draw (2,5.2) circle(0.2); \node at (2,5.2) {C};
\draw (5,4.3) circle(0.2); \node at (5,4.3) {D};
\draw (1.6,2.9) circle(0.2); \node at (1.6,2.9) {E};
\draw[->,thick] (-1,6) -- (0,6);
\node at (2,6) {Kingman's transitions};
\draw[->,dashed] (4.5,6) -- (5.5,6);
\node at (8.5,6) {Additional $\Lambda$-coalescent transitions};
\end{tikzpicture}
\caption{Flow diagram for the case of four sequences in the $\Lambda$-coalescent model. The numbers in the circles refer to the topologies in Figure~\ref{Fig:TreesForFourSequences}.}
\label{Fig:LambdaFlowFourSeq}
\end{figure}

\begin{algorithm}
The state-space is determined as follows. The transition
 \[ \vect a=(a_1,\dots,a_n) \rightarrow \vect b=(b_1,\dots,b_n)  \]
is possible if the vector $\vect c=(c_1,\dots,c_n)=\vect b-\vect a$
fulfils the three conditions 
 \begin{itemize}
 \item[(i)] $\sum_{i=1}^n c_i{\bf1}_{\{c_i>0\}}=1$ (one new branch is created)
 \item[(ii)] $\sum_{i=1}^n c_i{\bf1}_{\{c_i<0\}}=-2$ (two branches are merged)
 \item[(iii)] $\sum_{i=1}^n ic_i=0$, (balance equation on the number of individuals of the sample involved).
 \end{itemize}
The transition rates between the states are 
 \begin{equation}
 \label{SKing}
 S_{\vect a\vect b}
 =\lambda_{\sum_{i=1}^n a_i,-\sum_{i=1}^n c_i
{\bf1}_{\{c_i<0\}}}
 \prod_{i:c_i<0} {a_i \choose -c_i}.
 \end{equation}
 \end{algorithm}
\noindent It is natural to start with $\vect{a}=(n,0,...,0)$ and identify the remaining states subsequently. In Figure~\ref{Fig:LambdaFlowFourSeq} we show the state space and possible transitions for the general $\Lambda$-coalescent.

For a general $\Lambda$-coalescent process, mutations on branches with one descendant give rise to singletons in the site frequency spectrum, while mutations with two or three descendants give rise to doubletons, tripletons and so on in the site frequency spectrum.  The quantities
\[ 
Y_i =  \int_0^\tau \mat{R}_{X_t,i} \;dt , \;\; i=1,2,3,...,n-1,   
\]
are the total branch lengths where a mutation is shared by exactly $i$ samples. If the mutation rate is~$\theta/2$, then the expected site frequency spectrum (SFS) is given by 
\[ \Exp (\xi_i ) = \frac{\theta}{2}\Exp (Y_i) , \]
where $\Exp (Y_i)$ is given by \eqref{eq:gen-mean}. Covariances are given by
\[  \mbox{Cov} (\xi_i,\xi_j) = \frac{\theta^2}{4}\mbox{Cov}(Y_i,Y_j) , \]
for which we use \eqref{eq:gen-cross-moment} and \eqref{eq:gen-cov}.

\begin{example}
Here we consider the variance, covariance and expected site frequency spectrum for the Psi-coalescent (Figure \ref{fig:SFS-Psi}) and the Beta-coalescent (Figure \ref{fig:SFS-Beta}).
The bumps for the Psi-coalescent can be explained by the fact that, at each coalescence event, a proportion $\psi$ of the branches are merged, giving a higher probability for branches with $n\psi$ descendants to appear.

\begin{figure}[H]
\centering
 \includegraphics[scale=0.28]{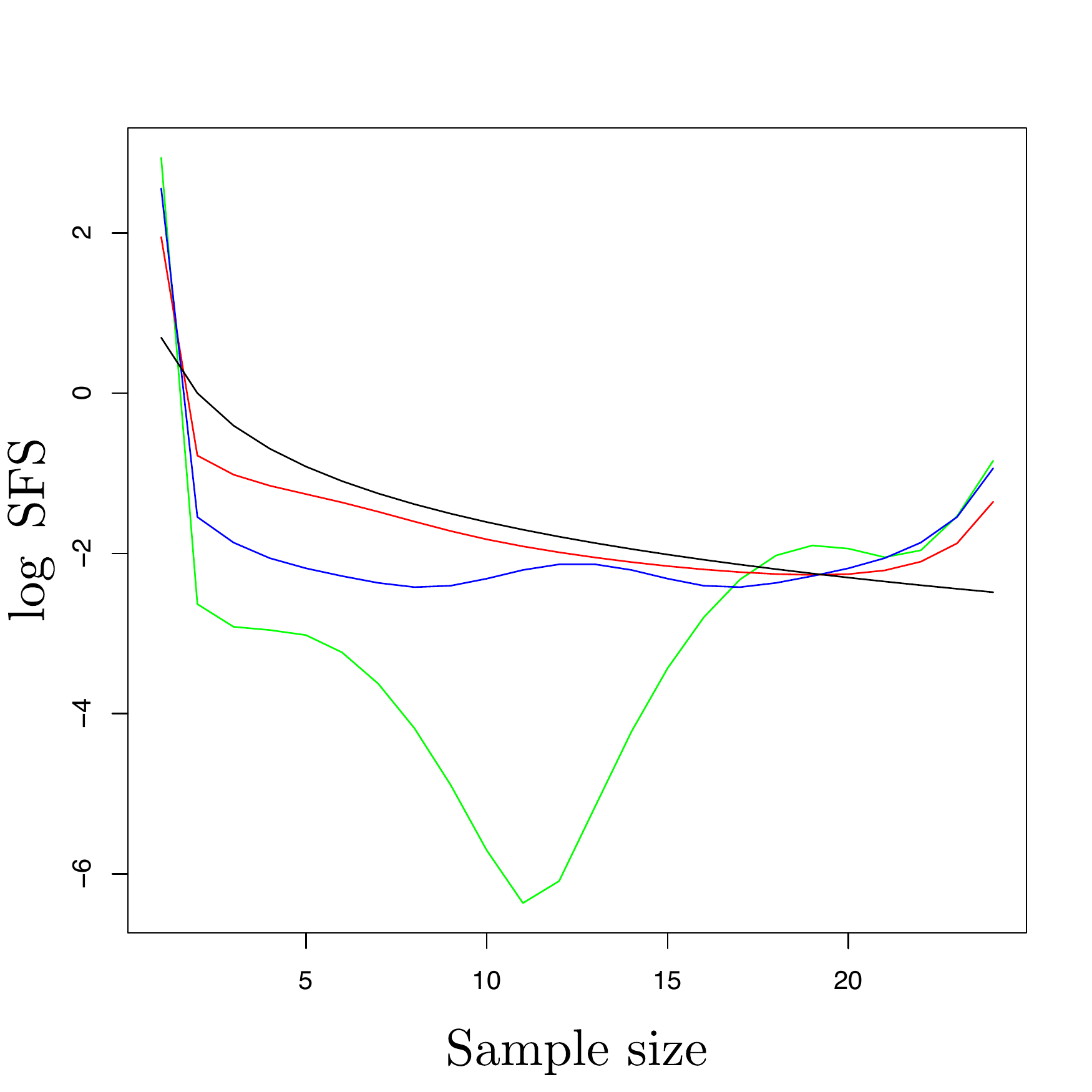}
 \includegraphics[scale=0.28]{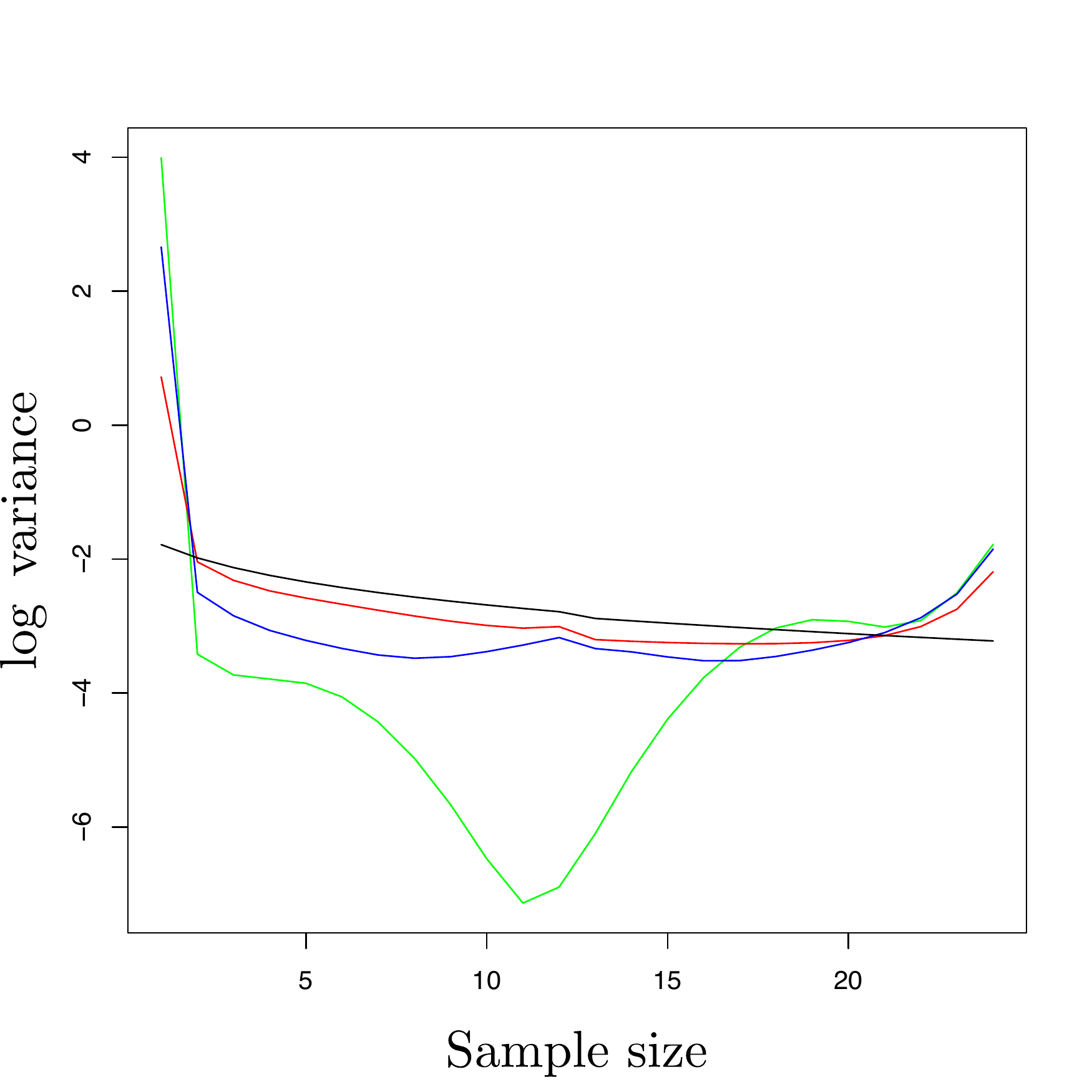}
\includegraphics[scale=0.28]{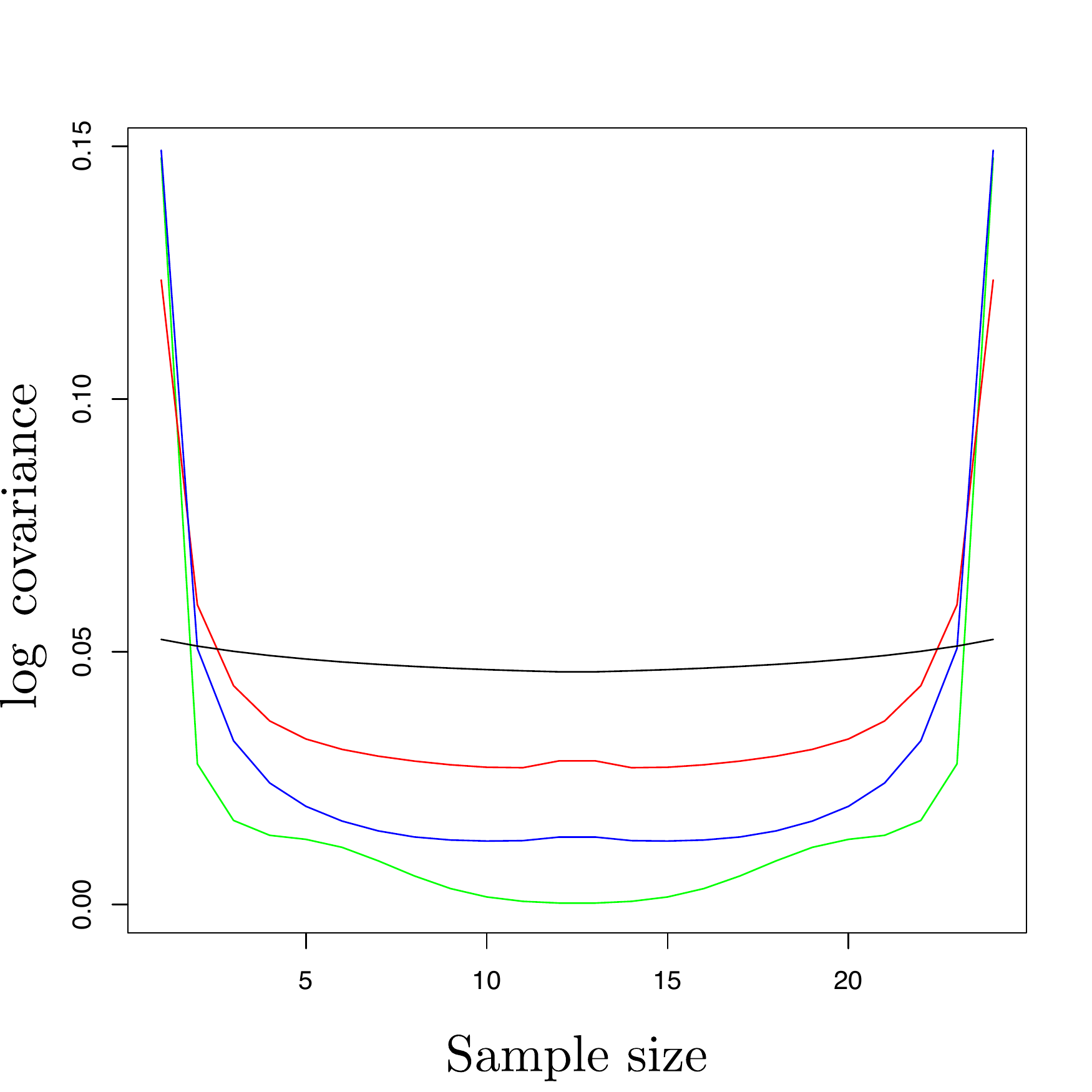} 
\caption{\label{fig:SFS-Psi} Psi-coalescence for $\psi=0.25$ (red), $\psi=0.5$ (blue), $\psi=0.75$ (green) compared to Kingman's coalescent (black). Left: logarithm of Expected SFS.  Middle: log variances of SFS. Right: log anti-diagonal values for the covariance matrix of the SFS }
\end{figure}

\begin{figure}[H]
\centering
 \includegraphics[scale=0.28]{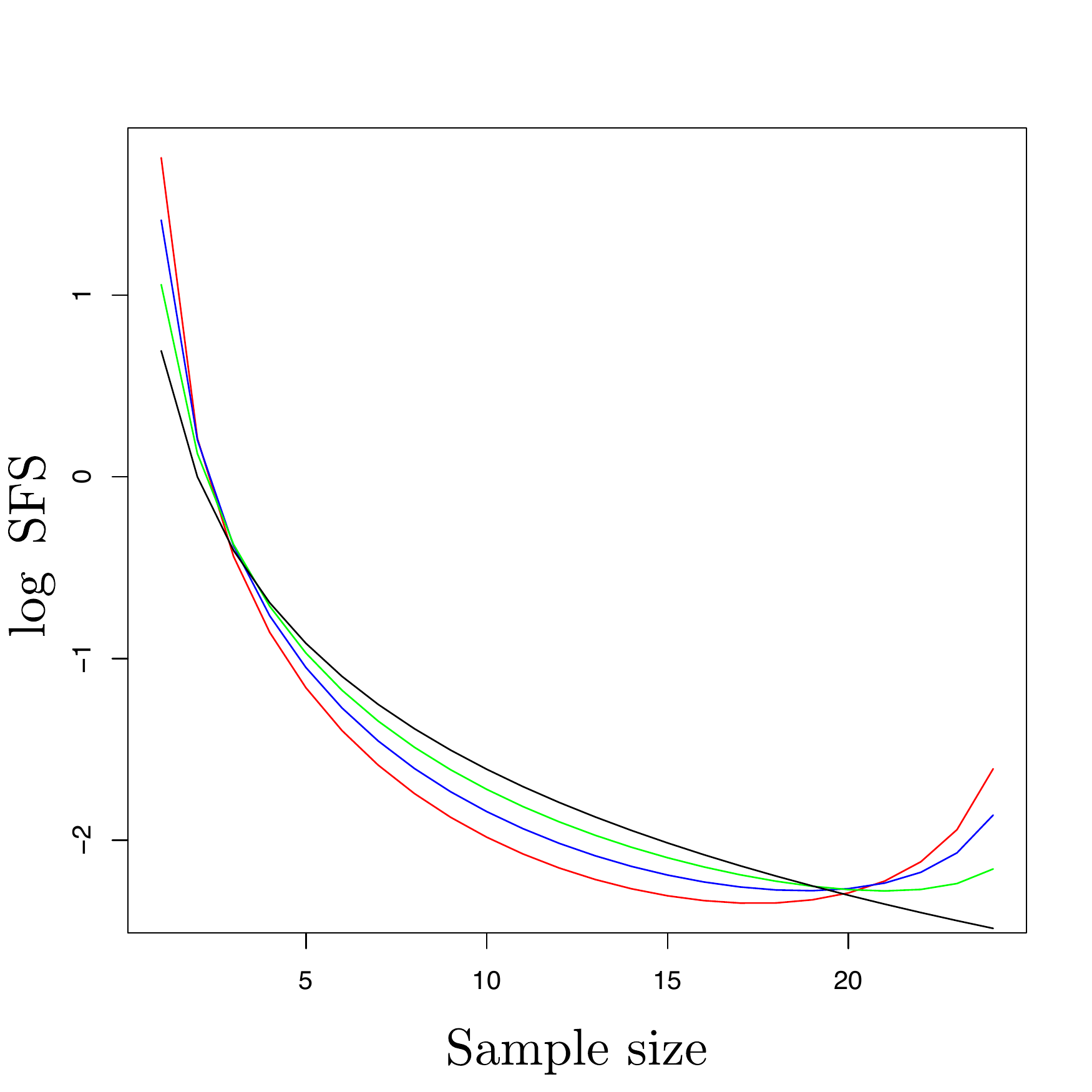}
 \includegraphics[scale=0.28]{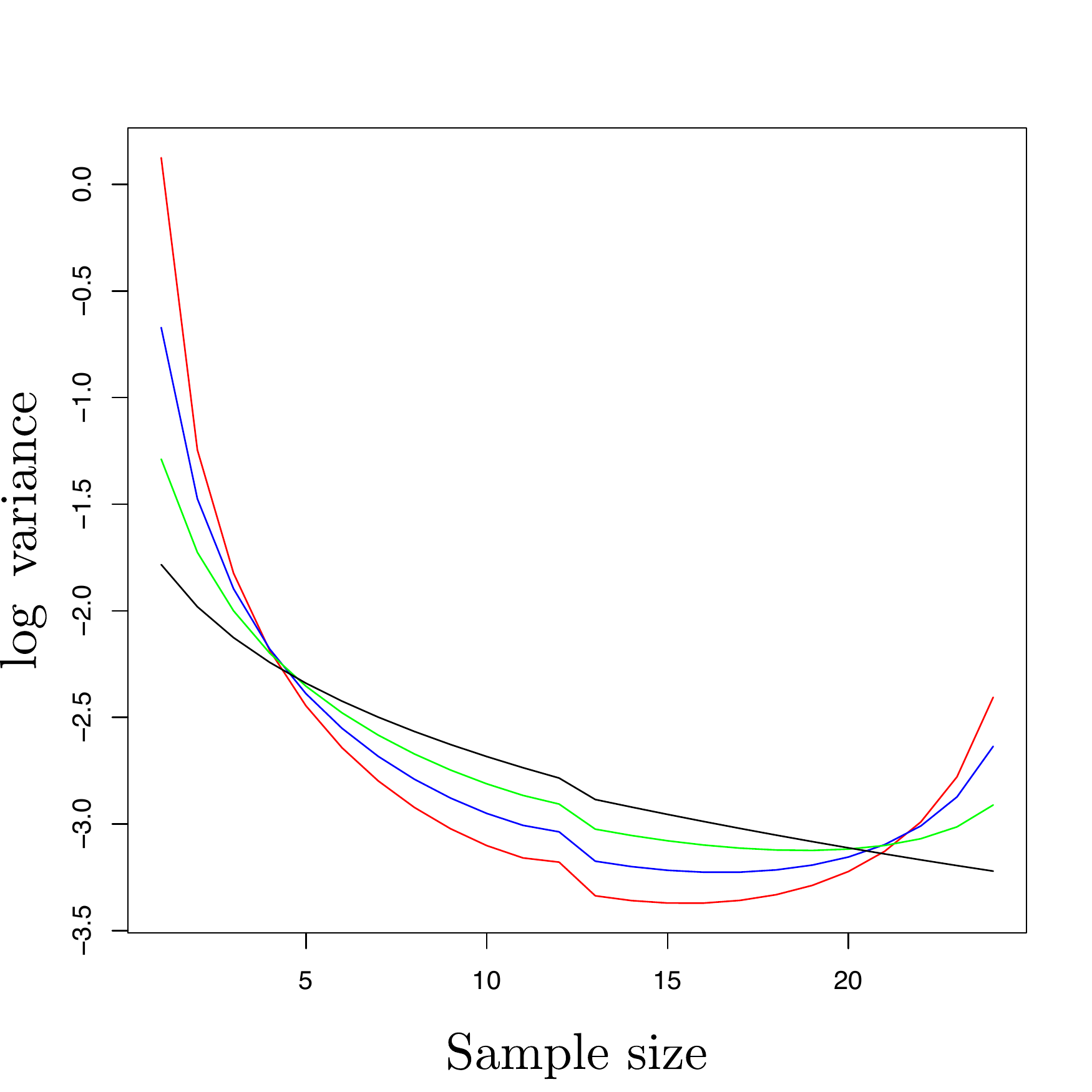}
\includegraphics[scale=0.28]{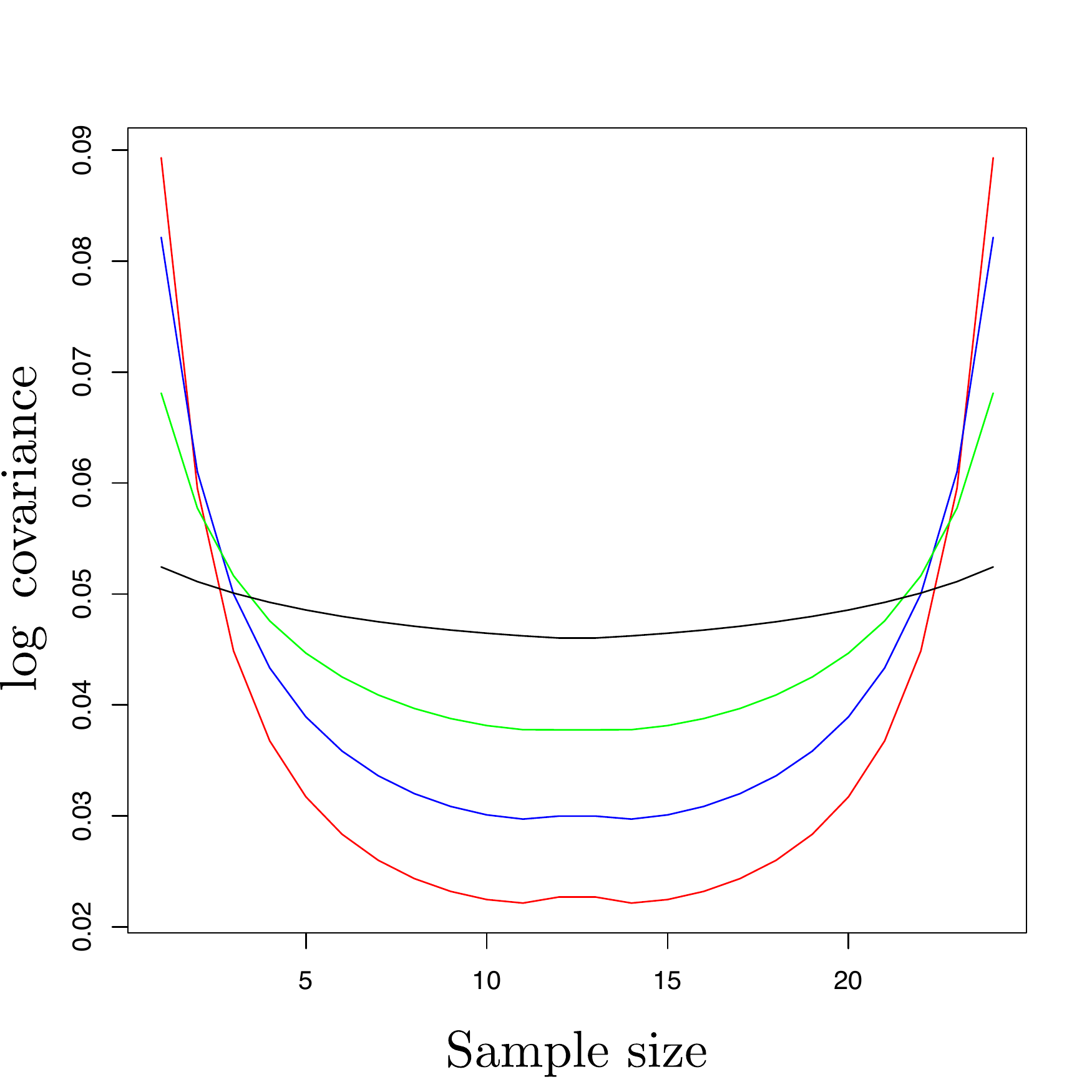} 
\caption{\label{fig:SFS-Beta} Beta-coalescence  for $\alpha=1.25$ (red), $\alpha=1.50$ (blue), $\alpha=1.75$ (green) and Kingman's coalescent (black). Left: logarithm of Expected SFS.  Middle: log variances of SFS. Right: log anti-diagonal values for the covariance matrix of the SFS }
\end{figure}
Concerning the covariances we only plot the anti-diagonal entries of the covariance matrix as in \cite{Durrett2008}~p.56. 
 Our results for the mean, variance and covariance agree with those obtained by the recursive formulae presented in \cite{Blath}. Higher order moments can also be calculated using  \eqref{eq:higher-order-moments}. For example, the result of the paper \cite{KLASSMANN} essentially reduces to calculating the formula \eqref{eq:third-order-moment}. 

 \end{example}

\section{Ancestral graph with recombination}\label{sec:recomb}
In this section we show how multivariate phase-type theory fits as a model for the distribution of branch length and can be used to express expected summary statistics for statistical associations of mutation patterns at different loci. We begin with a sample of size $n=2$ and then extend to larger sample sizes. 
\subsection{Sample size two}
\begin{figure}[H]
\begin{center}
\begin{tikzpicture}[scale=0.35,line width=2pt,every node/.style={scale=0.35}]

\draw(0,0) node[anchor=center]{\stateOne};
\draw(3,1) node[anchor=south east]{\scalebox{1.5}{\LARGE \bf 1}};

\draw(0,6) node[anchor=center]{\stateTwo};
\draw(3,7) node[anchor=south east]{\scalebox{1.5}{\LARGE \bf 2}};

\draw(0,12) node[anchor=center]{\stateThree};
\draw(3,13) node[anchor=south east]{\scalebox{1.5}{\LARGE \bf 3}};

\draw(-6,0) node[anchor=center]{\stateFour};
\draw(-3,-2) node[anchor=south east]{\scalebox{1.5}{\LARGE \bf 6}};

\draw(6,0) node[anchor=center]{\stateFive};
\draw(4,-2) node[anchor=south east]{\scalebox{1.5}{\LARGE \bf 4}};

\draw(-12,0) node[anchor=center]{\stateSix};
\draw(-14,-2) node[anchor=south east]{\scalebox{1.5}{\LARGE \bf 7}};

\draw(12,0) node[anchor=center]{\stateSeven};
\draw(15,-2) node[anchor=south east]{\scalebox{1.5}{\LARGE \bf 5}};

\draw(0,-12) node[anchor=center]{\scalebox{0.8}{\stateEight}};
\draw(0,-6) node[anchor=center]{\scalebox{0.8}{\stateNine}};

\draw(2.5,-13) node[anchor=north]{\scalebox{1.5}{\LARGE \bf 8}};

\draw(0,-9) node[anchor=center,scale=3]{or};
\draw[-,line width=1.5mm] (-2,-4)--(2,-4)--(2,-14)--(-2,-14)--(-2,-3.82);
\draw[->,>=stealth',thick] (1,2) to[bend right=15]
    node[midway,right] {\scalebox{1.4}{\LARGE $\rho$}} (1,4);
\draw[->,>=stealth',thick,shorten >=3pt] (0,-2) to[bend right=15]
    node[midway,left] {\scalebox{1.4}{\LARGE 1}} (0,-4);
\draw[->,>=stealth',thick] (-1,4) to[bend right=15]
    node[midway,left] {\scalebox{1.4}{\LARGE 1}} (-1,2);
\draw[->,>=stealth',thick] (1,8) to[bend right=15]
    node[midway,right] {\scalebox{1.4}{\LARGE $\rho/2$}} (1,10);
\draw[->,>=stealth',thick] (-2,6) to[bend right=15]
    node[midway,right,xshift=2mm] {\scalebox{1.4}{\LARGE 1}} (-6,2);
\draw[->,>=stealth',thick] (2,6) to[bend left=15]
    node[midway,left,xshift=-2mm] {\scalebox{1.4}{\LARGE 1}} (6,2);
\draw[->,>=stealth',thick] (-1,10) to[bend right=15]
    node[midway,left] {\scalebox{1.4}{\LARGE 4}} (-1,8);
\draw[->,>=stealth',thick] (-2,12) to[bend right=15]
    node[midway,right,xshift=2mm] {\scalebox{1.4}{\LARGE 1}} (-12,2);
\draw[->,>=stealth',thick] (2,12) to[bend left=15]
    node[midway,left,xshift=-2mm] {\scalebox{1.4}{\LARGE 1}} (12,2);
\draw[->,>=stealth',shorten >=3pt,thick] (-6,-2) to[bend right=15]
    node[midway,right,xshift=2mm] {\scalebox{1.4}{\LARGE 1}} (-2,-6);
\draw[->,>=stealth',thick] (-8,1) to[bend right=15]
    node[midway,above] {\scalebox{1.4}{\LARGE $\rho/2$}} (-10,1);
\draw[->,>=stealth',shorten >=3pt,thick] (6,-2) to[bend left=15]
    node[midway,left,xshift=-2mm] {\scalebox{1.4}{\LARGE 1}} (2,-6);
\draw[->,>=stealth',shorten >=3pt,thick] (8,1) to[bend left=15]
    node[midway,above] {\scalebox{1.4}{\LARGE $\rho/2$}} (10,1);
\draw[->,>=stealth',thick] (-10,-1) to[bend right=15]
    node[midway,below] {\scalebox{1.4}{\LARGE 2}} (-8,-1);
\draw[->,>=stealth',shorten >=3pt,thick] (-12,-2) to[bend right=15]
    node[midway,right,xshift=2mm] {\scalebox{1.4}{\LARGE 1}} (-2,-12);
\draw[->,>=stealth',thick] (10,-1) to[bend left=15]
    node[midway,below] {\scalebox{1.4}{\LARGE 2}} (8,-1);
\draw[->,>=stealth',shorten >=3pt,thick] (12,-2) to[bend left=15]
    node[midway,left,xshift=-2mm] {\scalebox{1.4}{\LARGE 1}} (2,-12);
\draw(-12,11) node[anchor=center]{\scalebox{1.4}{\LARGE Left tree taller}};
\draw(-12,10) node[anchor=center]{\scalebox{1.4}{\LARGE $\tau_a>\tau_b$}};
\draw(12,11) node[anchor=center]{\scalebox{1.4}{\LARGE Right tree taller}};
\draw(12,10) node[anchor=center]{\scalebox{1.4}{\LARGE $\tau_a<\tau_b$}};

\end{tikzpicture}
\caption{\label{fig:recomb} Flow diagram for the two-locus ancestral recombination graph. }\end{center}
\end{figure}
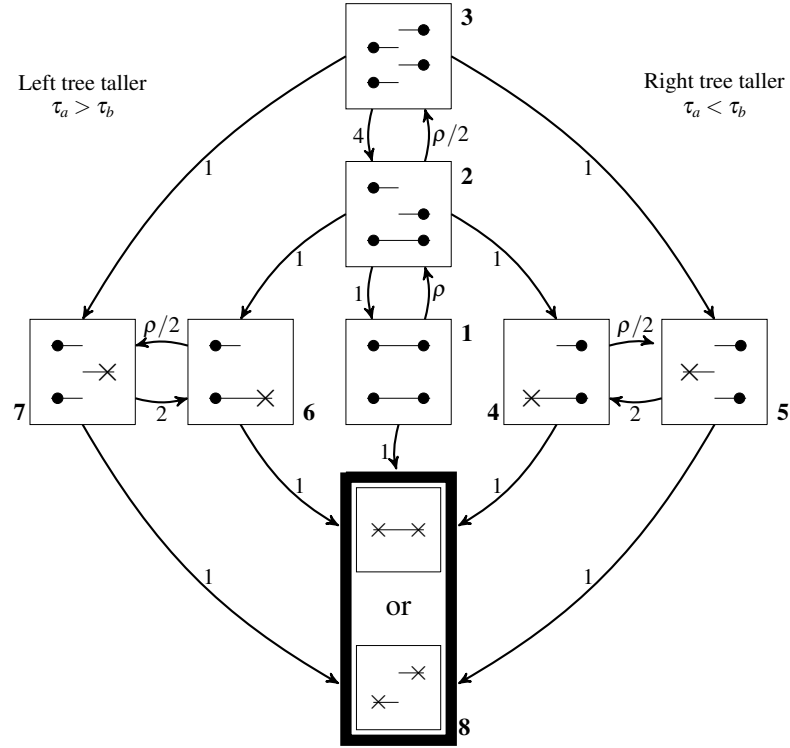

Recall the ancestral recombination graph for two loci and two samples originally presented in \cite{SimChu1997TPB}, and summarized as Figure 7.7 in \cite{wakeley2008coalescent}, and recently discussed in detail in~\cite{Hobolth201448}. For reference the graph is reproduced here in Figure~\ref{fig:recomb}. The filled circles represent material ancestral to the sample, and the crosses represent that the most common ancestor has been found. The lines between the circles or crosses indicate if the ancestral material is present in the same chromosome. The starting state is state~1 at present day with two samples from the same chromosome. 

The time when both loci have found their common ancestor is $\mbox{PH}_7(\vect{\alpha},\mat{S})$ distributed with $\vect{\alpha}=(1,0,0,0,0,0,0)$ and
\begin{equation}
 \mat{S}= \left( \begin{array}{ccc|cc|cc}
-1-\rho & \rho & 0 & 0 & 0 & 0 & 0\\
1 & -3 - \rho/2 &  \rho/2  & 1 & 0 & 1 & 0 \\
0 & 4 & -6 & 0 & 1 & 0 & 1 \\ \hline
0 & 0 & 0 & -1-\rho/2 &  \rho/2 & 0 & 0  \\
0 & 0 & 0 & 2 & -3 & 0 & 0 \\\hline
0 & 0 & 0 & 0 & 0 & -1-\rho/2 & \rho/2 \\
0 & 0 & 0 & 0 & 0 & 2 & -3
\end{array} \right), \ \ \vect{s}=
\begin{pmatrix}
1 \\
0 \\
0 \\ \hline
1 \\
1 \\ \hline
1 \\
1 
\end{pmatrix}.
 \label{eq:S-mat-jens-asger}
 \end{equation}
We observe that $\mat{S}$ has the natural block structure partitioning
\[  \mat{S}= \begin{pmatrix}
\mat{S}_{11} & \mat{S}_{12} & \mat{S}_{13}\\
  \mat{0} & \mat{S}_{22} & \mat{0} \\
  \mat{0} & \mat{0} & \mat{S}_{33} 
\end{pmatrix}, \ \ \vect{s}=\begin{pmatrix}
\vect{s}_1 \\
\vect{s}_2 \\
\vect{s}_3
\end{pmatrix}, \]
as already indicated in \eqref{eq:S-mat-jens-asger}. Also note the highly symmetric structure of the partitioning where $\mat{S}_{12}=\mat{S}_{13}$, $\mat{S}_{22}=\mat{S}_{33}$ and $\vect{s}_2=\vect{s}_3$.

\begin{figure}[htb]
  \centering
  \includegraphics[scale=0.37]{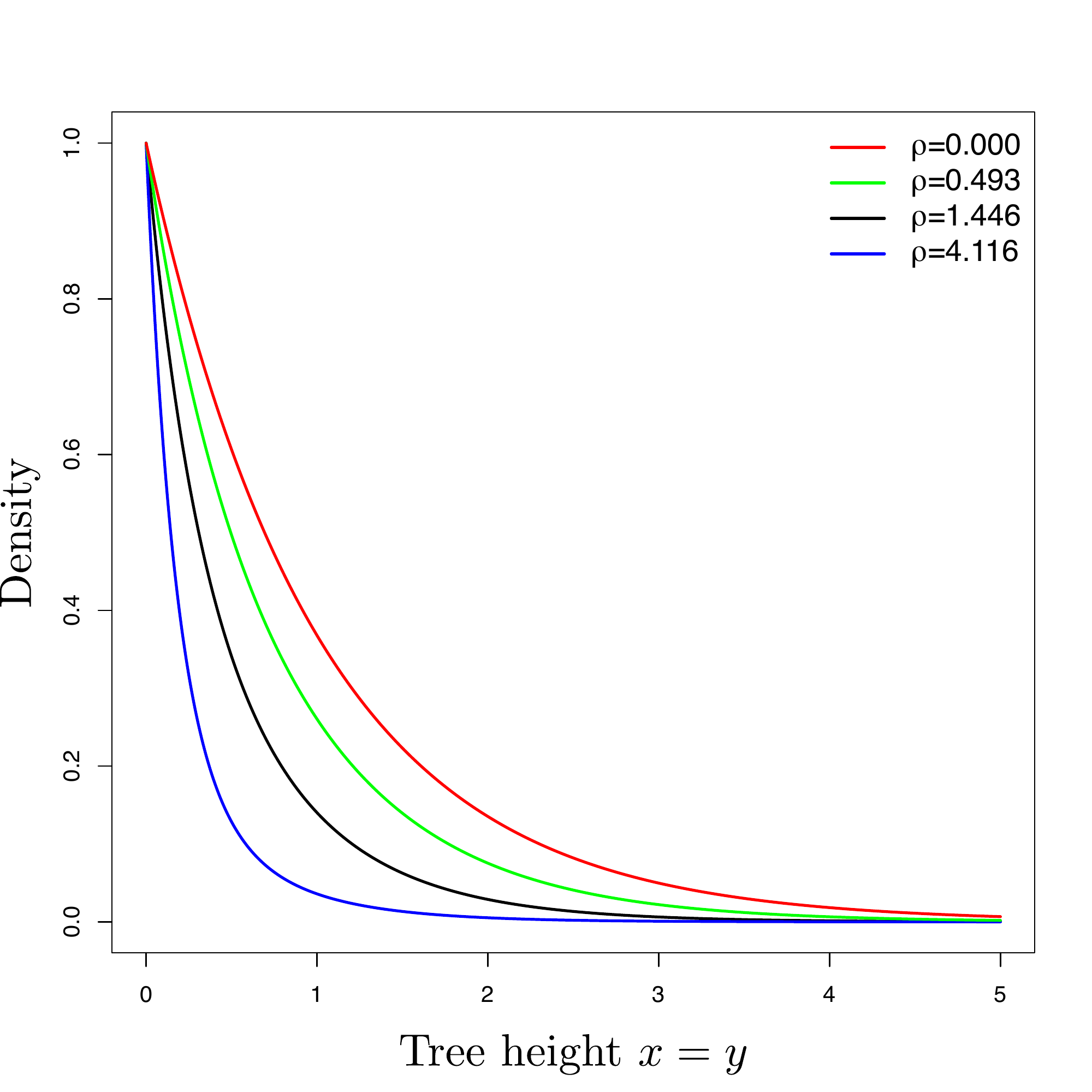}
   \includegraphics[scale=0.37]{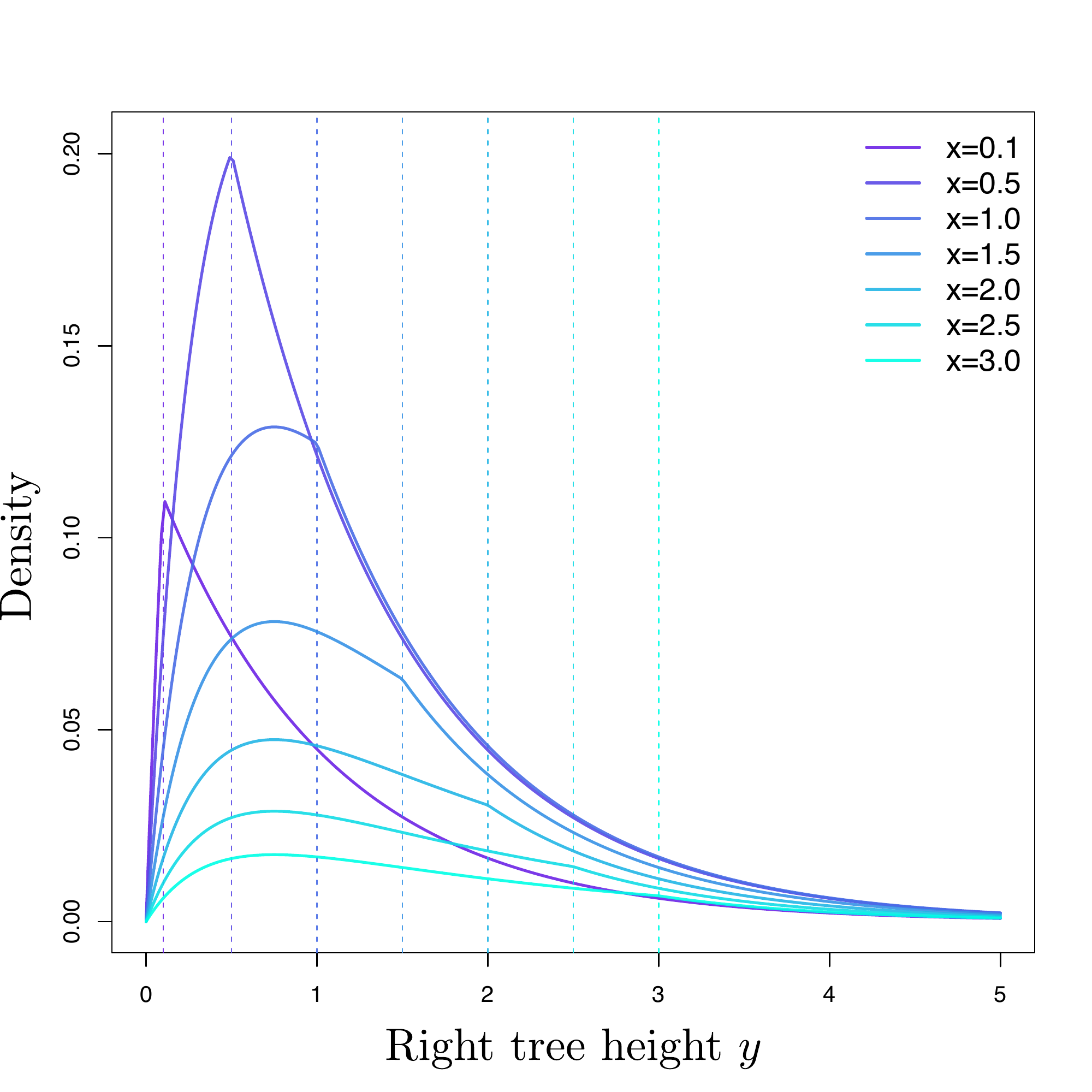}
 \caption{Left: The density for equal tree height~(\ref{eq:dens-of-common-tau}) for values of $\rho\in\{0,0.493,1.446,4.116\}$. The density integrates to $(1,0.75,0.50.25)$ for these values of $\rho$ such that for e.g. $\rho=1.446$ the probability for the two tree heights being equal is 0.5. Right: The joint density (\ref{JointTreeHeightForTwoSamples}) for a right tree height~$y$ for various values of a left tree height $x$.}
  \label{Fig:SimonsenChurchill}
\end{figure}

If the Markov jump process underlying $\mbox{PH}_7(\vect{\alpha},\mat{S})$  exits to the absorbing state from a state in $\{ 1,2,3\}$, then the height~$\tau_a$ of the left tree and the height~$\tau_b$ of the right tree are the same, i.e. $\tau_a=\tau_b$ with the common height being phase-type distributed with representation $\mbox{PH}_3(\vect{\alpha}_1,\mat{S}_{11})$ where $\vect{\alpha}_1=(1,0,0)$. The common distribution of $\tau_a=\tau_b$ then has density
 \begin{equation}
   f(x) =(1,0,0) \exp \left\{ \begin{pmatrix}
  -1-\rho & \rho & 0 \\
  1 & -3-\rho/2 & \rho/2 \\
  0 & 4 & -6
  \end{pmatrix} x\right\} \begin{pmatrix}
  1 \\
  0 \\
  0
  \end{pmatrix} . \label{eq:dens-of-common-tau} \end{equation}
The density for equal tree height is shown in the left plot in Figure~\ref{Fig:SimonsenChurchill}.
This is a defective distribution since second and third exit rates are set to zero prohibiting the process to jump to the left or right states of the diagram so
 \[  
 -\mat{S}_{11}\vect{e}\neq \vect{s}_1=\begin{pmatrix}
 1 \\
 0 \\
 0
 \end{pmatrix}. 
\]
The missing mass is exactly  the probability of this occurring and amounts to
\[ 1-\int_0^\infty f(x)dx = 1-\vect{\alpha}_1(-\mat{S}_{11})^{-1}\vect{s}_1 .\]
The density function \eqref{eq:dens-of-common-tau} can also be evaluated explicitly, i.e. expressed in terms of polynomials and exponentials involving $\rho$ and $x$. However, this expression is lengthy and messy since the eigenvalues of the intensity matrix are not particularly nice functions. On the other hand, for specific numeric values of $\rho$ the numeric calculation of \eqref{eq:dens-of-common-tau} is straightforward and efficient. Thus there seems to be no reason for pursuing a non-matrix representation of \eqref{eq:dens-of-common-tau} in practice. 

Now let us consider the case where $\tau_a\neq \tau_b$. Assume that $x=\tau_a<y=\tau_b$. Then the right tree is taller, and we must exit from states $\{1,2,3\}$ to $\{4,5\}$ at time $x$. The $3$-dimensional row vector
\[ 
  \vect{\alpha}_1 e^{\mat{S}_{11}x}  
\]
contains the probabilities of being in state $1$, $2$ or $3$ when exiting while the $2$-dimensional row vector
 \[  \vect{\pi}_1= \vect{\alpha}_1 e^{\mat{S}_{11}x}\mat{S}_{12} \]
contains the probabilities that states $4$ and $5$ are entered. 
Thus $ \vect{\pi}_1$ serves as the initial (defective) distribution of entering states $\{4,5\}$, and the remaining time spent in states $\{4,5\}$ prior to absorption is hence phase-type distributed $\mbox{PH}_2(\vect{\pi}_1,\mat{S}_{22})$. Hence we conclude that the joint density for $(\tau_a,\tau_b)$, $f_{(\tau_a,\tau_b)}(x,y)$, for the case of $x<y$ is
\[  
   f_{(\tau_a,\tau_b)}(x,y) = 
   \vect{\alpha}_1e^{\mat{S}_{11} x}
   \mat{S}_{12}
   e^{\mat{S}_{22}(y-x)}\vect{s}_2, 
   \ \ \ 
   x<y. 
\]
Similarly, for the case of $x>y$ we get that 
\[ 
   f_{(\tau_a,\tau_b)}(x,y) = 
   \vect{\alpha}_1e^{\mat{S}_{11} y}
   \mat{S}_{13}
   e^{\mat{S}_{33}(x-y)}\vect{s}_2, 
   \ \ \
   x>y,
\]
and since $\mat{S}_{22}=\mat{S}_{33}$ and $\mat{S}_{12}=\mat{S}_{13}$ we get that the two densities are identical. 

We can perform a reduction of the state-space. The exit rates are 
\[  \vect{s}_2=\vect{s}_3= \begin{pmatrix}
1 \\
1
\end{pmatrix}, \]
and therefore the phase-type distributions corresponding to the states $\{4,5\}$ and $\{6,7\}$ are both exponential distributions with rate $1$ (recall equation \eqref{eq:PH-exponential-dist} and the following remark). Thus the direct inter-action between states $4$ and $5$ (respectively $6$ and $7$) has no practical effect and we can reduce $\mat{S}$ to 
\begin{equation}
\tilde{\mat{S}}= 
\left( \begin{array}{ccc|c|c}
-1-\rho & \rho & 0 & 0  & 0 \\
1 & -3 - \rho/2 &  \rho/2  & 1  & 1  \\
0 & 4 & -6 & 1 &  1 \\ \hline
0 & 0 & 0 & -1 & 0   \\ \hline
0 & 0 & 0 & 0  & -1  \\
\end{array} \right), \ \ \tilde{\vect{s}}=
\begin{pmatrix}
1 \\
0 \\
0 \\ \hline
1 \\ \hline
1 
\end{pmatrix} .
 \label{eq:S-mat-jens-asger-1}
 \end{equation}
The corresponding joint densities are then given by
 \begin{equation}
 f_{(\tau_a,\tau_b)}(x,y)= (1,0,0) \exp 
 \left\{ \begin{pmatrix}
 -1-\rho & \rho & 0 \\
 1 & -3-\rho/2 & \rho/2 \\
 0 & 4 & -6
 \end{pmatrix} 
 x\right\} 
 \begin{pmatrix}
 0 \\
 1 \\
 1
 \end{pmatrix} e^{-(y-x)}  
 \label{JointTreeHeightForTwoSamples}
  \end{equation}
for $x<y$ and vice versa for $x>y$. 
The density is illustrated in the right plot in Figure~\ref{Fig:SimonsenChurchill}.

\cite{wakeley2008coalescent} notes that the inter-actions between states $4,5$ and $6,7$ are not needed. This remark results in the reduction 
\[  \mat{S}_{22}=\mat{S}_{33}= \begin{pmatrix}
-1 & 0 \\ 0 & -1
\end{pmatrix}  \]
where the two states are preserved instead of collapsing them into a single one as in our case where $\mat{S}_{22}=\mat{S}_{33}=\{ -1 \}$. This representation of course results in the same joint density as above.

\subsection{General sample size}
\begin{figure}[htb!]
  \centering
  \scalebox{.24}{\includegraphics{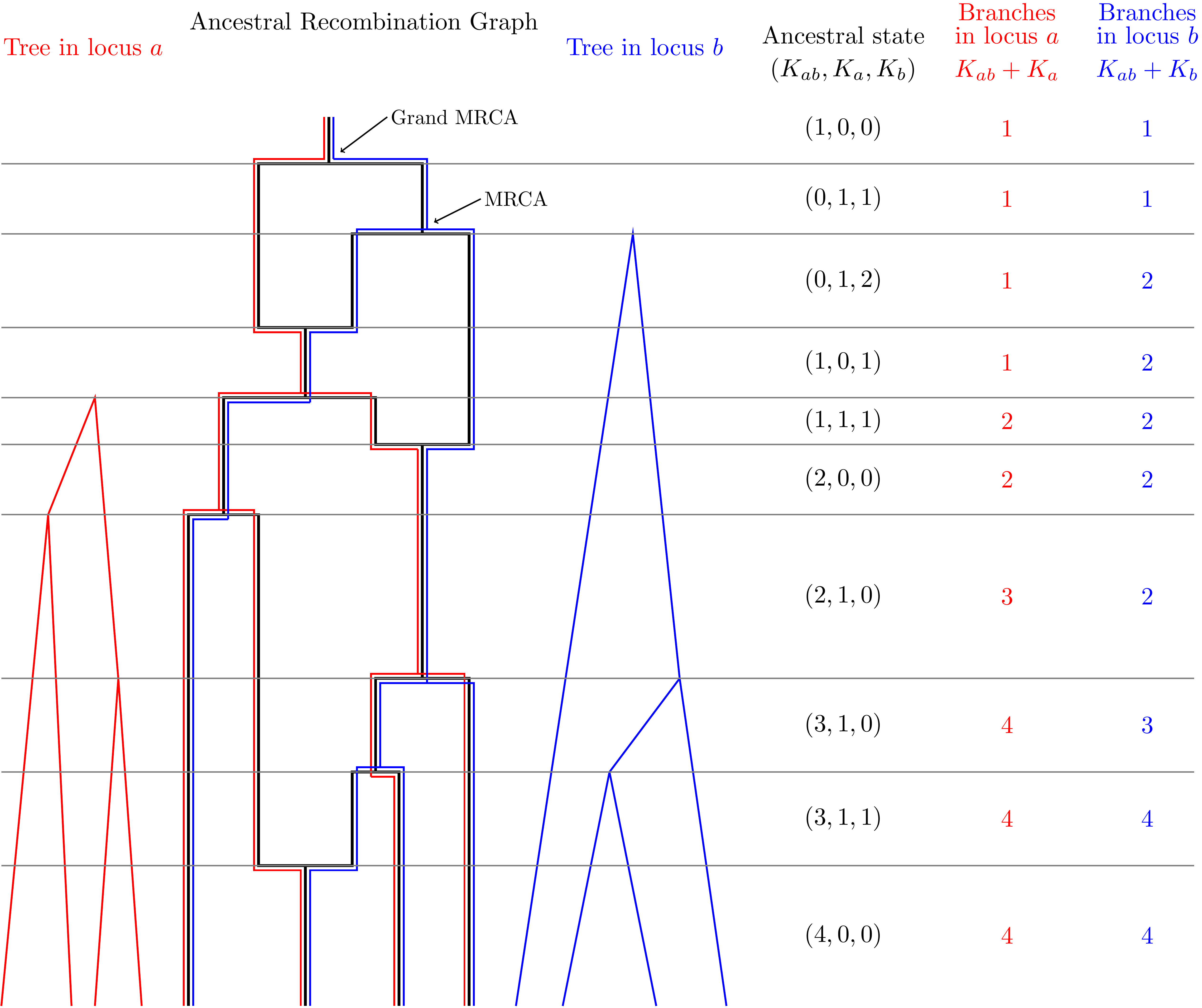}}
  \caption{Ancestral recombination graph (in black) for two loci and four sequences and the corresponding trees in the left locus (in red) and right locus (in blue). The figure is adapted from Figure~7.5 in \cite{wakeley2008coalescent}.}
  \label{WakeleyFig75}
\end{figure}

In Figure~\ref{WakeleyFig75} we recapitulate Figure~7.5 page~226 in \cite{wakeley2008coalescent} and introduce the notation. Four linked sequences have evolved back in time according to the ancestral recombination graph. We are interested in the joint distribution of the total branch length $\mathcal{L}_a$ in locus~$a$ and the total branch length $\mathcal{L}_b$ in locus~$b$. This process was recently studied using a rather complex hyperbolic system of partial differential equations \cite{MiSt2017} . We avoid labelling the sequences and consider the number of sequences $K_{ab}$ with ancestral material in both loci, the number of sequences $K_a$ with ancestral material in locus~$a$ only, and the number of sequences $K_b$ with ancestral material in locus~$b$ only.     

Define the state of the ancestral recombination graph at time $t$ to be $A(t)=(K_{ab}(t),K_a(t),K_b(t))$. The number of branches in the two loci at time $t$ is then $L_a(t)=K_{ab}(t)+K_a(t)$ and $L_b(t)=K_{ab}(t)+K_b(t)$. The time to the most recent common ancestor (the tree height) in each locus is given by
\begin{eqnarray*}
  \tau_a=\inf\{t\geq 0:L_a(t)=1\} \;\;
  {\rm and} \;\;
  \tau_b=\inf\{t\geq 0:L_b(t)=1\}.
\end{eqnarray*}
The total branch length in each locus is
\begin{eqnarray*}
  \mathcal{L}_a=\int_0^{\tau_a} L_a(t)dt \;\;
  {\rm and} \;\;
  \mathcal{L}_b=\int_0^{\tau_b} L_b(t)dt.
\end{eqnarray*}
Similarly as for two samples we want to study the joint distribution of $(\mathcal{L}_a,\mathcal{L}_b)$ as a function of the recombination rate~$\rho$. 

The ancestral process for two loci and a sample of $n$ unlabelled sequences has a state-space given by triplets $(k_{ab},k_a,k_b)$
where entries are non-negative integers with $k_{ab}+\max\{k_a,k_b\} \leq n$ and with triplets $(0,k_a,0)$ for $0\leq k_a\leq n$ and $(0,0,k_b)$ for $0\leq k_b \leq n$ removed.
The grand MRCA $(1,0,0)$ is defined to be the absorbing state because at that time all the ancestral sequences have found common ancestry. 

The rates between the states are given by 
\begin{eqnarray}
 \mat Q=\mat Q^c+\frac{\rho}{2} \mat Q^r,
 \label{coal+recomb}
\end{eqnarray}  
where the transitions that correspond to coalescent events are 
\begin{eqnarray*}
  q^c_{(k_{ab},k_a,k_b),(k_{ab}-1,k_a,k_b)}&=&
    {k_{ab} \choose 2}\\
  q^c_{(k_{ab},k_a,k_b),(k_{ab},k_a-1,k_b)}&=&
    {k_a \choose 2}+k_{ab}k_a\\
  q^c_{(k_{ab},k_a,k_b),(k_{ab},k_a,k_b-1)}&=&
    {k_b \choose 2}+k_{ab}k_b,
\end{eqnarray*}
and the transitions that correspond to recombination events are
\begin{eqnarray*}
  q^r_{(k_{ab},k_a,k_b),(k_{ab}-1,k_a+1,k_b+1)}&=&k_{ab}.\\
\end{eqnarray*}

\begin{figure}[htb!]
  \centering
  \scalebox{.36}{\includegraphics{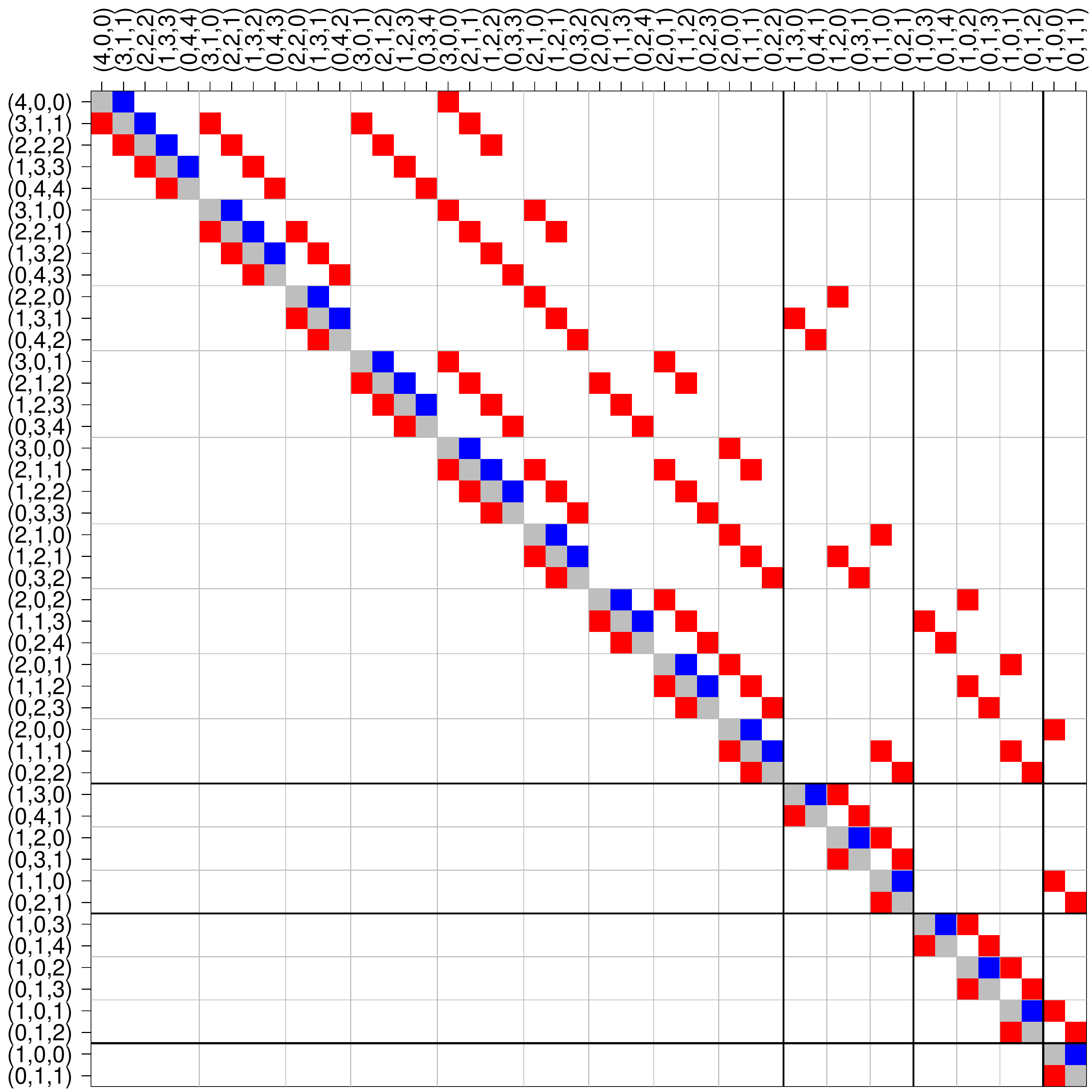}}
  \caption{State-space and rates for four sequences. Red entries in the rate matrix correspond to coalescent events and blue entries in the rate matrix correspond to recombination events.}
  \label{StateDiagramFourFig}
\end{figure}

Consider the case $n=4$. In Figure~\ref{StateDiagramFourFig} we illustrate the state space and the rates between states. 
The intensity matrix is indexed in the order of $(K_{ab}+K_a,K_{ab}+K_b)$ such that we begin with the 9 blocks
\[ (4,4),(4,3),(4,2),(3,4), (3,3),(3,2),(2,4),(2,3),(2,2) , \]
where both loci have at least two lineages. The next 3 blocks are 
$ (4,1),(3,1),(2,1)$,
where the tree in locus $b$ is finished. Then we have the 3 blocks  
$(1,4),(1,3),(1,2)$
where the tree in locus $a$ is finished.
The final block $(1,1)$ is the overall absorbing state. 
In a block-partioned form we write the intensity matrix as follows:
\begin{eqnarray*}
&&
\hspace{-1cm}\left(\begin{array}{ccccccccc|ccc|ccc|c}
\bf{A}_{44}^{44} & \bf{A}_{44}^{43} & \bf{0} & \bf{A}_{44}^{34} & \bf{A}_{44}^{33} & \bf{0} & \bf{0} & \bf{0} & \bf{0} & \textcolor{red}{\bf{0}} & \textcolor{red}{\bf{0}} & \textcolor{red}{\bf{0}} & \textcolor{blue}{\bf{0}} & \textcolor{blue}{\bf{0}} & \textcolor{blue}{\bf{0}} & \bf{0}\\
   \bf{0} & \bf{A}_{43}^{43} & \bf{A}_{43}^{42} & \bf{0} & \bf{A}_{43}^{33} & \bf{A}_{43}^{32} & \bf{0} & \bf{0} & \bf{0} & \textcolor{red}{\bf{0}} & \textcolor{red}{\bf{0}} &\textcolor{red}{\bf{0}} &\textcolor{blue}{\bf{0}} &\textcolor{blue}{\bf{0}} &\textcolor{blue}{\bf{0}} & \bf{0}  \\    
   \bf{0} & \bf{0} & \bf{A}_{42}^{42} & \bf{0} &\bf{0} &\bf{A}_{42}^{32} & \bf{0} &\bf{0} &\bf{0} & \textcolor{red}{\bf{A}_{42}^{41}} & \textcolor{red}{\bf{A}_{42}^{31}} & \textcolor{red}{\bf{0}} &\textcolor{blue}{\bf{0}} &\textcolor{blue}{\bf{0} }&\textcolor{blue}{\bf{0}} &\bf{0} \\
   \bf{0} &\bf{0} &\bf{0} & \bf{A}_{34}^{34} & \bf{A}_{34}^{33} & \bf{0} & \bf{A}_{34}^{24} & \bf{A}_{34}^{23} & \bf{0} & \textcolor{red}{\bf{0}} &\textcolor{red}{\bf{0}} &\textcolor{red}{\bf{0}} &\textcolor{blue}{\bf{0}} &\textcolor{blue}{\bf{0}} & \textcolor{blue}{\bf{0}} &\bf{0} \\ 
   \bf{0} &\bf{0} &\bf{0} &\bf{0} & \bf{A}_{33}^{33} & \bf{A}_{33}^{32} & \bf{0} & \bf{A}_{33}^{23} & \bf{A}_{33}^{22} &   \textcolor{red}{\bf{0}} &\textcolor{red}{\bf{0}} &\textcolor{red}{\bf{0}} & \textcolor{blue}{\bf{0}} &\textcolor{blue}{\bf{0}} &\textcolor{blue}{\bf{0}} &\bf{0} \\
   \bf{0} &\bf{0} &\bf{0} &\bf{0} &\bf{0} & \bf{A}_{32}^{32} & \bf{0} &\bf{0} & \bf{A}_{32}^{22} & \textcolor{red}{\bf{0}} &\textcolor{red}{\bf{A}_{32}^{31}} &\textcolor{red}{\bf{A}_{32}^{21}} & \textcolor{blue}{\bf{0}} &\textcolor{blue}{\bf{0}} &\textcolor{blue}{\bf{0}} &\bf{0} \\
   \bf{0} &\bf{0} &\bf{0} &\bf{0} &\bf{0} &\bf{0} &\bf{A}_{24}^{24} &  \bf{A}_{24}^{23} & \bf{0} & \textcolor{red}{\bf{0}} &\textcolor{red}{\bf{0}} &\textcolor{red}{\bf{0}} &\textcolor{blue}{\bf{A}_{24}^{14}} & \textcolor{blue}{\bf{A}_{24}^{13}} & \textcolor{blue}{\bf{0}} & \bf{0} \\
   \bf{0} &\bf{0} &\bf{0} &\bf{0} &\bf{0} &\bf{0} &\bf{0} & \bf{A}_{23}^{23} & \bf{A}_{23}^{22} & \textcolor{red}{\bf{0}} & \textcolor{red}{\bf{0}} &\textcolor{red}{\bf{0}} & \textcolor{blue}{\bf{0}} & \textcolor{blue}{\bf{A}_{23}^{13}} & \textcolor{blue}{\bf{A}_{23}^{12}} & \bf{0}  \\
   \bf{0} &\bf{0} &\bf{0} &\bf{0} &\bf{0} &\bf{0} &\bf{0} &\bf{0} & \bf{A}_{22}^{22} & \textcolor{red}{\bf{0}} &\textcolor{red}{\bf{0}} &\textcolor{red}{\bf{A}_{22}^{21}} & \textcolor{blue}{\bf{0}} &\textcolor{blue}{\bf{0}} & \textcolor{blue}{\bf{A}_{22}^{12}} & \bf{A}_{22}^{11}\\\hline
    \bf{0} &\bf{0} &\bf{0} &\bf{0} &\bf{0} &\bf{0} &\bf{0} &\bf{0} &\bf{0} & \textcolor{red}{\bf{A}_{41}^{41}} & \textcolor{red}{\bf{A}_{41}^{31}} & \textcolor{red}{\bf{0}} &\bf{0} &\bf{0} &\bf{0} &\textcolor{red}{\bf{0}} \\
    \bf{0} &\bf{0} &\bf{0} &\bf{0} &\bf{0} &\bf{0} &\bf{0} &\bf{0} &\bf{0} &\textcolor{red}{\bf{0}} & \textcolor{red}{\bf{A}_{31}^{31}} & \textcolor{red}{\bf{A}_{31}^{21}} & \bf{0} &\bf{0} &\bf{0} &\textcolor{red}{\bf{0}} \\
    \bf{0} &\bf{0} &\bf{0} &\bf{0} &\bf{0} &\bf{0} &\bf{0} &\bf{0} &\bf{0} &\textcolor{red}{\bf{0}} &\textcolor{red}{\bf{0}} & \textcolor{red}{\bf{A}_{21}^{21}} & \bf{0} &\bf{0} &\bf{0} & \textcolor{red}{\bf{A}_{21}^{11}} \\\hline
   \bf{0} &\bf{0} &\bf{0} &\bf{0} &\bf{0} &\bf{0} &\bf{0} &\bf{0} &\bf{0} &\bf{0} &\bf{0} &\bf{0} &\textcolor{blue}{\bf{A}_{14}^{14}} & \textcolor{blue}{\bf{A}_{14}^{13}} & \textcolor{blue}{\bf{0}} & \textcolor{blue}{\bf{0}} \\
   \bf{0} &\bf{0} &\bf{0} &\bf{0} &\bf{0} &\bf{0} &\bf{0} &\bf{0} &\bf{0} &\bf{0} &\bf{0} &\bf{0} &\textcolor{blue}{\bf{0}} & \textcolor{blue}{\bf{A}_{13}^{13}} & \textcolor{blue}{\bf{A}_{13}^{12}} & \textcolor{blue}{\bf{0}}\\
  \bf{0} &\bf{0} &\bf{0} &\bf{0} &\bf{0} &\bf{0} &\bf{0} &\bf{0} &\bf{0} &\bf{0} &\bf{0} &\bf{0} & \textcolor{blue}{\bf{0}} & \textcolor{blue}{\bf{0}} & \textcolor{blue}{\bf{A}_{12}^{12}} & \textcolor{blue}{\bf{A}_{12}^{11}} \\\hline
   \bf{0} &\bf{0} &\bf{0} &\bf{0} &\bf{0} &\bf{0} &\bf{0} &\bf{0} &\bf{0} &\bf{0} &\bf{0} &\bf{0} &\bf{0} &\bf{0} &\bf{0} &\bf{0} 
   \end{array}\right)  . \\
   &=&\begin{pmatrix}
\mat{S}_{ab} & \textcolor{red}{\mat{S}_{ab}^a} & \textcolor{blue}{\mat{S}_{ab}^b} & \mat{S}_{ab}^0 \\
\mat{0} & \textcolor{red}{\mat{S}_a} & \mat{0} & \textcolor{red}{\mat{S}_a^0} \\
\mat{0} & \mat{0} & \textcolor{blue}{\mat{S}_{b}} & \textcolor{blue}{\mat{S}_b^0} \\
\vect{0} & \vect{0}& \vect{0}&  \vect{0} 
\end{pmatrix},\\
  \end{eqnarray*}
where the block matrices are define in the obvious way. 

If $\tau_a>\tau_b$ it is because there is a transition from the $\mat{S}_{ab}$ block to the red square $\textcolor{red}{\mat{S}_{a}}$, the transition of which is performed by the matrix $\textcolor{red}{\mat{S}_{ab}^{a}}$ in the red rectangle.
From $\textcolor{red}{\mat{S}_{a}}$ the remaining time is phase-type distributed with exit rate vector $\textcolor{red}{\mat{S}_a^0}$, denoted by the red rectangle at the level of $\mat{S}_{a}$. The situation where $\tau_b>\tau_a$ is entirely symmetrical.  
The density for $(\tau_a,\tau_b)$ is hence given by
 \begin{equation}
f_{(\tau_a,\tau_b)}(x,y)= \left\{  
\begin{array}{ll}
\vect{e}_1^\prime e^{\mat{S}_{ab}y}\textcolor{red}{\mat{S}_{ab}^a e^{\mat{S}_a(x-y)}\mat{S}_a^0} &  \; {\rm for} \; x>y \\
\vect{e}_1^\prime e^{\mat{S}_{ab}x}\mat{S}_{ab}^0 & \; {\rm for} \; x=y \\
\vect{e}_1^\prime e^{\mat{S}_{ab}x}\textcolor{blue}{\mat{S}_{ab}^b e^{\mat{S}_b(y-x)}\mat{S}_b^0} &  \; {\rm for} \; y>x
\end{array}
 \right. 
 \label{FourSeqDns}
 \end{equation}   
where $\vect{e}_1^\prime=(1,0,\ldots,0)$ because the first state (indexed by~(4,0,0)) is the starting state. 
\begin{figure}[htb]
  \centering
  \includegraphics[scale=0.37]{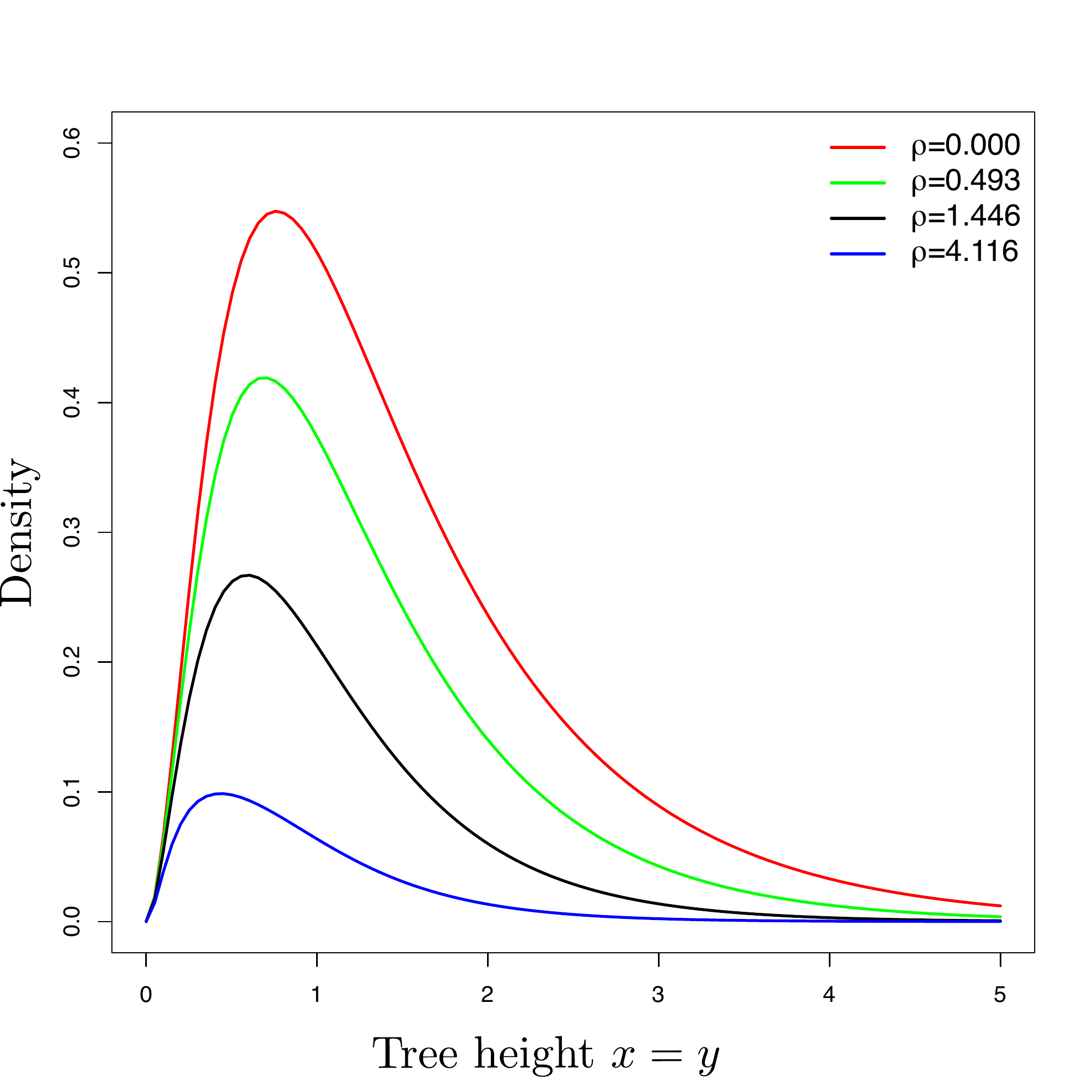}
   \includegraphics[scale=0.37]{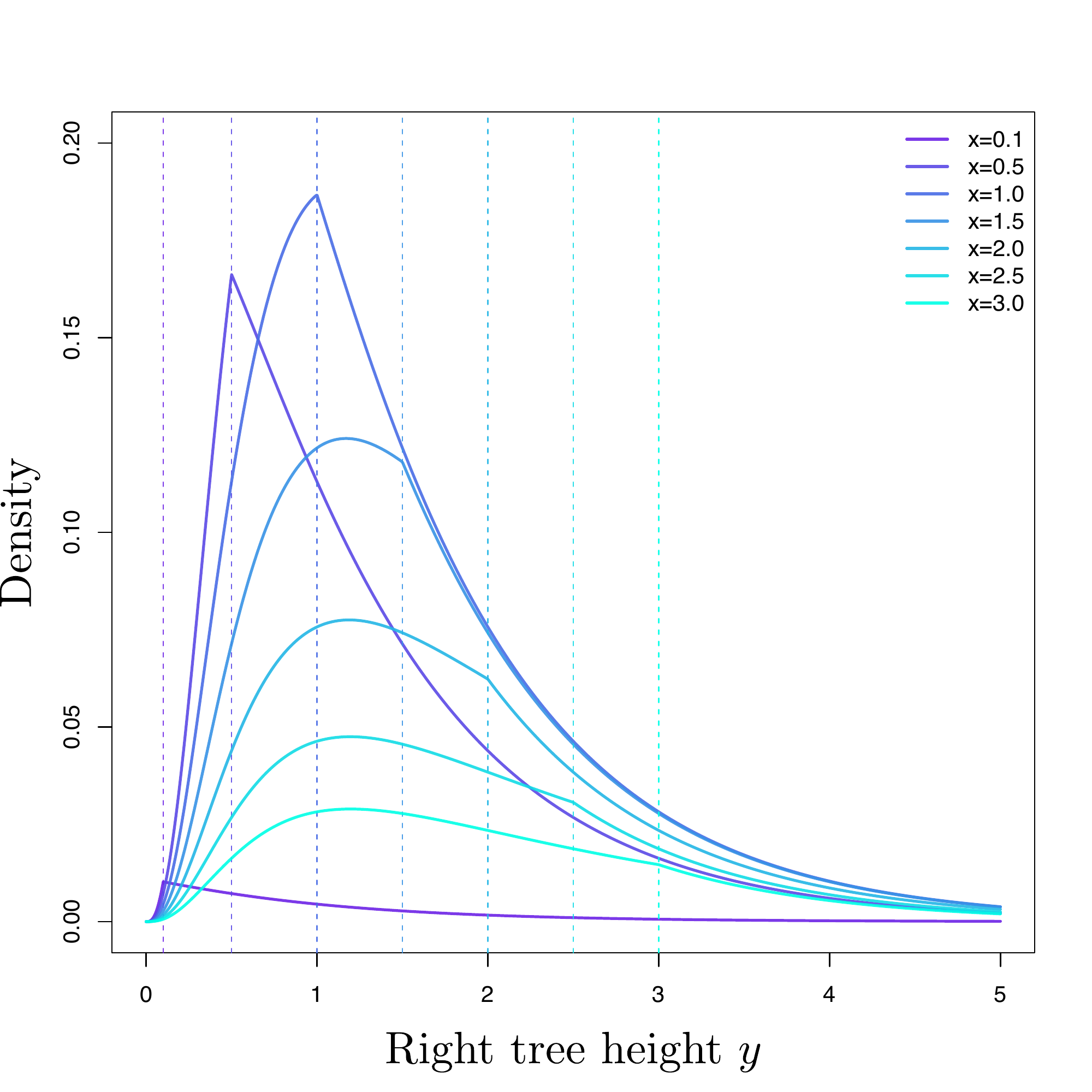}
 \caption{Left: The density for equal tree height~(\ref{FourSeqDns}) for values of $\rho\in\{0,0.493,1.446,4.116\}$. 
Right: The joint density (\ref{FourSeqDns}) for a right tree height~$y$ for various values of a left tree height $x$.}
  \label{Fig:FourSeqDensities}
\end{figure} 

Next we consider the total branch lengths. The reward matrix $\mat{R}$ is given by
\[ 
\mat{R} = \begin{pmatrix} 
4\vect{e} & 4\vect{e} \\
4\vect{e} & 3\vect{e}\\
4\vect{e} & 2\vect{e}\\
3\vect{e} & 4\vect{e}\\
3\vect{e} & 3\vect{e}\\
3\vect{e} & 2\vect{e} \\
2\vect{e} & 4\vect{e}\\
2\vect{e} & 3\vect{e}\\
2\vect{e} & 2\vect{e} \\ \hline
4\vect{e} & \vect{0}\\
3\vect{e} & \vect{0} \\
2\vect{e} & \vect{0}\\ \hline
\vect{0} & 4\vect{e}\\
\vect{0} & 3\vect{e}\\
\vect{0} & 2\vect{e} 
\end{pmatrix} ,
 \]
 where $\vect{e}$ are column vectors of ones and $\vect{0}$ zero vectors, all of appropriate dimensions. Then 
 \[ (\mathcal{L}_a,\mathcal{L}_b)\sim \mbox{MPH}^*(\vect{e}_1^\prime,\mat{S},\mat{R}),  \]
 where
 \[ \mat{S}= \begin{pmatrix}
 \mat{S}_{ab} & \mat{S}_{ab}^a & \mat{S}_{ab}^b \\
 \mat{0} & \mat{S}_a & \mat{0} \\
 \mat{0} & \mat{0} & \mat{S}_{ab}^b 
 \end{pmatrix}  . \]
 While the joint Laplace transform and (cross) moments have explicit forms, this is in general not the case for the densities and distribution functions in the $MPH^*$ class, and the case of $(\mathcal{L}_a,\mathcal{L}_b)$ presents such an example.

Let $\mat{R}_{\cdot i}$ denote the column number $i=1,2$ of $\mat{R}$.
Then 
\[ 
  \Exp (\mathcal{L}_a) =\vect{e}_1^\prime (-\mat{S})^{-1}\mat{R}_{\cdot 1} 
  \ \  \mbox{and} \ \ 
  \Exp (\mathcal{L}_b) =\vect{e}_1^\prime (-\mat{S})^{-1}\mat{R}_{\cdot 2}, 
\]
and
\[  
  \Exp (\mathcal{L}_a\mathcal{L}_b ) = \vect{e}_1^\prime (-\mat{S})^{-1}\mat{\Delta}(\mat{R}_{\cdot 1})(-\mat{S})^{-1}\mat{R}_{\cdot 2}  + \vect{e}_1^\prime (-\mat{S})^{-1}\mat{\Delta}(\mat{R}_{\cdot 2})(-\mat{S})^{-1}\mat{R}_{\cdot 1}  ,\]
 where $\mat{\Delta}(\vect{r})$ denotes the diagonal matrix with the vector $\vect{r}$ as diagonal. In particular,
\[\Exp (\mathcal{L}_a^2)=2  \vect{e}_1^\prime (-\mat{S})^{-1}\mat{\Delta}(\mat{R}_{\cdot 1})(-\mat{S})^{-1}\mat{R}_{\cdot 1} \]
and similarly for $\mathcal{L}_b$. Thus we can explicitly find the variances and covariance by substitution into the formulae
\[\mbox{Var}(\mathcal{L}_a)=\Exp (\mathcal{L}_a^2)-\Exp (\mathcal{L}_a)^2 \ \ \mbox{and}\ \   \mbox{Cov}(\mathcal{L}_a,\mathcal{L}_b)= \Exp (\mathcal{L}_a\mathcal{L}_b )- \Exp (\mathcal{L}_a ) \Exp (\mathcal{L}_b ) , \]
and correlation 
\[  
\mbox{Corr}(\mathcal{L}_a,\mathcal{L}_b) =\frac{\mbox{Cov}(\mathcal{L}_a,\mathcal{L}_b)}{\sqrt{\mbox{Var}(\mathcal{L}_a)}\sqrt{\mbox{Var}(\mathcal{L}_b)}}. 
\]

\begin{figure}[htb]
 \centering
   \includegraphics[scale=0.4]{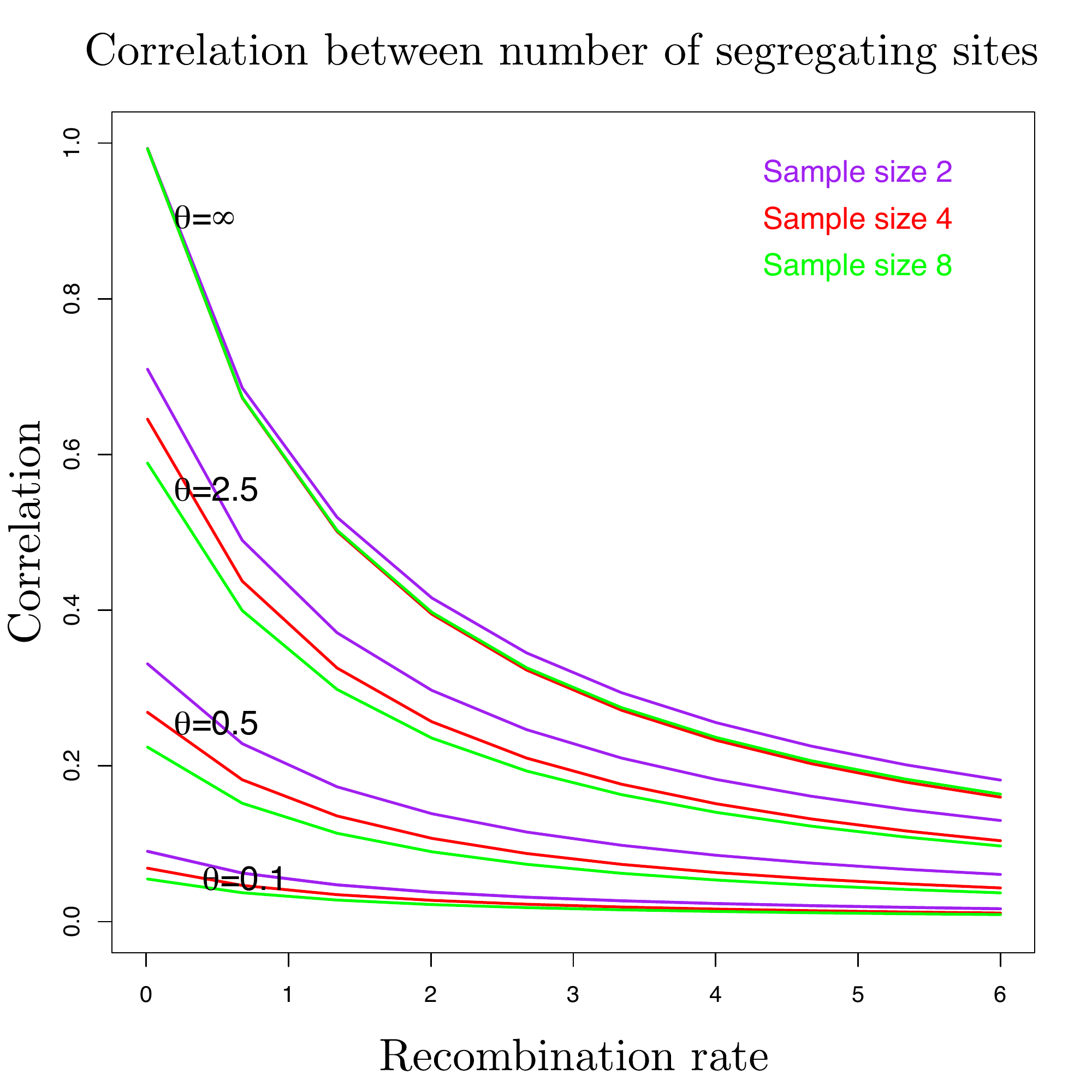}
   \caption{Correlation between the number of segregating sites~(\ref{Eq:CorrSegr}) at two loci for sample sizes $n=(2,4,8)$ and mutation rates $\theta=(0.1,0.5,2.5)$ and $\theta\rightarrow \infty$.}
   \label{Fig:SegregatingSitesCorr}
 \end{figure}

Now let $S_a$ and $S_b$ denote the number of segregating sites in locus~$a$  and locus~$b$, and let the mutation rates in the two loci be $\theta_a/2$ and $\theta_b/2$. 
Recall that 
$S_a|\mathcal{L}_a\sim{\rm Pois}(\mathcal{L}_a\theta_a/2)$ and 
$S_b|\mathcal{L}_b\sim{\rm Pois}(\mathcal{L}_b\theta_b/2)$ and 
$S_a|(\mathcal{L}_a,\mathcal{L}_b)$ is independent of  
$S_b|(\mathcal{L}_a,\mathcal{L}_b)$.
We have 
\begin{equation*}
  \Exp[S_a]=\Exp\Big[\Exp[S_a|\mathcal{L}_a]\Big]
  =\Exp\big[\mathcal{L}_a\theta_a/2\big]
  =\frac{\theta_a}{2}{\Exp}\big[\mathcal{L}_a\big],
\end{equation*}
and similarly $\Exp[S_b]=(\theta_b/2)\Exp\big[\mathcal{L}_b\big]$.
We get
\begin{equation*}
  \Exp[S_aS_b]
  =\Exp\Big[\Exp[S_aS_b|(\mathcal{L}_a,\mathcal{L}_b)]\Big]
  =\Exp\Big[\Exp[S_a|\mathcal{L}_a]
   \Exp[S_b|\mathcal{L}_b]\Big]
  =\frac{\theta_a\theta_b}{4}
   \Exp\big[\mathcal{L}_a\mathcal{L}_b\big],
\end{equation*}
and
\begin{equation*}
  {\rm Cov}[S_a,S_b]=
  \Exp[S_aS_b]-{\rm E}[S_a]{\rm E}[S_b]=
  \frac{\theta_a\theta_b}{4}
  {\rm Cov}\big[\mathcal{L}_a,\mathcal{L}_b\big].
\end{equation*}
Furthermore
\begin{eqnarray*}
  {\rm Var}[S_a]=
  {\rm Var}\Big[\Exp[S_a|\mathcal{L}_a]\Big]+
  \Exp\Big[{\rm Var}[S_a|\mathcal{L}_a]\Big]=
  {\rm Var}\big[\mathcal{L}_a\theta_a/2\big]+
  \Exp\big[\mathcal{L}_a\theta_a/2\big]=
  \frac{\theta_a^2}{4}{\rm Var}\big[\mathcal{L}_a\big]+
  \frac{\theta_a}{2}{\Exp}\big[\mathcal{L}_a\big],
\end{eqnarray*}
and similarly 
${\rm Var}[S_b]=
  (\theta_b^2/4){\rm Var}\big[\mathcal{L}_b\big]+
  (\theta_b/2)\Exp\big[\mathcal{L}_b\big]$.
Finally we have
\begin{eqnarray*}
  {\rm Corr}[S_a,S_b]=
  \frac{{\rm Cov}[S_a,S_b]}{\sqrt{{\rm Var}[S_a]{\rm Var}[S_b]}}=
  \frac{{\rm Cov}[\mathcal{L}_a,\mathcal{L}_b]}
  {\sqrt{{\rm Var}[\mathcal{L}_a]+
         \frac{2}{\theta_a}{\Exp}[\mathcal{L}_a]}
   \sqrt{{\rm Var}[\mathcal{L}_b]+
         \frac{2}{\theta_b}{\Exp}[\mathcal{L}_b]}}.
\end{eqnarray*}
Note that 
\begin{description}
\item{(i)} The correlation is a separable function of $\theta_a$ and $\theta_b$.
\item{(ii)} The correlation is increasing as a function of $\theta_a$ or $\theta_b$. 
\item{(iii)} ${\rm Corr}[S_a,S_b]<{\rm Corr}[\mathcal{L}_a,\mathcal{L}_b]$ for any $(\theta_a,\theta_b)$.
\item{(iv)} ${\rm Corr}[S_a,S_b]\rightarrow{\rm Corr}[\mathcal{L}_a,\mathcal{L}_b]$ for $\theta_a\rightarrow\infty$ and $\theta_b\rightarrow\infty$.
\item{(v)} ${\rm Corr}[S_a,S_b]\rightarrow 0$ for $\theta_a \rightarrow 0$ or $\theta_b\rightarrow 0$.
\item{(vi)} For $\theta_a=\theta_b=\theta$ we have ${\rm Var}[S_a]={\rm Var}[S_b]$ and
\begin{eqnarray}
  {\rm Corr}[S_a,S_b]=
  \frac{{\rm Cov}[S_a,S_b]}{{\rm Var}[S_a]}=
  \frac{{\rm Cov}[\mathcal{L}_a,\mathcal{L}_b]}
       {{\rm Var}\big[\mathcal{L}_a\big]+
        \frac{2}{\theta}{\rm E}\big[\mathcal{L}_a\big]}.
   \label{Eq:CorrSegr}
\end{eqnarray}
\end{description}
In Figure~\ref{Fig:SegregatingSitesCorr} we show the correlation~(\ref{Eq:CorrSegr}) between the number of segregating sites in two loci for sample sizes $n=(2,4,8)$, mutation rates $\theta=(0.1,0.5,2.5)$ and $\theta \rightarrow \infty$, and as a function of the recombination rate~$\rho$. For $n=2$ and $\theta\rightarrow \infty$ we recover the well known result $(\rho+18)/(\rho^2+13\rho+18)$ (e.g. \cite{wakeley2008coalescent} equation (7.17) page 231).

\section{Discussion}
We have demonstrated how phase-type theory is a useful framework for calculating distributions and summary statistics in basic models in population genetics. The coalescent models that we have analysed are time-homogeneous. The structured coalescent analysed in \cite{KumagaiUyenoyama2015} is another example of a time-homogeneous model that can be explored in the phase-type framework. A future research direction could be to extend the analysis to time-inhomogeneous evolutionary models. \cite{MiSt2017} recently computed the joint distribution of the total branch length in two loci with variable population size. It could be interesting to extend our constant population size analysis in Section \ref{sec:norecomb} and Section~4 to the variable population size model. A first approach could be to consider a piecewise constant population size model, handle each epoch of constant size separately, and finally merge the various epochs. Such an approach requires calculations of moments in end-point conditioned continuous Markov chains, and these can be found using results from \cite{HobolthJensen2011}.   

Another important coalescent model is the isolation-with-migration model with multiple populations (e.g. \cite{Hey2010}). This model is characterized by times in the past where populations merge, and migration rates between the present and ancestral populations. Statistical inference in this model is very challenging, but \cite{LohseEtAl2011} and \cite{LohseEtAl2016} have developed a efficient and general method for likelihood inference using generating functions. Perhaps phase-type theory could provide an alternative framework for robust and reliable parameter estimation in isolation-with-migration models. 
On the other hand, a multiple species coalescent for which phase-type theory is practicable is the simple nested coalescent \cite{SNEC}, although the state-space and the rate matrix are more tedious to be set up.

Statistical inference in phase-type distributions has traditionally been based on observations of the time of absorption of the stochastic process and maximum likelihood inference. This situation is in stark contrast to genetic data which most often consists of DNA sequences from samples of present day individuals. Likelihood inference for coalescent models that have a phase-type structure needs to be developed.

\section*{Acknowledgements}
ASJ is partially supported by CONACyT Grant CB-2014/243068.
We are grateful to Lars N{\o}rvang Andersen, Johanna Bertl, Svend Nielsen, Paula Tataru and Kai Zeng for discussions, comments and suggestions on an earlier version of this manuscript. 

\section*{Supplementary Information}
In the Supplementary Information we provide R code for the reproduction of selected figures in the paper:
Figure~\ref{3moments-length-betaCoalescent} 
(the three first moments of the Beta-coalescent), 
Figure~\ref{fig:SFS-Beta}
(the mean and covariance of the SFS for the Beta-coalescent),
Figure~\ref{Fig:SimonsenChurchill}
(the tree height densities for two loci and two samples),
Figure~\ref{Fig:FourSeqDensities}
(the tree height densities for two loci and four samples),
and finally
Figure~\ref{Fig:SegregatingSitesCorr}
(the correlation between the number of segregating sites in two loci).

\end{document}